\documentclass[12pt]{article}
\usepackage{graphicx}
\usepackage{amsmath}
\usepackage{hhline}
\usepackage{amssymb}
\usepackage{times}
\usepackage{array}
\usepackage{float}
\usepackage{cite}

\newcolumntype{L}[1]{>{\raggedright}p{#1}}
\newcolumntype{R}[1]{>{\raggedleft}p{#1}}

\newlength{\dinwidth}
\newlength{\dinmargin}
\newcommand{\GeV}{{\rm\,GeV}}

\newcommand{\pperp}{T}

\newcommand{\pt}{p_{_{T}}}
\newcommand{\tc}{{\em f+2c}}
\newcommand{\tf}{{\em 2f+c}}

\begin{document}
%\begin{linenumbers}

\begin{titlepage}
\begin{flushleft}
DESY 07-200 \hfill ISSN 0418-9833 \\
November 2007
\end{flushleft}
\vspace*{2cm}

\begin{center}
  \Large
  {\bfseries
Three- and Four-jet Production at Low x at HERA
  }

  \vspace*{1cm}
    {\Large H1 Collaboration}
\end{center}

\begin{abstract}

\noindent
Three- and four-jet production is measured in deep-inelastic $ep$ scattering
at low $x$ and $Q^2$ with the H1 detector using
an integrated luminosity of $44{.}2\ {\rm pb}^{-1}$.
Several phase space regions are selected for the three-jet analysis
in order to study the underlying
parton dynamics from global topologies to the more restrictive regions of forward
jets close to the proton direction.
The measurements of cross sections for events with at least three jets
are compared to fixed order QCD predictions of ${\mathcal{O}}(\alpha_{\rm s}^2)$ and
${\mathcal{O}}(\alpha_{\rm s}^3) $
and with Monte Carlo simulation programs where
higher order effects are approximated by parton showers.
A good overall description
is provided by the 
${\mathcal{O}}(\alpha_{\rm s}^3) $ calculation. 
Too few events are predicted 
at the lowest  $x \sim 10^{-4}$, especially for topologies with two forward jets.
This hints to large contributions at low $x$ 
from initial state radiation of gluons close to the proton direction and unordered
in transverse momentum. 
The Monte Carlo program in which gluon radiation is generated by the
colour dipole model gives a good
description of both the three- and the four-jet data  in absolute normalisation and shape.
\end{abstract}

  \vspace*{1cm}
\begin{center}
Submitted to Eur. Phys. J. {\bf C}
\end{center}
\end{titlepage}
\begin{flushleft}
  %-- H1AUTS Author list by names 
%-- Status: Tue May 15 09:11:45 CEST 2007  Number of authors = 289 

F.D.~Aaron$^{5,49}$,           %BUCH-PD        11/06           Aaron               
A.~Aktas$^{11}$,               %DESY-LEFT      09/06           Aktas               
C.~Alexa$^{5}$,                %BUCH-PD        06/06           Alexa               
V.~Andreev$^{25}$,             %LPI -PD        8/88            Andreev             
B.~Antunovic$^{26}$,           %MPIM-ST        09/03           Antunovic           
S.~Aplin$^{11}$,               %DESY-PD        01/04           Aplin               
A.~Asmone$^{33}$,              %ROME-ST        07/2            Asmone              
A.~Astvatsatourov$^{4}$,       %BRUX-PD        07/04           Astvatsatourov      
S.~Backovic$^{30}$,            %PODG-PD        03/2            Backovic            
A.~Baghdasaryan$^{38}$
% ...

%-- H1 Institutes 
\bigskip{\it
% ...
$ ^{4}$ Inter-University Institute for High Energies ULB-VUB, Brussels;
          Universiteit Antwerpen, Antwerpen; Belgium$^{ c}$ \\
 $ ^{5}$ National Institute for Physics and Nuclear Engineering (NIPNE) ,
          Bucharest, Romania \\
% ...
$ ^{11}$ DESY, Hamburg, Germany \\
% ...
$ ^{25}$ Lebedev Physical Institute, Moscow, Russia$^{ e}$ \\
 $ ^{26}$ Max-Planck-Institut f\"ur Physik, M\"unchen, Germany \\
% ...
$ ^{30}$ Faculty of Science, University of Montenegro,
          Podgorica, Montenegro$^{ e}$ \\
% ...
$ ^{33}$ Dipartimento di Fisica Universit\`a di Roma Tre
          and INFN Roma~3, Roma, Italy \\
% ...
$ ^{38}$ Yerevan Physics Institute, Yerevan, Armenia \\
% ...
$ ^{49}$ Also at Faculty of Physics, University of Bucharest,
          Bucharest, Romania \\
% ...
$ ^c$ Supported by FNRS-FWO-Vlaanderen, IISN-IIKW and IWT
      and  by Interuniversity
Attraction Poles Programme,
      Belgian Science Policy \\
% ...
$ ^e$ Supported by the Deutsche Forschungsgemeinschaft \\
% ...
}
\end{flushleft}
\newpage

\pagestyle{plain}
%
%%%%%%%%%%%%%%%%%%%%%%%%%%%%%%%%%%%%%%%%%%%%%%%%%%%%%%%%%%%%%%%%%%%%%%%%%%%%%%%%%%%%%
\section{Introduction}
%
%%%%%%%%%%%%%%%%%%%%%%%%%%%%%%
%
% HERA low x phasespace intro:
%
%%%%%%%%%%%%%%%%%%%%%%%%%%%%%%
%
%
The HERA electron-proton collider has extended significantly the available
kinematic range for tests of Quantum Chromodynamics. 
The high centre of mass
energy of 319\;GeV allows for deep-inelastic scattering (DIS)
at a large negative four momentum transfer squared 
\mbox{$ Q^2\ge4\,{\rm GeV}^2$}
on partons which carry a very small fraction $x$ of the
proton momentum down to values of $10^{-4}$. 
This is the domain of high parton densities 
in the proton dominated by
gluons and sea quarks. 
In addition, DIS at
low $x$ corresponds to scattering at high 
$\gamma^*p$ centre of mass energies and is therefore
intimately linked to the high energy behaviour of QCD. 
%

\begin{figure}[htb]
\begin{center}
\includegraphics[width=0.40\linewidth]{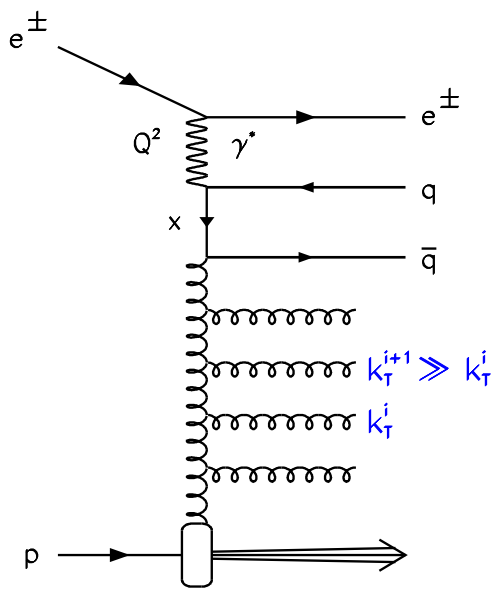}
\caption[]
{Exemplary parton cascade diagram for DIS at low $x$:
in the approximation of the DGLAP leading log $Q^2$ resummation the emitted gluons are strongly ordered
in their transverse momenta $k_T$.
}
\label{fig:gluonl}
\end{center}
\end{figure}

%%%%%%%%%%%%%%%%%%%%%%%%%%%%%%
%
% Collinear Factorisation
%
%%%%%%%%%%%%%%%%%%%%%%%%%%%%%%
%
Many of the available calculations for DIS processes    
make use of collinear factorisation~\cite{dglapref}:
the cross sections are expressed
as a convolution of hard partonic subprocesses
with proton parton density functions (PDFs).
The latter describe the probabilities to find partons
in the proton which carry a fraction $x$ of the proton momentum. 
The separation of the calculation into two pieces is specified
by the factorisation scale $\mu_f^2$: 
initial state radiations from the proton 
%partonic subprocesses 
with virtualities above this
scale are treated in the hard partonic part while
those below are absorbed in the PDFs.
For inclusive DIS, $Q^2$ provides the natural scale for $\mu_f$, i.e.
$\mu_f^2=Q^2$.
The evolution of the PDFs with 
$\mu_f^2$ is generally described by the DGLAP~\cite{dglapref} equations.
To leading logarithmic accuracy this is equivalent to the exchange of a parton
cascade, with the exchanged partons strongly ordered in
virtuality up to $Q^2$.
For low $x$ this becomes approximately an ordering in
$k_T$, the transverse momentum of the partons in the cascade
as shown in figure \ref{fig:gluonl}.
However, at low $x$ the collinear factorisation scheme may break down.
The DGLAP leading logarithmic approximation neglects topologies
of gluon radiation with unordered $k_T$.
These appear in the full perturbative expansion as 
$(\mbox{log}\, 1/x)^n$ terms which are naturally expected to become
large at low $x$.
At very low values of $x$ it is
believed that the theoretically most appropriate description is given by the BFKL
evolution equations~\cite{bfkl}, which resum large logarithms of $1/x$.
The BFKL resummation imposes no restriction on the
ordering of the transverse momenta within the parton cascade. 
Compared to the DGLAP approximation more gluons with sizable transverse momentum 
are  emitted near the proton direction, referred to in the
following as the forward direction. 
This should lead to a significantly increased
rate of forward jets~\cite{mueller}.
A promising approach to parton evolution at low and larger values of $x$ is
given by the CCFM~\cite{ccfm} evolution equation, 
which, because it uses angular-ordered parton emission, is
equivalent to the BFKL ansatz for $x
\rightarrow 0$, while reproducing the DGLAP equations at large $x$.

Higher order calculations in the collinear factorisation scheme can also
improve the treatment for the low $x$ region, 
since the log $1/x$ terms
can be treated up to the given order of $\alpha_s$.
It is an interesting question how well such
fixed order approximations can work 
or if one would still need a full log 1/x resummation to all orders as
provided by the BFKL approach.
For inclusive DIS $ep \rightarrow eX$, QCD 
analyses (see e.g.~\cite{Adloff:2003uh,Chekanov:2005nn,Pumplin:2002vw,Alekhin:2005gq,Martin:2007bv})
were performed using collinear factorisation
calculations in next-to-leading order
(NLO) $\mathcal{O}(\alpha_{\rm s})$\footnote{The notation used throughout this paper 
is that for any calculation labelled $\mathcal{O}(\alpha_{\rm s}^{n})$,
the prediction of a cross section includes all orders up to $n$.}
and/or 
next-to-next-to-leading (NNLO)
$\mathcal{O}(\alpha_{\rm s}^{2})$ for the hard subprocess
with parton densities matched to that order.
These calculations are able to describe the inclusive DIS data from HERA
and fixed target experiments over a large range in $Q^2$ down to 
$Q^2\sim 2\;\mbox{GeV}^2$ from   
the largest to the lowest covered $x$ values.

Final states with jets in DIS are an ideal tool to investigate low $x$ dynamics:
the jets can be used to tag higher order processes and furthermore
provide direct access to the outgoing hard partons.
The H1 and ZEUS measurements of dijet production 
at low $x$~\cite{Breitweg:2001rq,Aktas:2003ja,Aktas:2004px}
and of inclusive forward jet production \cite{Aktas:2005up,Chekanov:2005yb,Chekanov:2007pa} show that the leading order (LO) $\mathcal{O}(\alpha_{\rm s})$
calculations based on DGLAP greatly underestimate the data.
NLO $\mathcal{O}(\alpha_{\rm s}^2)$
calculations can account for some of the LO deficiencies,
but the description remains unsatisfactory at low $x$ and $Q^2$.
%
%%%%%%%%%%%%%%%%%%%%%%%%%%%%%%%%%%%%%%%%%%%%%%%%%%%%%%%%%%%%%%
%
% Three-jet introduction
%
%%%%%%%%%%%%%%%%%%%%%%%%%%%%%%%%%%%%%%%%%%%%%%%%%%%%%%%%%%%%%% 
%
%
In the present paper events with at least three- or four-jets in
the final state are investigated.
%
%Examples of leading order (LO) and next to leading order (NLO) processes
%for three-jets are shown in figure \ref{feynman}.
%
In contrast to inclusive jets and dijets,
three- and four-jet final states require the radiation of at
least one and two hard gluons respectively in addition to the
$q\bar{q}$ pair from the dominating hard boson-gluon-fusion
scattering process $\gamma^{*}g \rightarrow q\bar{q}$
(see figure \ref{feynman}).
Therefore three- and four-jet processes are ideally suited 
to study the gluon emissions and the underlying parton 
dynamics in the proton.
Three-jet cross sections in DIS have been measured previously,
both by the H1\cite{wobisch} and the ZEUS \cite{zeuspaper05,Chekanov:2007dx} collaborations.
In these analyses the leading jets were required to have
a large transverse momentum of at least about $7$ GeV.
All measured cross sections were found to be well described 
by NLO $\mathcal{O}(\alpha_{\rm s}^3)$ predictions in the collinear factorisation scheme.

This paper presents a new measurement of three-jet production. 
The analysis
is performed in an extended phase space, covering jets with
low transverse momenta down to $4\;$GeV, and based on a three times larger
luminosity than used in the previous H1 publication \cite{wobisch}. 
The analysis reaches values of $x$ as low as $x=10^{-4}$.
In addition, 
cross sections for events with at least four jets 
are measured for the first time in DIS.
The data are compared with the NLOJET++ \cite{NLOjet} fixed order calculations
in the collinear factorisation scheme.
This program provides LO 
$\mathcal{O}(\alpha_{\rm s}^2)$ predictions for the three-jet case.
In addition NLOJET++ is the only available program 
which provides perturbative calculations for jet cross sections
in hadronic collisions to $\mathcal{O}(\alpha_{\rm s}^3)$ accuracy. 
This corresponds to NLO and LO precision 
for the three- and four-jets cases, respectively.
The two Monte
Carlo generators RAPGAP \cite{RAPGAP}
and  DJANGOH \cite{DJANGOH},
which were able to describe reasonably well
inclusive forward jet and dijet production at low $x$,
are also tested. 
%
%Both programs implement 2$\rightarrow$2 hard QCD subprocesses (e.g. $\gamma^{*} g \rightarrow q\bar{q}$).
%
%In RAPGAP additional processes are generated using a resolved photon component.
%
%Higher order processes that produce further hard outgoing partons
%are generated in both RAPGAP and DJANGOH by initial and final state parton showers.
%
%In RAPGAP the showers are generated according to
%the leading log $Q^2$ approximation following DGLAP evolution. 
%
%DJANGOH is used with the colour dipole model (CDM) \cite{CDM} to produce
%additional gluon radiation.
%

The sensitivity to deviations from the DGLAP approach
may be increased by selecting kinematic regions
where gluon radiation is  suppressed for this approximation.
This is the case for events with a hard
forward jet and a large separation in rapidity to a central parton system. Two different
subsets of the inclusive three-jet sample are
studied: one sample with one forward jet and two central
jets and one with two forward jets and one central jet. 
 
%%%%%%%%%%%%%%%%%%%%%%%%%%%%%%%%%%%%%%%%%%%%%%%%%%%%%%%%%%%%%%%%%%%%%%%%%%%%%%%%%%%%%
%\input{kinematics}
%%%%%%%%%%%%%%%%%%%%%%%%%%%%%%%%%%%%%%%%%%%%%%%%%%%%%%%%%%%%%%%%%%%%%%%%%%%%%%%%%%%%%
\section{Kinematics and Measurement Observables}
\label{sec:kinobs}
\begin{figure}[htb]
%\begin{center}
%\hspace*{0.1\linewidth}
%$\mathcal{\pmb  O}(\pmb \alpha_{\rm \bf s}^2)$\hspace*{0.4\linewidth}$\mathcal{\pmb O}(\pmb \alpha_{\rm\bf  s}^3)$
%\end{center}
%\vspace*{-0.1\linewidth}
\begin{center}
\includegraphics[width=0.40\linewidth]{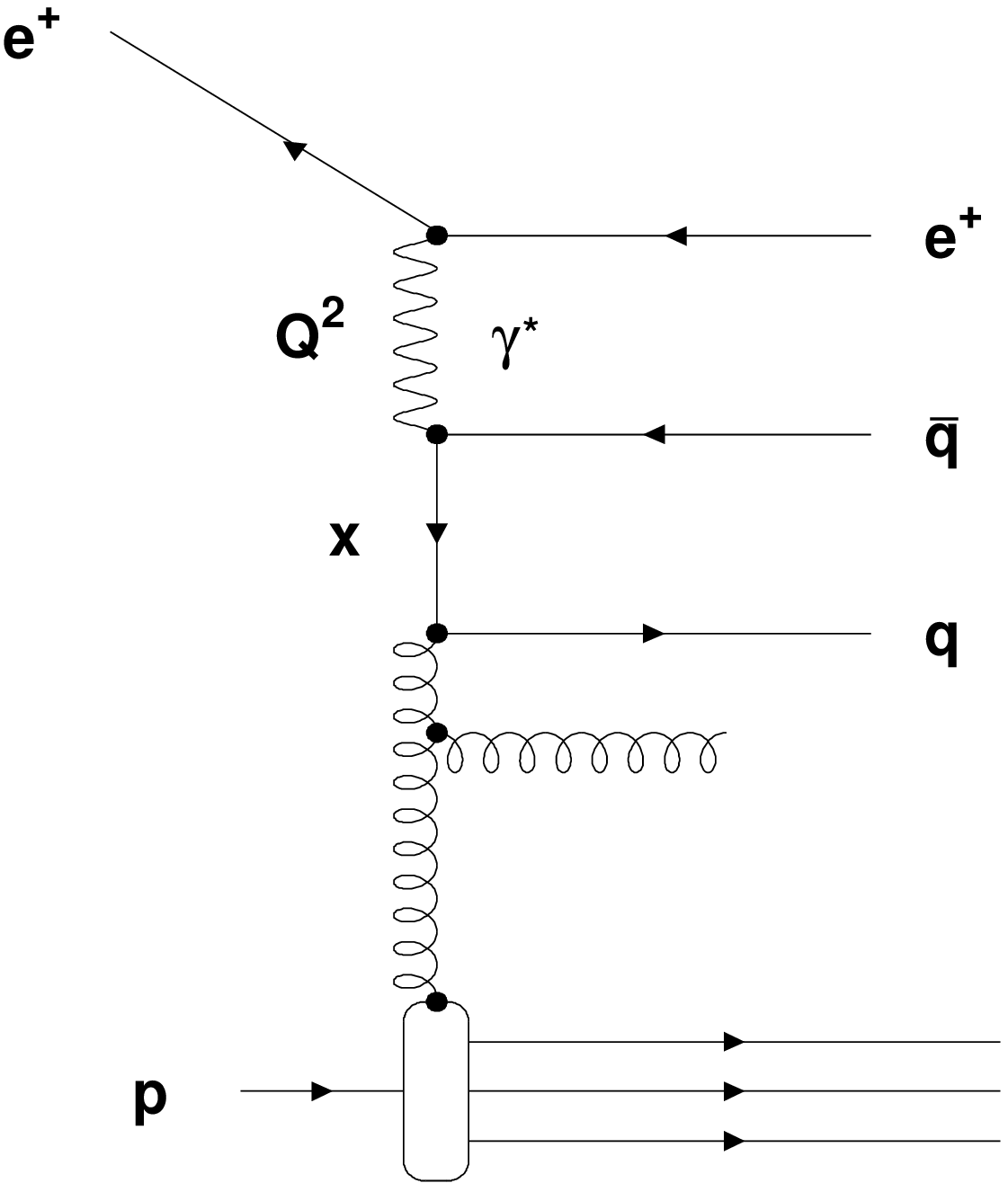}
\hspace*{5mm}
\includegraphics[width=0.40\linewidth]{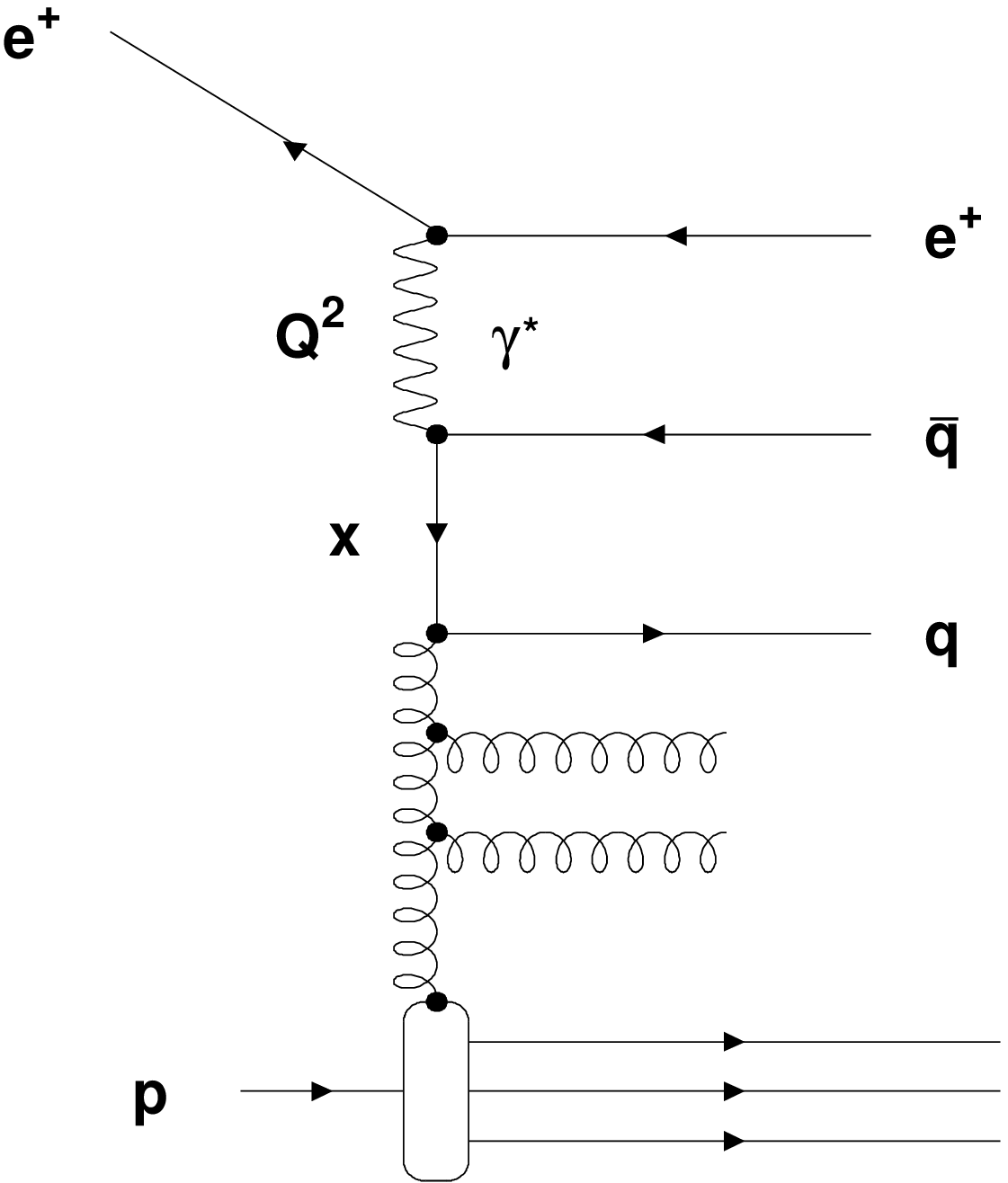}
\vspace{0.5 cm}
\caption[]
{Examples of leading order (left) and next to leading order (right) diagrams for three-jet
production in DIS  at HERA with one and two radiated gluons, respectively. 
}
\label{feynman}
\end{center}
\end{figure}

Figure \ref{feynman} shows two examples of
DIS processes dominating the production of
three or more jets at low $x$.
The diagrams contribute to order
$\alpha_{\rm s}^2$ and $\alpha_{\rm s}^3$ to the cross section, 
respectively. The radiated
gluons are predominantly emitted
in the forward direction whereas the quarks from the hard scattering process are
mostly central. 
%
%(i.e. have pseudorapidities around zero).
%

\begin{figure}[htb]
\begin{center}
\includegraphics[width=0.50\linewidth]{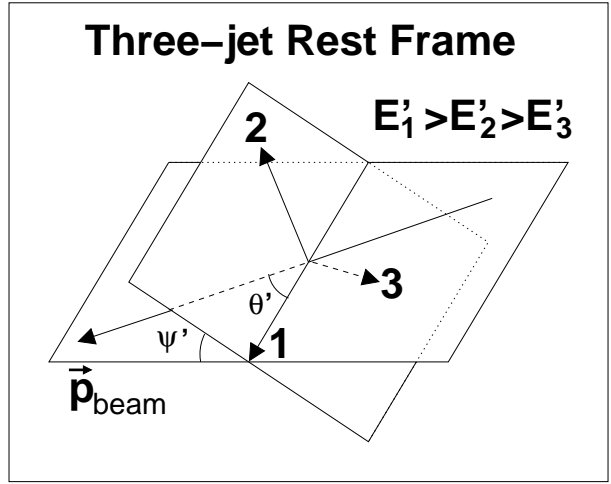}
\vspace{0.5 cm}
\caption[]
{Definition of the angles $\theta'$ and $\psi'$
in the three-jet rest system
\cite{tevatron}. The 3-vector $\vec{p}\;'_{\rm beam}$ is defined by $\vec{p}\;'_{\rm beam}=\vec{p}\;'_{{\rm in},\;1}-
\vec{p}\;'_{{\rm in},\;2}$ where $\vec{p}\;'_{{\rm in},\;1}$
and  $\vec{p}\;'_{{\rm in},\;2}$
are the 3-momenta of the two incoming
 interacting particles
in the three-jet centre of mass frame.
They are sorted with respect to their energy in the laboratory frame: $E_{{\rm in},\;1}>E_{{\rm in},\;2}$.
The incoming interacting particles are the exchanged virtual boson and
the parton from the proton side (predominantly a gluon). The latter
is assumed to move parallel with
the proton and to carry a fraction of its momentum reconstructed as
$x_{\rm gluon}=x\cdot(1+(\hat{s}/Q^2))$, where $\hat{s}$
denotes the squared centre of mass energy of the 
three-jet system.}
\label{Tevatronvar}
\end{center}
\end{figure}

The kinematic variables which describe the hard electron-quark scattering process
are the negative four
momentum transfer squared $Q^2=-q^2=(k-k')^2$ of
the exchanged virtual photon ($\gamma^{*}$), the Bj{\o}rken variable
$x = Q^2/(2pq)$, and the inelasticity $y=(qp)/(kp)$, 
where $k$, $k'$, $p$ and $q$ denote the four momenta of the
incoming and outgoing positron, the incoming proton and the exchanged photon, respectively.
The three variables are related by $Q^2=xys$, where $s$ denotes the fixed
$ep$ centre of mass energy squared.
Jets are defined in this analysis in the
$\gamma^*p$ centre of mass
system. The observables used to characterise the jets
are their transverse momentum $p_\pperp^*$
in the $\gamma^*p$ centre of mass frame and their
pseudorapidity $\eta$ 
in the laboratory system. The 
topology of a three-jet system
is fully specified by the following four canonical variables~\cite{tevatron}:
the scaled energy of the
jets $X_i' = 2E_i'/(E_1'+E_2'+E_3')$ ($i=1,2$; $E_1'>E_2'>E_3'$) and the two three-jet angles $\theta'$ and $\psi'$ as defined in figure~\ref{Tevatronvar}. These variables are measured in the three-jet centre of mass frame.
%%%
Four-jet events have additional degrees of freedom.
The two jets with the lowest dijet mass are combined in order to use
the same variables as in the three-jet case.

\section{QCD Predictions}
The  RAPGAP~\cite{RAPGAP} and DJANGOH~\cite{DJANGOH}
Monte Carlo event generator programs are used in this analysis
to estimate the corrections that must be applied to the data for
the finite acceptance, efficiency and resolution of the detector.
The two programs are also used
to provide predictions that can be compared with the data.
Both programs generate
hard QCD 2$\rightarrow$2
subprocesses (e.g. $\gamma^{*} g \rightarrow q\bar{q}$) which are convoluted
with the CTEQ5L \cite{cteq5} set of parton distributions for the proton.
The factorisation and renormalisation
scales are set to
$\mu_{\rm f}^2=\mu_{\rm r}^2=Q^2$ for DJANGOH and
\ $\mu_{\rm f}^2=\mu_{\rm r}^2=Q^2+\hat{p}_\pperp^2$ for RAPGAP,
where $\hat{p}_\pperp$ is the transverse momentum of the outgoing hard partons.
RAPGAP includes
resolved photon processes using the SaS 2D
photon parton distribution functions \cite{SAS-D}, which were
found to give a good description of the effective photon structure
function as measured by H1~\cite{H1resolved}.
Higher order QCD effects that produce 
further hard outgoing partons 
are generated in both RAPGAP and \mbox{DJANGOH} 
by parton showers: in RAPGAP the showers are 
ordered in the transverse momenta ($k_\pperp$) of the emissions, according
to the DGLAP leading log $Q^2$ approximation.
DJANGOH uses the colour dipole model (CDM) \cite{CDM}, in which partons
are generated by colour dipoles,
spanned between the partons in the cascade. Since
the dipoles radiate independently, there is no $k_\pperp$ ordering.
For the hadronisation, the Lund string fragmentation \cite{lund} is used 
both for RAPGAP and DJANGOH. 
QED radiative corrections are 
applied in DJANGOH using the HERACLES \cite{Heracles} program
and are neglected for RAPGAP. 
The DJANGOH predictions are referred to
as CDM in the following.

Fixed order QCD predictions at parton level are calculated using the NLOJET++ \cite{NLOjet} program, which
is able to predict three-jet cross sections in LO $\mathcal{O}(\alpha_{\rm s}^2)$ or
NLO $\mathcal{O}(\alpha_{\rm s}^3)$ and
 four-jet cross sections in LO $\mathcal{O}(\alpha_{\rm s}^3)$.
The renormalisation and factorisation scales are set to
\begin{equation*}
\mu_{\rm r}=\mu_{\rm f}=\frac{1}{m}\sum\limits_{i=1}^{N_{\rm jet}}{p^*_\pperp}_i,
\end{equation*}
with $m=3$ for the three-jet and $m=4$ for the four-jet cross sections and $N_{\rm jet}$ being the
number of jets fulfilling the applied jet cuts. The
value of $\alpha_{\rm s}(M_{Z})$ is fixed to 0.118 and the 
CTEQ6M~\cite{Pumplin:2002vw} proton parton density parameterisations are used.
The NLOJET++ parton level cross sections are corrected bin-by-bin
for hadronisation effects using the CDM simulation as discussed in detail in section~\ref{sec:xsec}.
Two uncertainties are considered for the NLOJET++ cross sections:
The uncertainty due to missing higher orders is
estimated by recalculating the cross sections with the scales $\mu_f$
and $\mu_r$ varied by a common factor of 2 or 0.5.
Hadronisation uncertainties are estimated by determining
the corrections to the hadron level alternatively with RAPGAP 
and taking 50\% of the difference between the corrections from CDM and RAPGAP as
systematic error.
%
%
%%%%%%%%%%%%%%%%%%%%%%%%%%%%%%%%%%%%%%%%%%%%%%%%%%%%%%%%%%%%%%%%%%%%%%%%%%%%%%%%%%%%%
%\input{expproc}
%%%%%%%%%%%%%%%%%%%%%%%%%%%%%%%%%%%%%%%%%%%%%%%%%%%%%%%%%%%%%%%%%%%%%%%%%%%%%%%%%%%%%
\section{Experimental Procedure}
\subsection{The H1 Detector} 
A detailed description of the H1 detector can be found in  \cite{H1det1}.
Here, a brief account of the components most relevant to the present 
analysis is given. The H1 coordinate system convention defines the
outgoing proton beam direction as the positive $z$ axis, also referred to as the `forward'
direction.
The polar angle $\theta$ is
defined with respect to this direction. The pseudorapidity is given
by $\eta=-\ln\tan(\theta/2)$.

The central {\it ep\/} interaction region is surrounded by two large concentric drift
chambers (CJCs), operated inside a $1{.}16\,{\rm T}$ solenoidal magnetic field. 
Charged particles are measured 
in the pseudorapidity range
$-1.5< \eta <1.5$ with a transverse momentum resolution of 
\mbox{$\sigma(\pt) / \pt \simeq 0.005 \, \pt \, /\GeV \, \oplus 0.015$.}
%~\cite{Kleinwort}.
Two additional drift chambers (CIZ, COZ) complement the CJCs by precisely
measuring the $z$ coordinates of track segments and hence improve the
determination of the polar angle.
Multi-wire proportional chambers (MWPC) provide fast signals for triggering
purposes.

A finely segmented electromagnetic
and hadronic liquid argon calorimeter (LAr) covers the
range $-1.5 < \eta < 3.4$. The energy resolution is
$\sigma(E)/E=0.11/\sqrt{E/{\rm GeV}}$
 for electromagnetic showers and $\sigma(E)/E=0.50/\sqrt{E/{\rm GeV}}$ for hadrons, as
measured in test beams \cite{testbeam}. A lead/scintillating fibre calorimeter (SpaCal, \cite{spacal}) 
covers the backward region $-4 < \eta < -1.4$.

The data sample of this analysis was collected using a trigger which requires the
scattered positron to be measured in the SpaCal,
at least one high transverse momentum track ($p_\pperp>800\ {\rm MeV}$)
to be reconstructed in the central tracking chambers and an event vertex
to be identified by the MWPCs. The trigger efficiency is higher than
85\% for the whole analysis phase space.

The {\it ep\/} luminosity is measured via the Bethe-Heitler Bremsstrahlung
process {\it ep\/} $\rightarrow$ {\it ep} $\gamma$, the final state photon being detected in
a crystal calorimeter at $z=-103$ m.

%%%%%%%%%%%%%%%%%%%%%%%%%%%%%%%%%%%%%%%%%%%%%%%%%%%%%%%%%%%%%%%%%%%%%%%%%%%%%%%%%%%%%
%\input{reconstruction}
%%%%%%%%%%%%%%%%%%%%%%%%%%%%%%%%%%%%%%%%%%%%%%%%%%%%%%%%%%%%%%%%%%%%%%%%%%%%%%%%%%%%%
\subsection{Event Selection and Kinematic Reconstruction}
A detailed account of this analysis can be found in \cite{cwerner}.
The data used in this analysis were taken in the 1999 and 2000 running
periods, in which HERA collided protons with an energy of $920\ {\rm GeV}$ with $27{.}5\ {\rm
GeV}$
positrons, corresponding to a centre of mass energy of $\sqrt{s}=319\;{\rm GeV}$.
The integrated luminosity of the data is 44.2 pb$^{-1}$.
DIS events are preselected requiring a scattered positron measured in the SpaCal
with an energy $E_e>9\;{\rm GeV}$.
The polar angle $\theta_e$ of the scattered positron is determined
from the cluster position
in the SpaCal and the $z$ position of the event vertex reconstructed
with the central tracking chambers.
The observables $y$, $Q^2$ and $x$ are derived from the electron kinematics

\begin{equation*}
y = 1-\frac{E_e}{E_{e,\,{\rm beam}}}\sin^2\frac{\theta_e}{2}\;,
\qquad Q^2=4E_{e,\,{\rm beam}}E_e\cos^2\frac{\theta_e}{2}\;,\qquad x=\frac{Q^2}{ys}\;,
\end{equation*}
where $E_{e,\,{\rm beam}}$ is the positron beam energy.
The kinematic range is chosen
to be $5\ {\rm GeV}^2<Q^2<80\ {\rm GeV}^2$, $0.1 < y < 0.7$, $ 10^{-4}<x<10^{-2}$ and
$ 156^\circ<\theta_e<175^\circ$.

The hadronic system, containing the jets,
is measured with the LAr and SpaCal calorimeters
and the central tracking system. Calorimeter cluster energies and
track momenta are combined using algorithms which avoid double
counting \cite{combhfs}.
Jets are formed from the hadronic final state particles boosted to the 
$\gamma^*p$ rest frame.
The inclusive $k_\pperp$ cluster algorithm \cite{inclkt} is applied
with a separation parameter of $1{.}0$. 
The $p_\pperp$ weighted recombination scheme is used in which the jets are treated 
massless. 
The jets are ordered with respect to their transverse momentum in the
$\gamma^*p$ rest frame (${p^*_\pperp}_i>{p^*_\pperp}_{i+1}$).
Only jets with a transverse momentum  ${p_\pperp^*}_i$ of at least 4 GeV and a pseudorapidity
in the range $-1<\eta_i< 2{.}5$ are considered for the analysis.
The latter cut ensures the jets to lie
well within the acceptance of the LAr calorimeter. 
At least three jets are required which fulfil these cuts.
It is demanded in addition that
${p_\pperp^*}_1+{p_\pperp^*}_2>9\ {\rm GeV}$.
The applied cuts ensure a good correlation between jets at detector level
and hadron or parton level and allow for comparison of the data to
the NLO $\mathcal{O}(\alpha_{\rm s}^3)$ calculation.
In addition to the above selections one of the three leading ${p_\pperp^*}$ jets
has to lie in the central
region $-1<\eta_i<1{.}3$.
This ensures a good trigger efficiency. 
After all cuts, 38400 events are selected
with at least three jets. 
5900 of these events contain at least four jets.
%
%\input{xsmeasurement}
%%%%%%%%%%%%%%%%%%%%%%%%%%%%%%%%%%%%%%%%%%%%%%%%%%%%%%%%%%%%%%%%%%%%%%%%%%%%%%%%%%%%%

\subsection{Cross Section Measurement}
\label{sec:xsec}
%\vspace{0.5cm}
\setlength{\tabcolsep}{0.5cm}
\begin{table}                                                                  
\begin{center}
{\textbf{Phase Space Definition}}\\[1em] 
\begin{tabular}{|c|}
\hline
$0.1 < y < 0.7$ \\
$ 5 \,{\rm GeV}^2 < Q^2 < 80 \,{\rm GeV}^2$ \\
$ 10^{-4}<x<10^{-2}$\\
$ 156^\circ<\theta_e<175^\circ$\\
$ E_e>9\,{\rm GeV}$\\
\hline
$N_{\rm jet} \ge 3$\\
${p_\pperp^*}_i > 4\ {\rm GeV}$ \\
${p_\pperp^*}_1 +{p_\pperp^*}_2>9\ {\rm GeV}$ \\
$-1 < \eta_i < 2.5$\\
\hline
at least one jet in the range \\
 $-1 < \eta_i < 1.3$ \\
\hline
\end{tabular}
\end{center}
\caption[Definition of Cross Sections]{The kinematic domain in which the
cross sections are measured.}
\label{tab:xsdef}
\end{table}
%\vspace{0.5cm}
The kinematic region for which the cross sections are measured
is given in table \ref{tab:xsdef}.
All cross sections are given as bin-averaged differential 
cross-sections defined at the level
of stable hadrons. 
Therefore the data are corrected for all detector effects, using 
Monte Carlo simulations.
For each generated event
the response of the H1 detector is simulated in detail 
including trigger effects. The events
are then subjected to the same reconstruction and analysis programs as the data. 
For each measurement bin
a correction factor is calculated as the ratio
of simulation entries at stable hadron level to that at detector level.
The same inclusive $k_\pperp$ algorithm is applied at the hadron and detector levels.
The detector correction factors are determined using the 
CDM simulation which is found to give a better description of the jet topologies than RAPGAP.
The Monte Carlo events are weighted in a few 
variables to adjust their kinematic distributions to the data. These variables are the $p^*_{\pperp}$ of the
leading jet,
$\eta_1 - \eta_2$, $\eta_1 + \eta_2 $ and $Q^2$.
After weighting the simulations provide a reasonable description of the shapes of all data  distributions.
The detector correction factors have been studied
in detail for all distributions. 
They vary between $0{.}6$ and $1{.}2$ for events with at least three-jets ($0{.}4$ and
$1{.}2$ for events with at least four-jets) and show a smooth behaviour.
Further small corrections are applied to the data to take QED radiative effects into account.
The data are corrected to the QED Born level using the CDM simulation.
A correction factor is determined for each measurement bin separately.

For comparison with the data, the fixed order NLOJET++ parton level calculations 
are corrected to the stable hadron level
by application of hadronisation correction factors $c_{\rm had}$.
These corrections are estimated bin-by-bin using the weighted CDM simulation.
Jets are obtained at the parton level 
using the inclusive $k_\pperp$ algorithm, 
both in NLOJET++ and CDM. 
For CDM the algorithm is applied to the
partons after the parton showering step.
As just mentioned,
the detector and hadronisation corrections are calculated using the weighted
Monte Carlo simulation events.
However, the unweighted Monte Carlo predictions are compared to the data, as will
be shown in section \ref{results}.

The correlations between the jets at 
the different levels have been studied in detail using Monte Carlo simulated events.
According to CDM, for the phase space given in table \ref{tab:xsdef},
\mbox{73-85\%} of the selected detector level jets
can be associated with a hadron level jet 
within a cone $\Delta R \ = \ \sqrt(\Delta \eta ^2 + \Delta \phi ^2) \le \ 0.4$
around the detector level jet and with a hadron level transverse momentum 
of $p_\pperp^* \ge 1.5 \;{\rm GeV}$.
This fraction of ''matched'' detector jets decreases to 65-75\% at the parton level.
The lowest matching fractions are observed for the more forward jet pseudorapidities.
These migrations dilute the 
interpretability of the data in terms of the underlying
partonic dynamics and must be well controlled.
The fraction of unmatched jets observed in RAPGAP 
agrees with CDM to better than 30\% everywhere.
Taking the differences from RAPGAP and CDM 
as the absolute uncertainty of the number of unmatched jets and
assuming this number to directly propagate into the measured
cross sections a maximal possible 
cross section error of 25\% is derived.
This possible error will be only considered in the discussion of the results
(section \ref{results}), whenever a large excess of data over NLO prediction 
is observed.
It is not included in the standard determination 
of systematic cross section errors 
which is described in the following.

\subsubsection{Estimate of Systematic Errors}
The errors of the measured differential cross sections
are separated into statistical errors of the data $\delta _{\rm stat}$, 
additional uncorrelated errors $\delta _{\rm uncorr}$, accounting for the
statistical errors of the Monte Carlo samples used to determine the various correction
factors, and systematic errors.
The latter are separated into two contributions: 
a global normalisation error $\delta_{\rm norm}$
and a correlated error  $\delta _{\rm corr}$ which affects only the shape of the
cross section distributions.

The effects of systematic uncertainties on the cross sections are evaluated
by applying appropriate variations to the Monte Carlo simulations.
The following sources of error are considered:
%\setitemindent{3}
\begin{itemize}
\item LAr hadronic energy scale: 
The absolute hadronic energy scale of the LAr calorimeter is known
to 4\% accuracy. This is the dominating uncertainty for the 
determination of the energy of the jets studied in this analysis.
\item SpaCal electromagnetic energy scale: The energy of the scattered positron
is known within a 2\% uncertainty.
 \item Positron angle measurement: The uncertainty in the measurement of the
polar angle of the scattered positron is $1\;{\rm mrad}$.
\item Track contribution to combined objects: The uncertainty of
this contribution is estimated by varying
the momenta of all contributing tracks by $\pm3\%$.
\item Trigger efficiency: The simulated trigger efficiencies are compared
with the efficiencies determined from data, using monitor trigger
samples. Agreement is found within 3\%.
\item Luminosity measurement: The measurement of the integrated luminosity is accurate within $1{.}5\%$.
\item Radiative correction: The uncertainty of the radiative correction factors is estimated to
be 2\% \cite{radiative}.
\item Model uncertainty: The cross sections are corrected to hadron level using 
the weighted CDM simulation. The uncertainty of these
corrections is estimated by calculating the correction factors with 
the weighted RAPGAP simulation and taking 50\% of the difference to CDM
as systematic uncertainty.
\end{itemize}
Typical values for the correlated uncertainties and the global normalisation
error on the cross sections for events with at least three jets
are given in table \ref{systerrors}.
\begin{table}
\renewcommand{\arraystretch}{1.24}
\setlength{\tabcolsep}{1mm}
\begin{center}
\begin{tabular}{|l||r|r|r|}
\hline
Error source & \multicolumn{1}{c|}{Systematic variation} & \multicolumn{1}{c|}{$\delta_{\rm corr}$} & \multicolumn{1}{c|}{$\delta_{\rm norm}$}\\
\hline
\hline
{LAr hadronic energy scale} & $\pm4\%$ & {$\pm3\%$} & {$^{+14}_{-12}\%$}\\
{SpaCal em energy scale} & $\pm2\%$ & {$\pm2\%$} & $^{+0{.}1}_{-0{.}8}\%$\\
Track contribution to comb.\ obj. & $\pm3\%$ & $\pm1\%$ & $\pm2{.}5\%$\\
Positron angle & $\pm1\;{\rm mrad}$ & $\pm0{.}5\%$ &$\pm1{.}3\%$\\
Trigger efficiency & $\pm3\%$ & $<0{.}1\%$ & $\pm3\%$\\
Luminosity & $\pm1{.}5\%$ & $<0{.}1\%$ & $\pm1{.}5\%$\\
Radiative correction & $\pm2\%$ & $<0{.}1\%$ & $\pm2\%$\\
{Model uncertainty} & $\pm$50\%{\raisebox{0.195mm}{\footnotesize (CDM-RAPGAP)}} & {$\pm8\%$} & {$^{+6}_{-5}\%$}\\
\hline
Sum & & & $^{+16}_{-14}\%$\\
\hline
\end{tabular}
\end{center}
\caption{Sources of correlated systematic uncertainties and the resulting
errors on the cross sections for events with at least three jets. 
The first column contains the
error source and the second the
range in which the quantity was varied to account for its uncertainty.
The two remaining 
columns give the typical correlated ($\delta_{\rm corr}$)
and global normalisation ($\delta_{\rm norm}$)
uncertainties on the cross sections which arise
from each error source.
\label{systerrors}}
\end{table}
%
%
% xxx why does latex produce extra horizontal space here?
The systematic errors are dominated by the LAr hadronic energy scale.
The second largest contribution stems from the model uncertainty.
The total global normalisation error is $^{+16}_{-14}\%$.
The systematic uncertainties for the cross sections of events with
at least four jets are
found to be of similar or somewhat larger size as those for events with at least 
three-jets; the total
global normalisation error is $^{+22}_{-19}$\%.

%%%%%%%%%%%%%%%%%%%%%%%%%%%%%%%%%%%%%%%%%%%%%%%%%%%%%%%%%%%%%%%%%%%%%%%%%%%%%%%%%%%%%
\section{Results}
The measured cross sections are shown in the figures \ref{fig:njet} to \ref{fourjet} 
and listed in the tables \ref{tab:3jet_1} to\ref{tab:4jet}.

\label{results}
\subsection{Cross Sections for Events with at Least Three Jets}
\label{sec:threej}
\begin{figure}[htbp]
\begin{center}
\hspace*{14mm}
\includegraphics[bb=103 70 493 529,width=6cm,keepaspectratio]{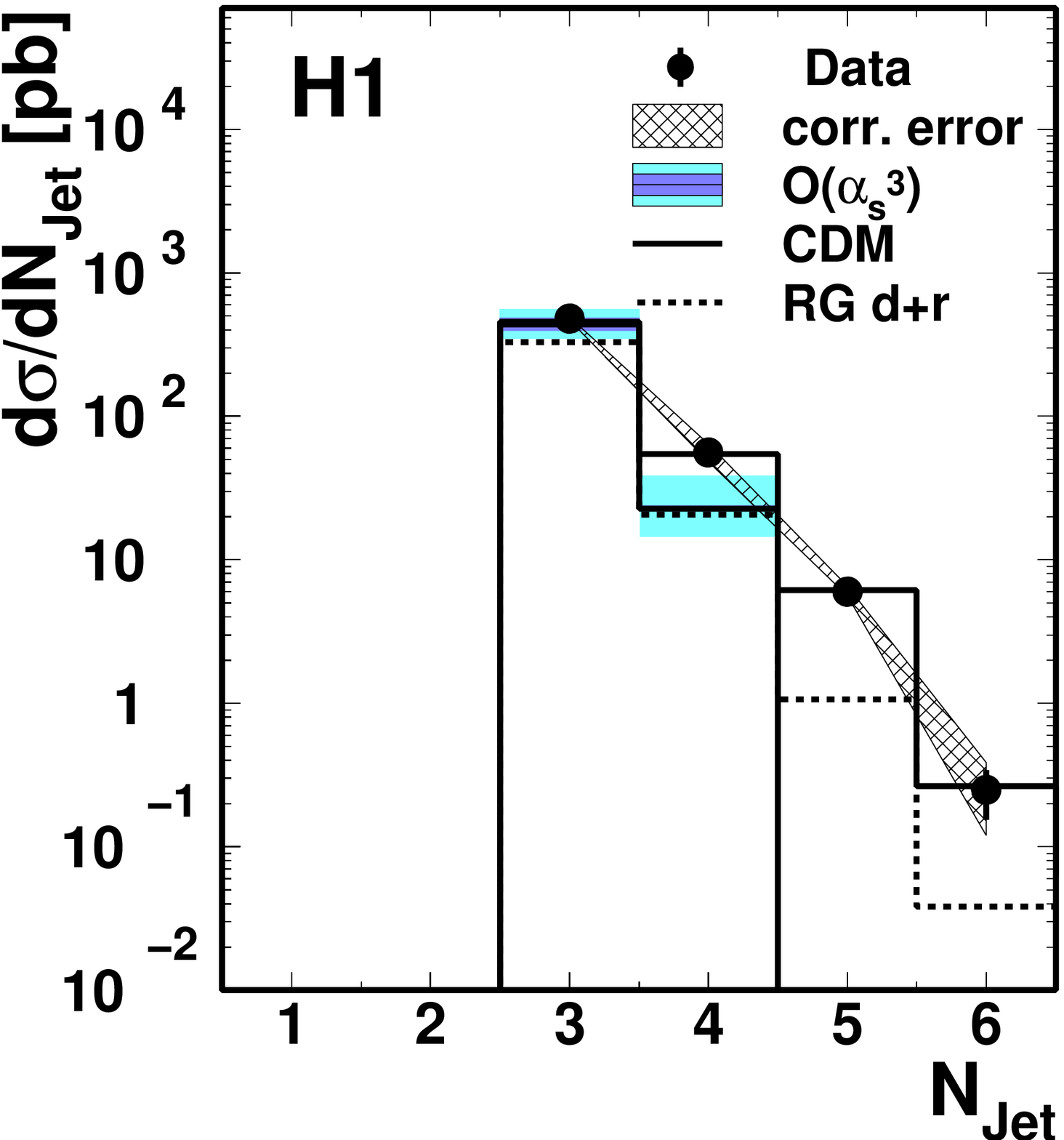}
\hspace*{20mm}
\includegraphics[bb=95 54 484 512,width=6cm,keepaspectratio]{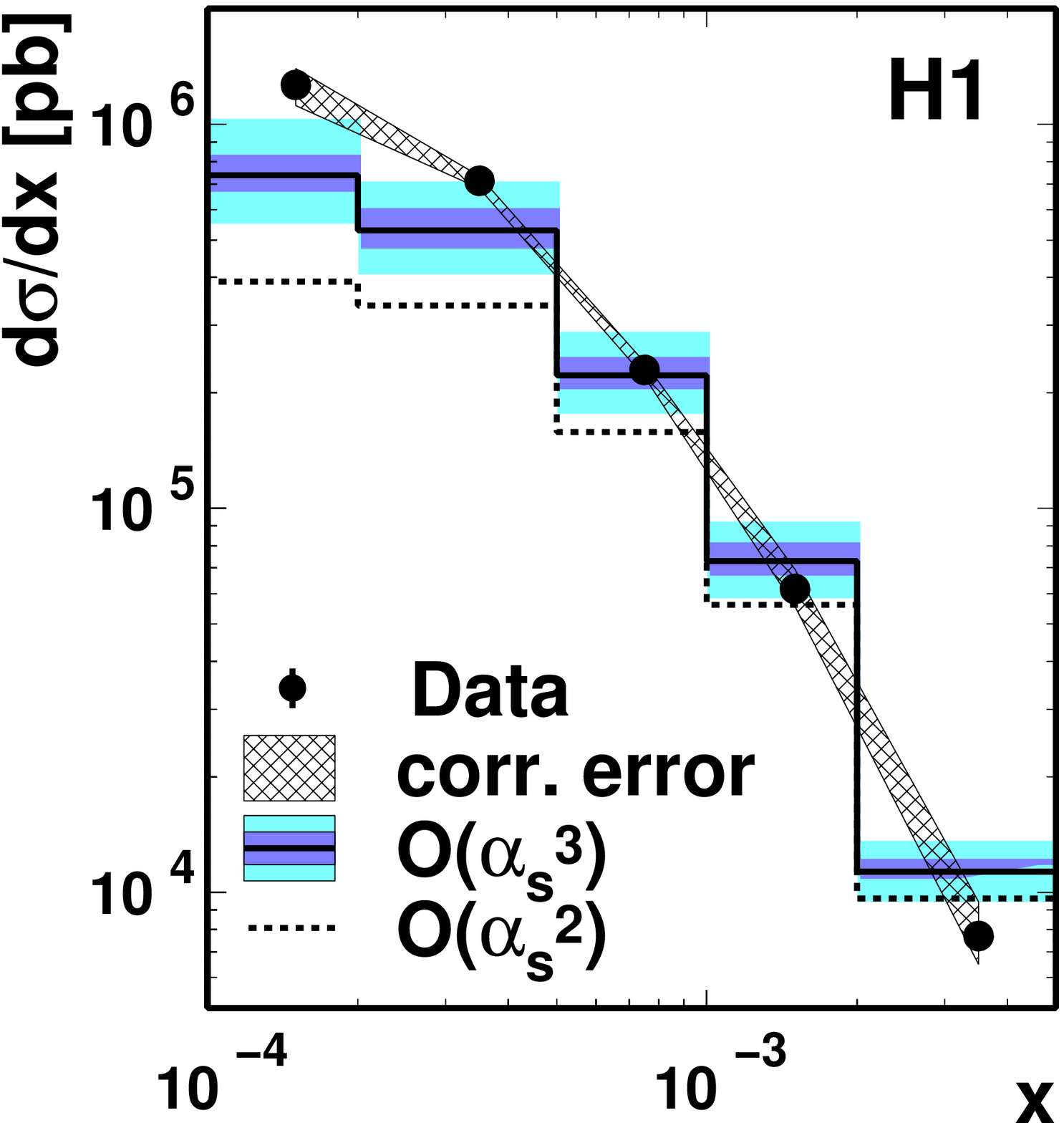}
\end{center}
\vspace*{11mm}
\caption{
Differential cross sections as a function of the number of jets $N_{\rm jet}$ 
found in the events and the Bj{\o}rken
scaling variable $x$.
The results are obtained for the selected events with at least three jets
in the kinematic range listed in table \ref{tab:xsdef}.
The cross sections are bin-averaged and plotted at the respective bin centers.
The inner error bars represent the 
statistical
 error of the data, the total error bars
correspond to the statistical and uncorrelated systematic uncertainties added in quadrature. 
The hatched
 error bands show the estimate of the correlated systematic uncertainties. The data have an 
additional
 overall
global normalisation error of $^{+16}_{-14}$\% (not shown). The dark shaded (inner) error band
 shows the NLO $\mathcal{O}(\alpha_{\rm s}^3)$ prediction
with the
uncertainty due to the hadronisation corrections,
the light shaded (outer) band includes the scale uncertainty added in quadrature. The
dashed line represents the LO $\mathcal{O}(\alpha_{\rm s}^2)$ prediction. The latter is not
shown in the
$N_{\rm jet}$ distribution which instead is compared with
the two Monte Carlo programs RAPGAP (direct + resolved, dashed line, labelled as 'RG d+r') and
 CDM
(solid  line).}
\label{fig:njet}
\end{figure}

\begin{figure}[htbp]
\begin{center}
\includegraphics[width=0.478\linewidth,keepaspectratio]{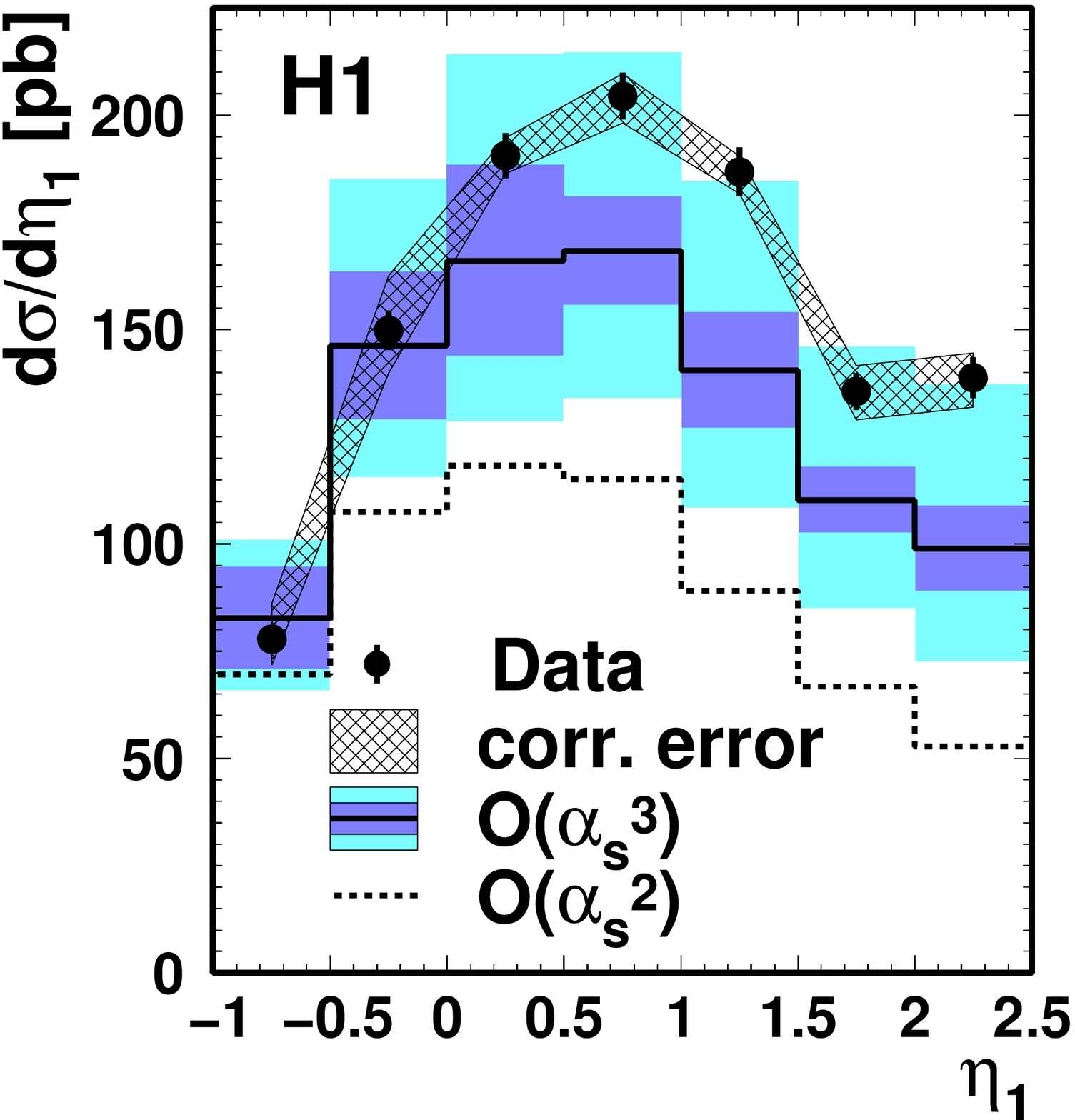}
\hspace*{5mm}
\includegraphics[width=0.478\linewidth,keepaspectratio]{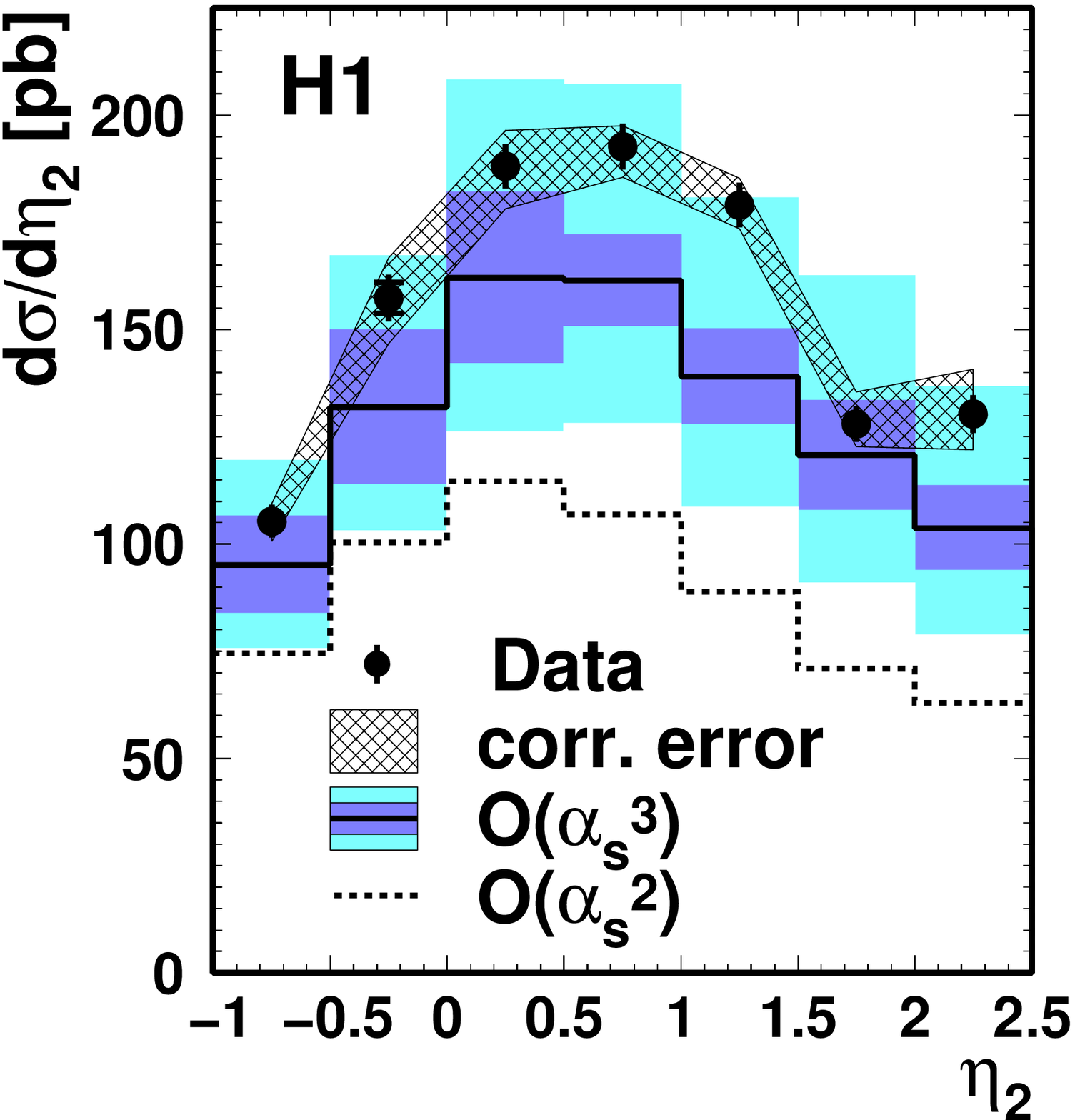}
\end{center}
\begin{center}
\includegraphics[width=0.478\linewidth,keepaspectratio]{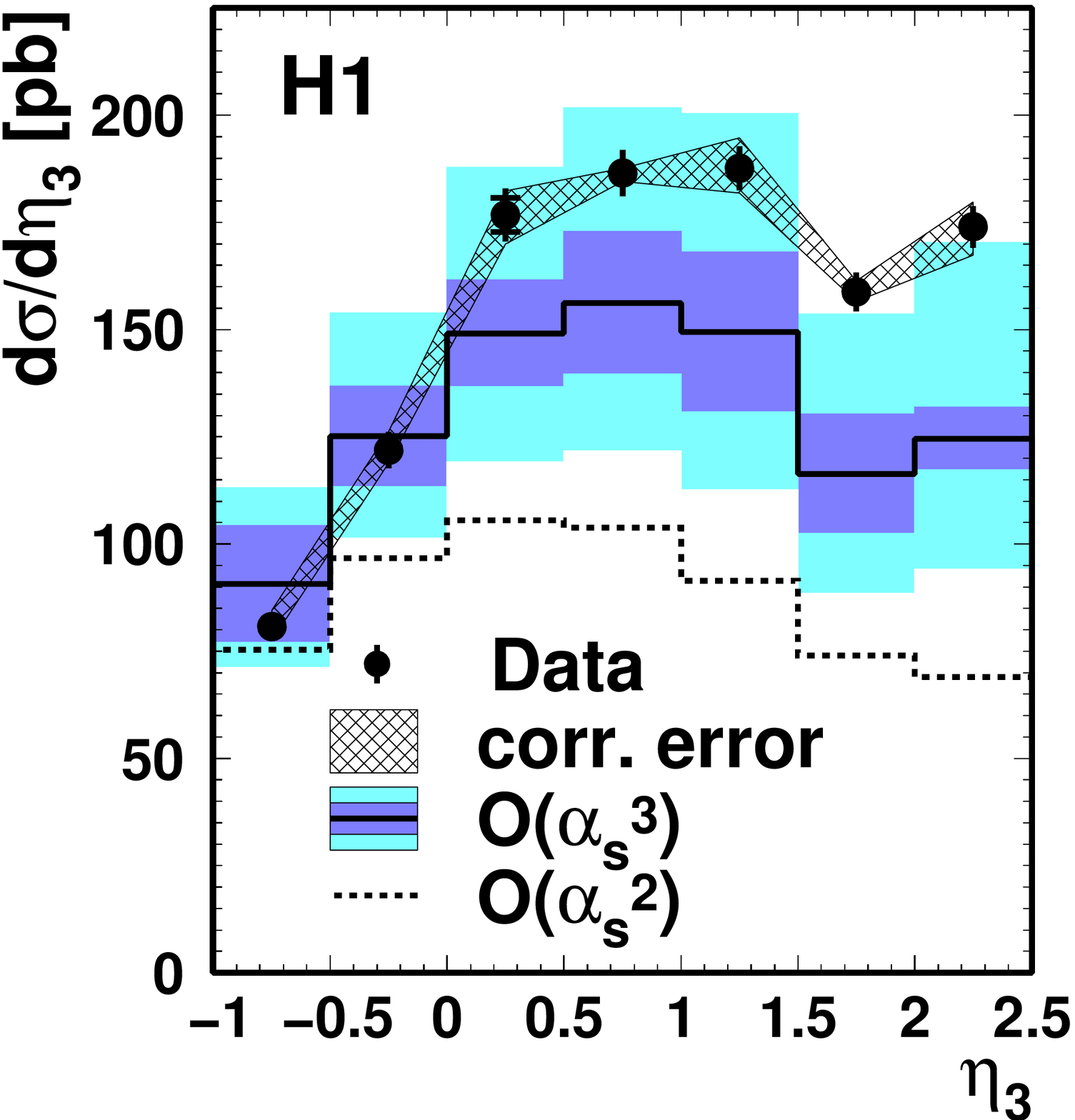}
\hspace*{5mm}
\hspace*{0.478\linewidth}
\end{center}
\caption{Differential cross sections as a function of the pseudorapidity $\eta_i$
for each of the three leading jets
(${p^*_\pperp}_1>{p^*_\pperp}_2>{p^*_\pperp}_3$).
The data are
compared to the LO $\mathcal{O}(\alpha_{\rm
s}^2)$ and to the NLO $\mathcal{O}(\alpha_{\rm s}^3)$
predictions.
See the caption of figure \ref{fig:njet} for further details.}
\label{fig:eta}
\end{figure}

\begin{figure}[htpb]
\begin{center}
\includegraphics[width=0.478\linewidth,keepaspectratio]{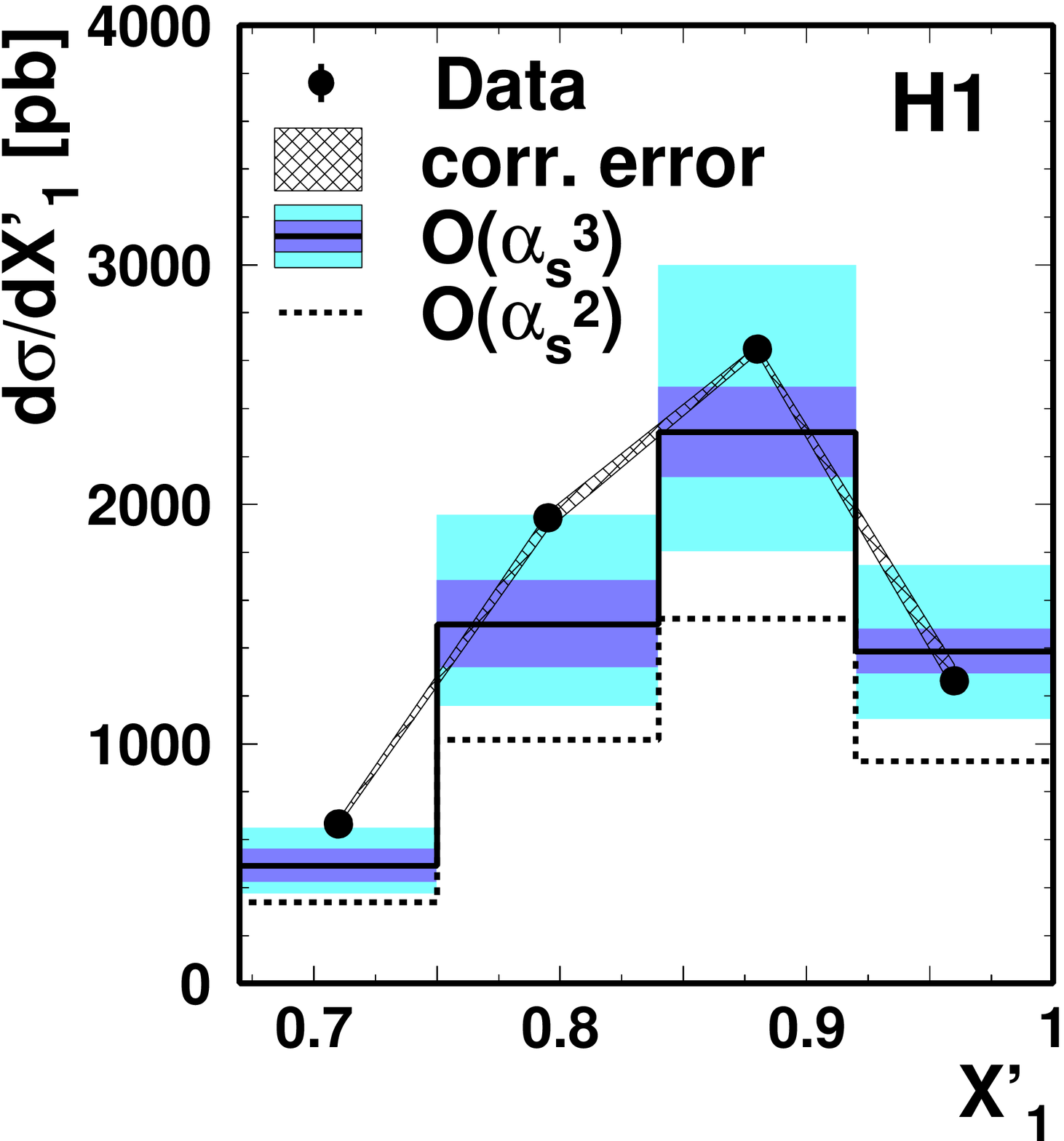}
\hspace*{5mm}
\includegraphics[width=0.478\linewidth,keepaspectratio]{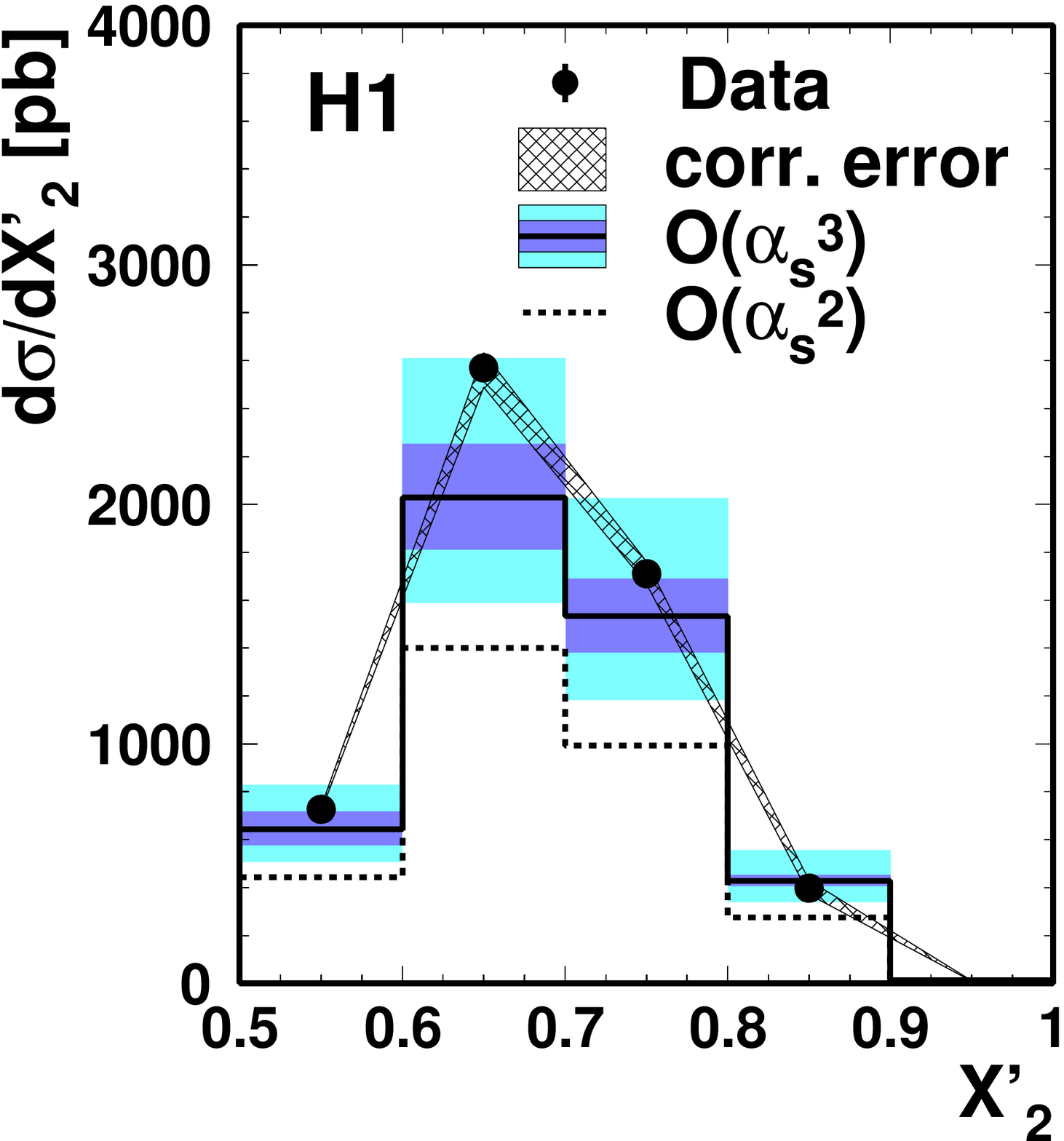}
\end{center}
\begin{center}
\hspace*{1mm}
\includegraphics[width=0.464\linewidth,keepaspectratio]{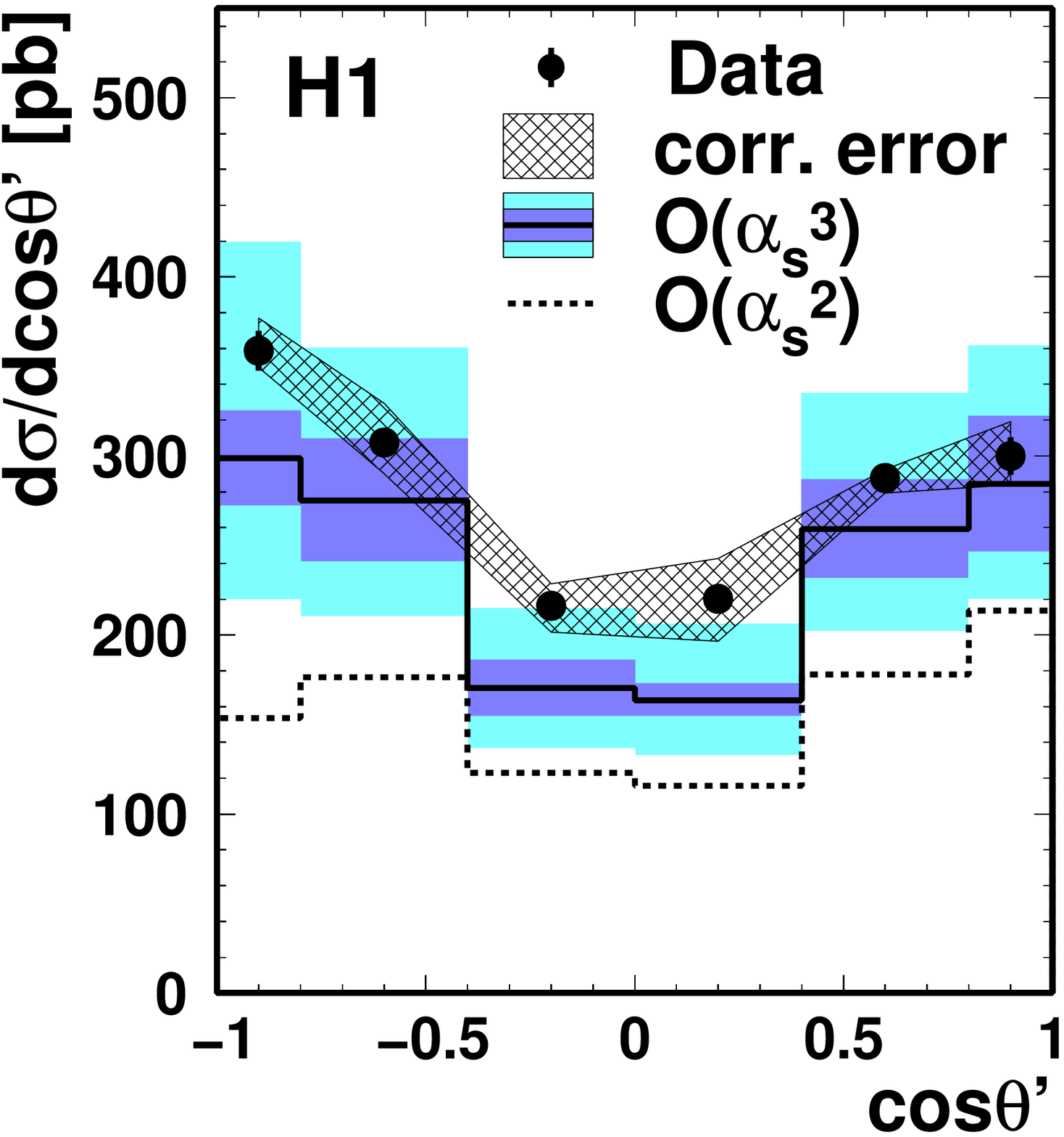}
\hspace*{7mm}
\includegraphics[width=0.464\linewidth,keepaspectratio]{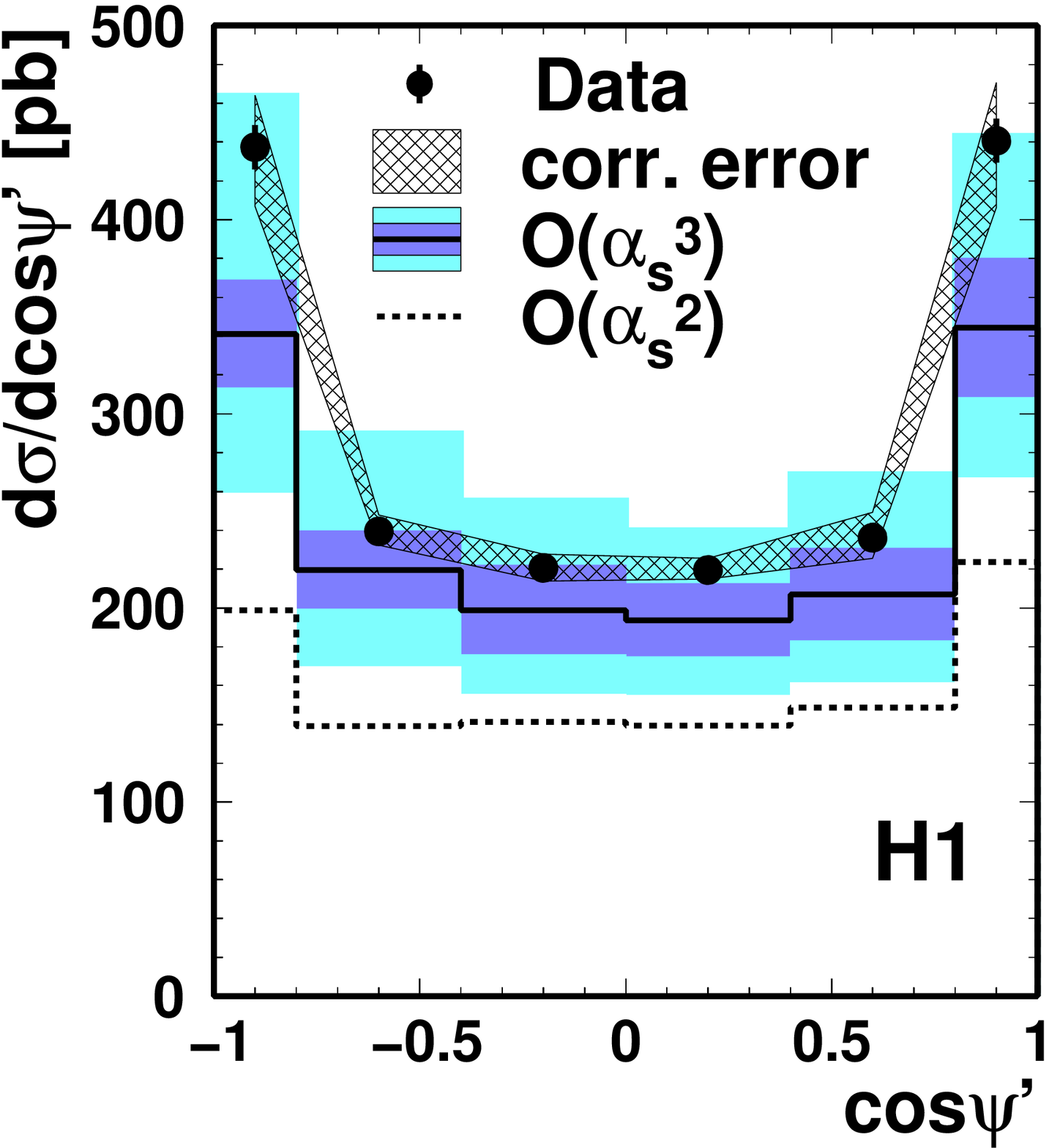}
\end{center}
\caption{Differential cross sections as a function of the scaled energies $X_1'$ and 
$X_2'$ of the two leading jets ($E_1'>E_2'>E_3'$ in the three-jet
centre of mass frame) and the two angles $\theta'$ and $\psi'$ as defined in figure
\ref{Tevatronvar}. The data are compared to the LO $\mathcal{O}(\alpha_{\rm s}^2)$ and the NLO 
$\mathcal{O}(\alpha_{\rm s}^3)$ predictions.
See the caption of figure \ref{fig:njet} for further details.}
\label{fig:cospsi}
\end{figure}

Differential cross sections are presented
in figures \ref{fig:njet} to \ref{fig:cospsi} 
for events with three or more jets as a function of the number of jets
($N_{\rm jet}$), the Bj{\o}rken variable $x$, the pseudorapidities of the three jets and 
the variables characterising the topology of the three jets in the three-jet centre of mass frame
(scaled jet energies and three-jet angles). The  kinematic
range for which the cross
sections are determined
is  specified in table \ref{tab:xsdef}.
The figures also show the predictions of the NLOJET++ fixed order QCD prediction in LO $\mathcal{O}(\alpha_{\rm s}^2)$
and NLO $\mathcal{O}(\alpha_{\rm s}^3)$, corrected for hadronisation effects.
The theory error, including scale variations and hadronisation correction
uncertainties added in quadrature, exceed the measurement uncertainty.
Figure \ref{fig:njet} (left) shows the distribution of the number
of jets found in the selected events which
extends up to $N_{\rm jet} = 6$. 
For this distribution the predictions of the NLO $\mathcal{O}(\alpha_{\rm s}^3)$ calculation
and of the two Monte Carlo programs RAPGAP and CDM are shown.
The NLO $\mathcal{O}(\alpha_{\rm s}^3)$ prediction agrees for $N_{\rm jet}=3$ and
underestimates the rate of events with 4 jets by a factor 2.6.
It does not produce any events with more than four jets as expected.
\par
The differential cross sections
shown in figures \ref{fig:njet} to \ref{fig:cospsi} are not described
by the LO ${\mathcal{O}}(\alpha_{\rm s}^2)$ 
QCD predictions neither in shape nor in magnitude. 
The main discrepancies are seen at low $x$ and
for forward 
jets (large positive $\eta$) where far too few events are predicted. 
The NLO $\mathcal{O}(\alpha_{\rm s}^3)$ 
prediction improves
the situation considerably in all regions where deviations from LO are observed.
A similar improvement was already noticed in the previous H1 three-jet
analysis \cite{wobisch}, 
which was restricted to the phase
space of large invariant three-jet masses above 25 GeV.
For that phase space the ${\mathcal{O}}(\alpha_{\rm s}^3)$
calculation could still describe the three-jet data down to the smallest
$x=10^{-4}$.
However, for the present analysis without such mass cuts applied,
at very small $x<2\cdot 10^{-4}$ the calculation undershoots the data, 
which lies approximately at the upper edge of the total theoretical
error band, by a factor of about $0{.}6$.
In the complementary region $x>2\cdot 10^{-4}$, the description is reasonable;
this was also observed in the recent ZEUS multi-jet analysis \cite{Chekanov:2007dx}
which was restricted to this phase space.
In summary, a large deficit of the NLO prediction persists only at low $x$ and for forward jets.
This is the kinematic region where  
unordered gluon radiation is expected to
enhance the jet production~\cite{mueller}.
The shapes of the cross section distributions for the 
three-jet topological variables shown in figure \ref{fig:cospsi}
are all well described by the NLO $\mathcal{O}(\alpha_{\rm s}^3)$ prediction,
only the global normalisation of the calculation is somewhat too low.
\subsection{Forward Jet Subsample}
\label{sec:forwj}
\begin{figure}[htbp]
\begin{center}
\includegraphics[width=0.478\linewidth]{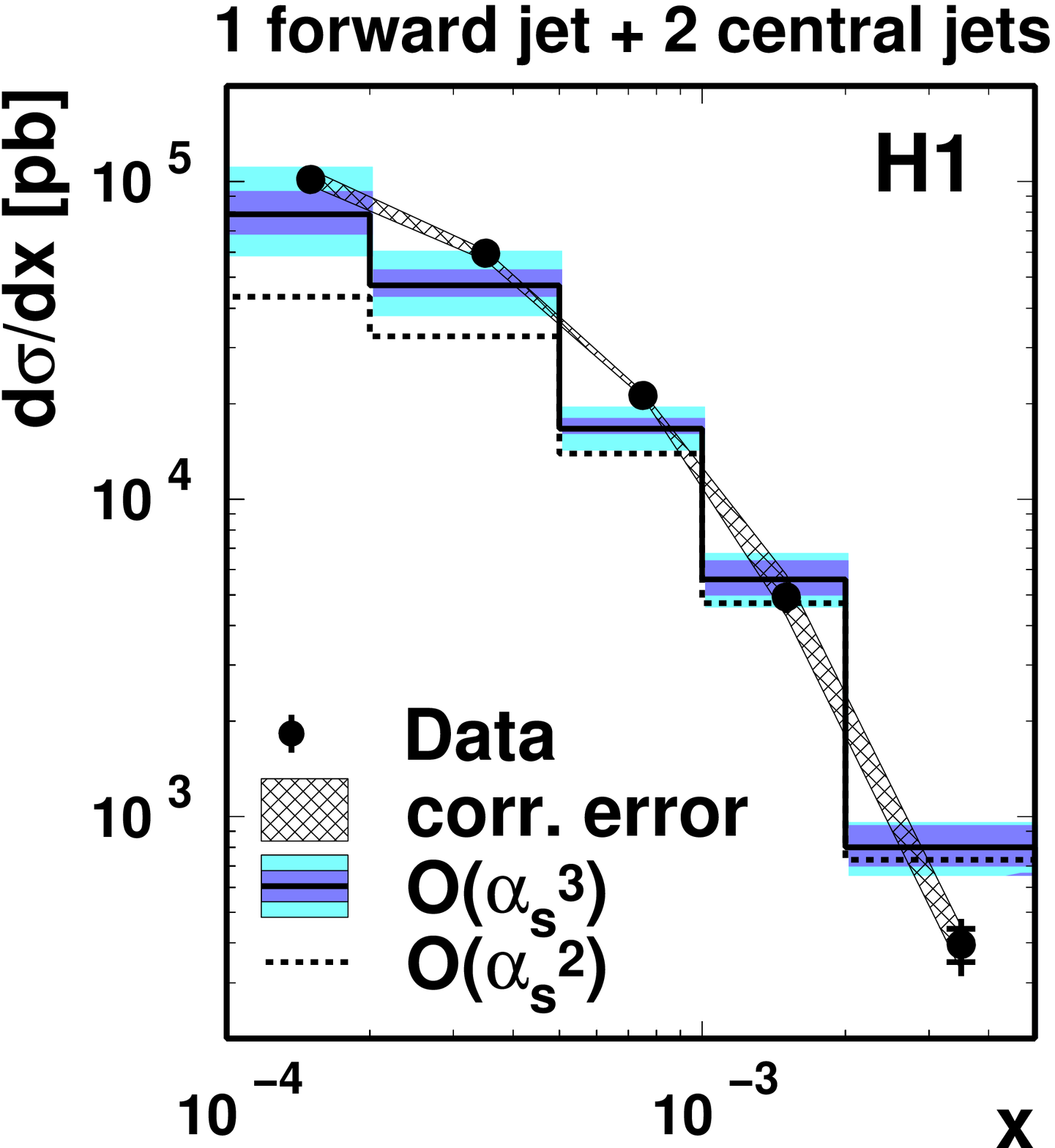}
\hspace*{5mm}
\includegraphics[width=0.478\linewidth]{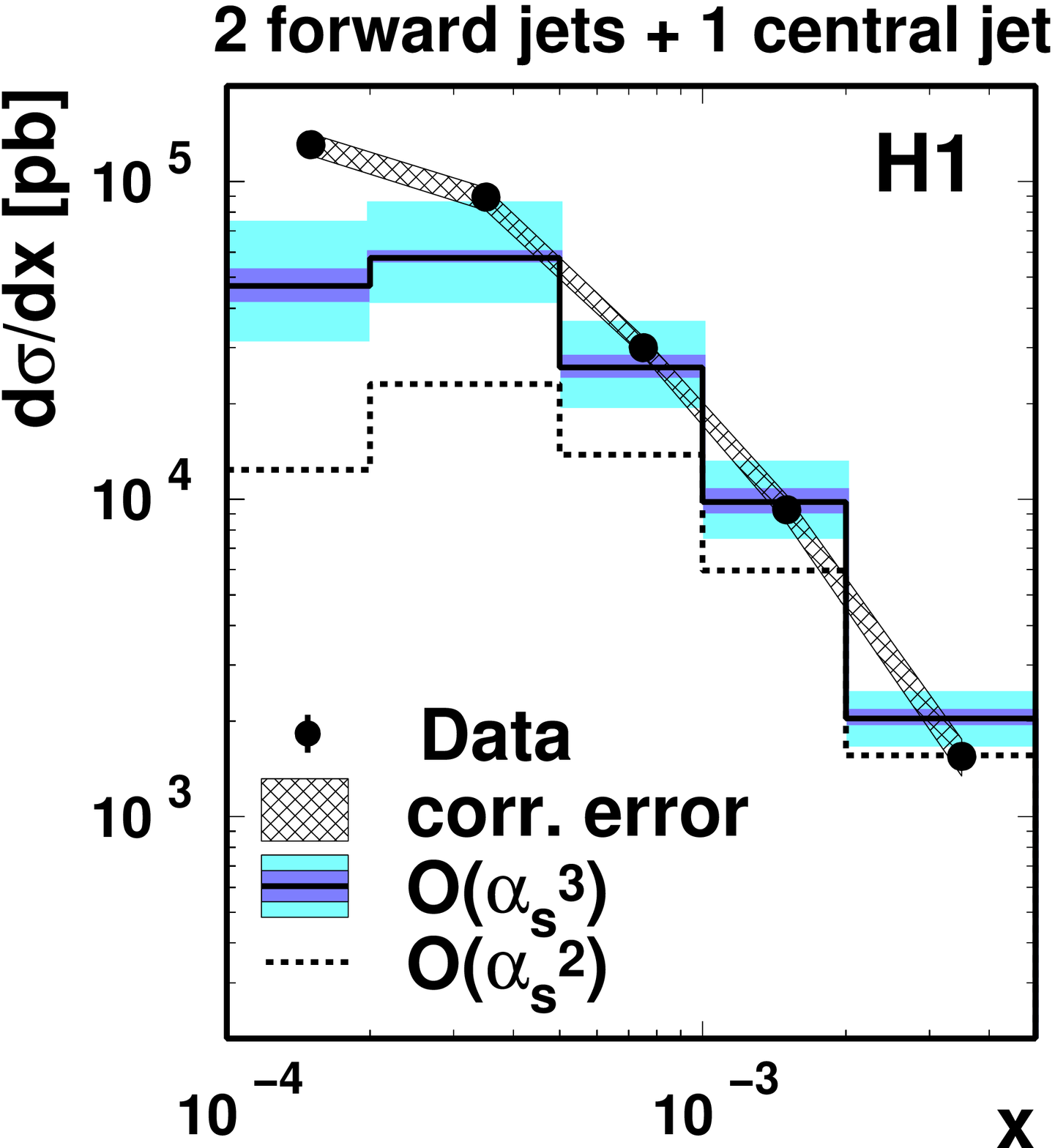}
\end{center}
\caption{Differential cross sections as a function of the Bj{\o}rken scaling variable $x$ for
the two selected subsamples of events with at least three jets:
the sample with one forward jet and two central jets (\tc) 
on the left, the sample with two
forward jets and one central jet (\tf) on the right.
The additional global normalisation errors
of the data ($^{+19}_{-14}\%$ for the \tc\/ sample and $^{+18}_{-15}\%$ for 
the \tf\/ selection)
are not displayed.
See the caption of figure \ref{fig:njet} for further details.}
\label{fig:fwdx}
\end{figure}

\begin{figure}[htbp]
\begin{center}
\hspace*{2.5mm}
\includegraphics[width=0.452\linewidth]{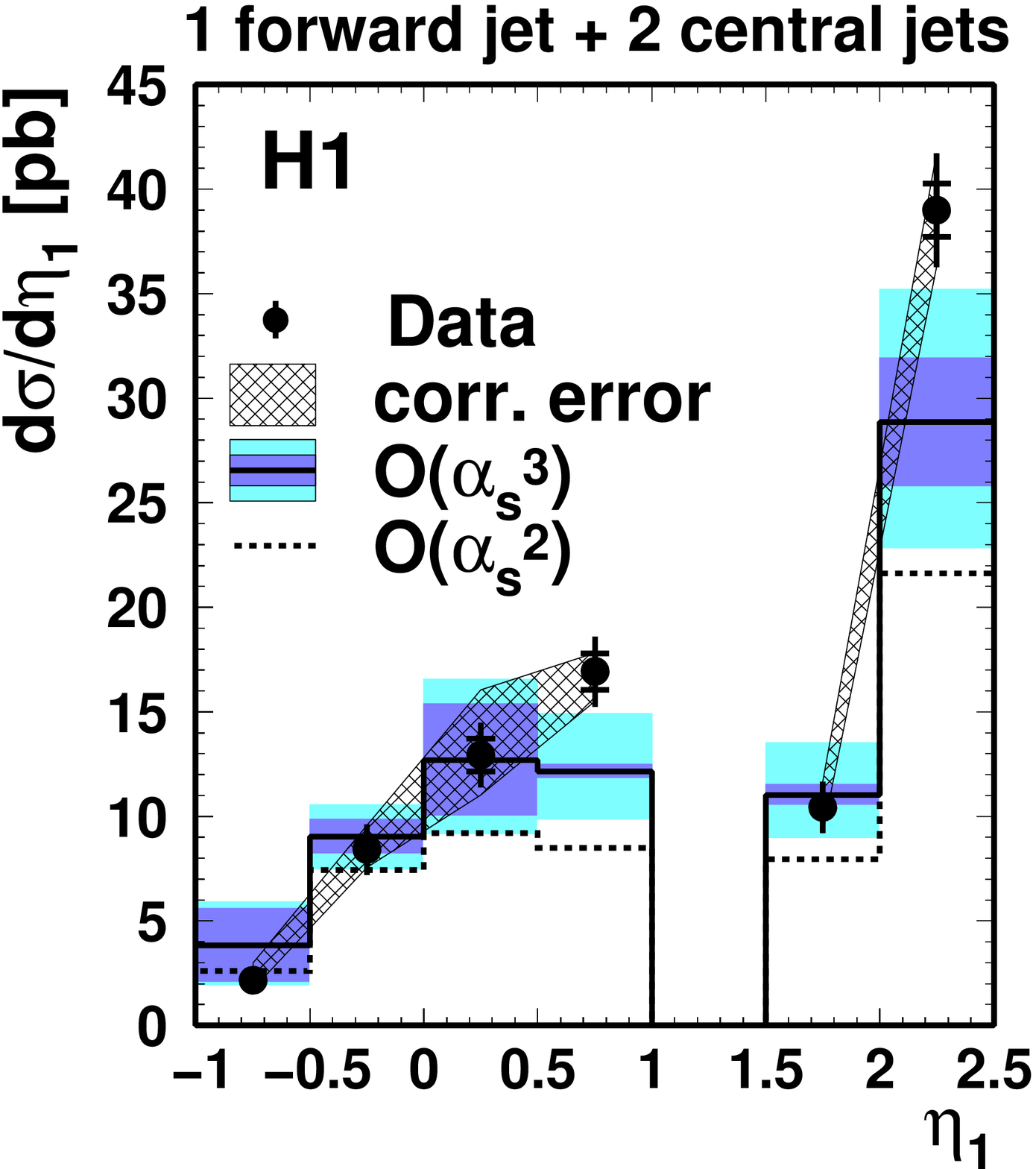}
\hspace*{9mm}
\includegraphics[width=0.452\linewidth]{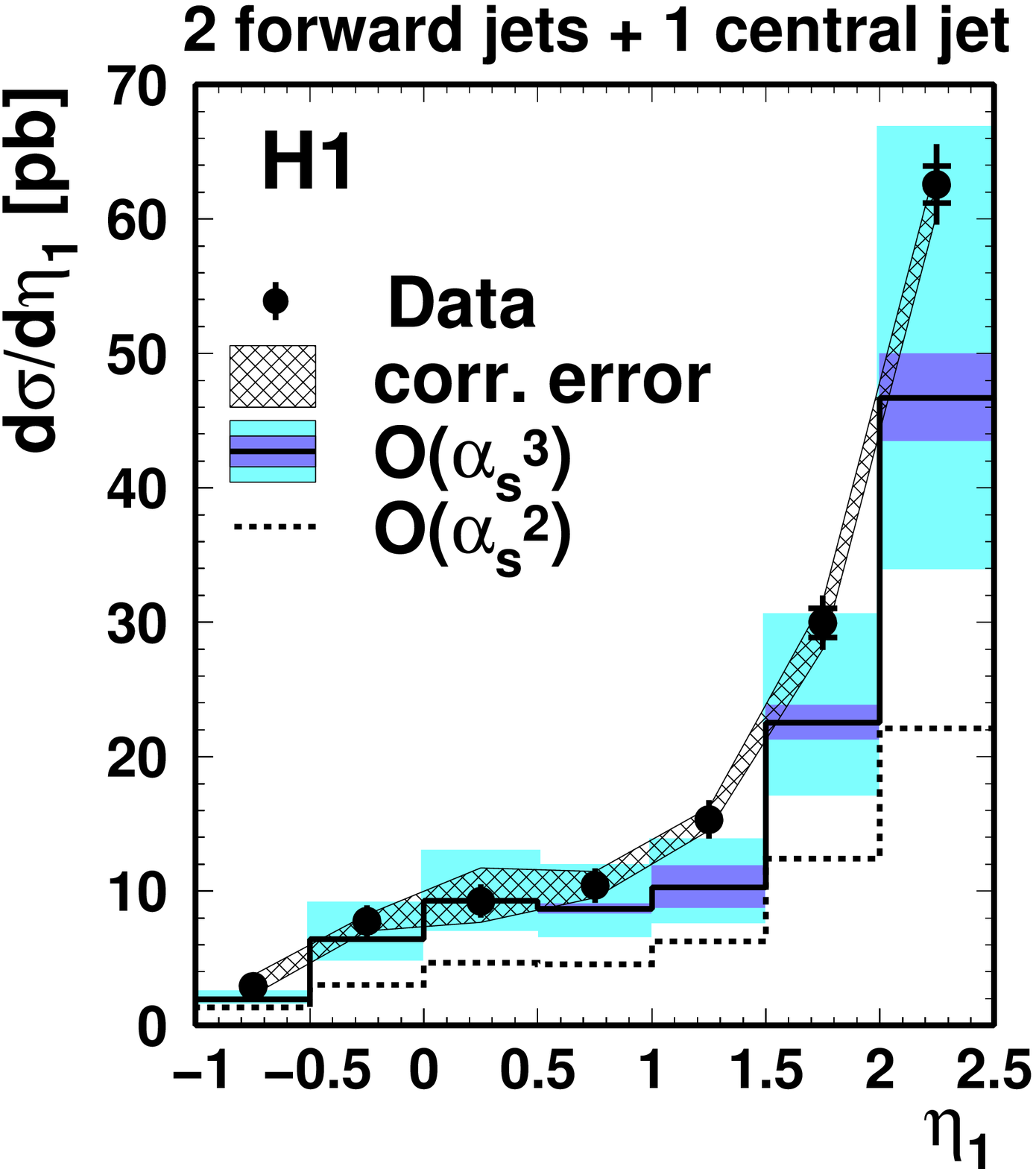}
\end{center}
\begin{center}
\includegraphics[width=0.478\linewidth]{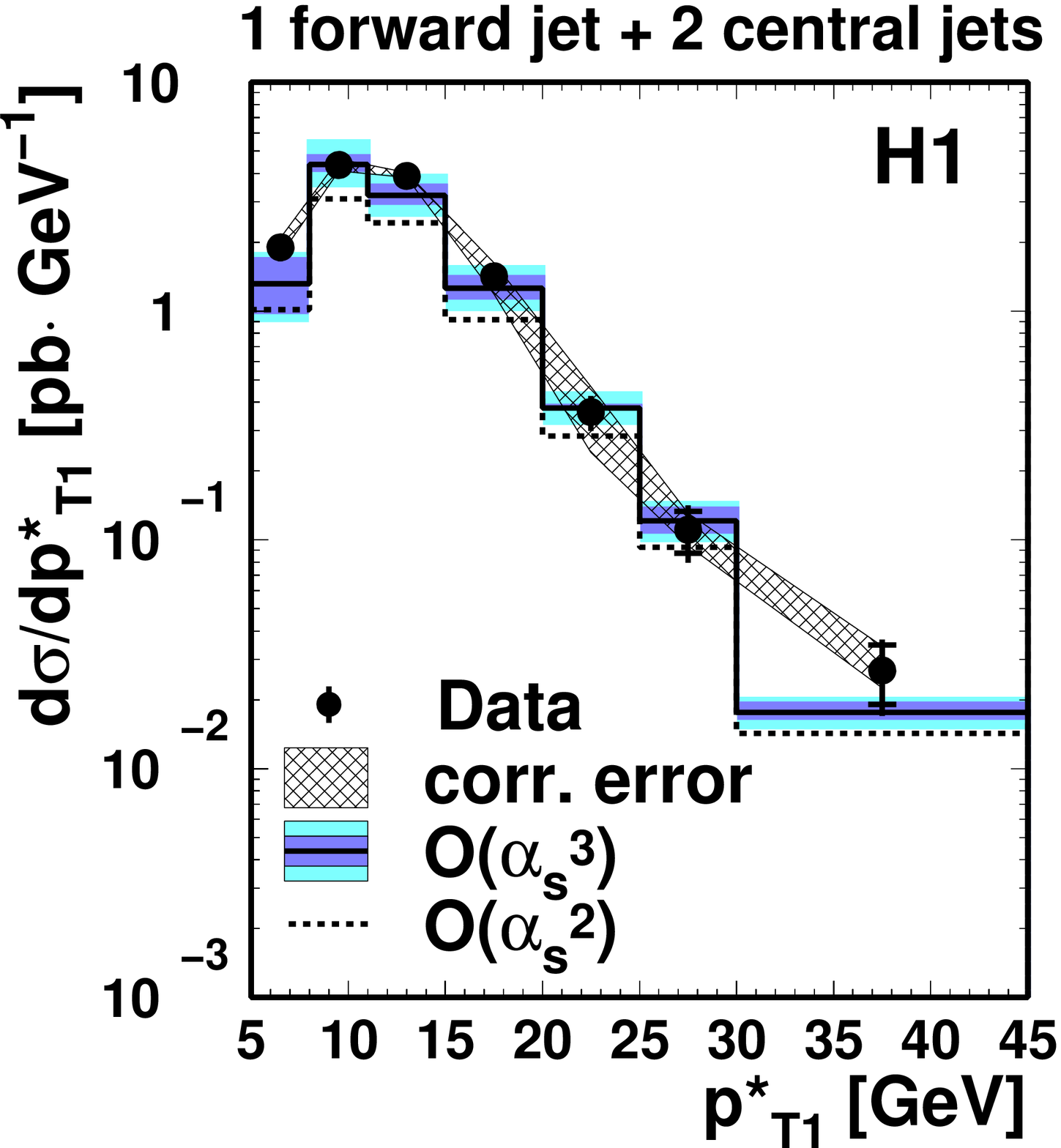}
\hspace*{5mm}
\includegraphics[width=0.478\linewidth]{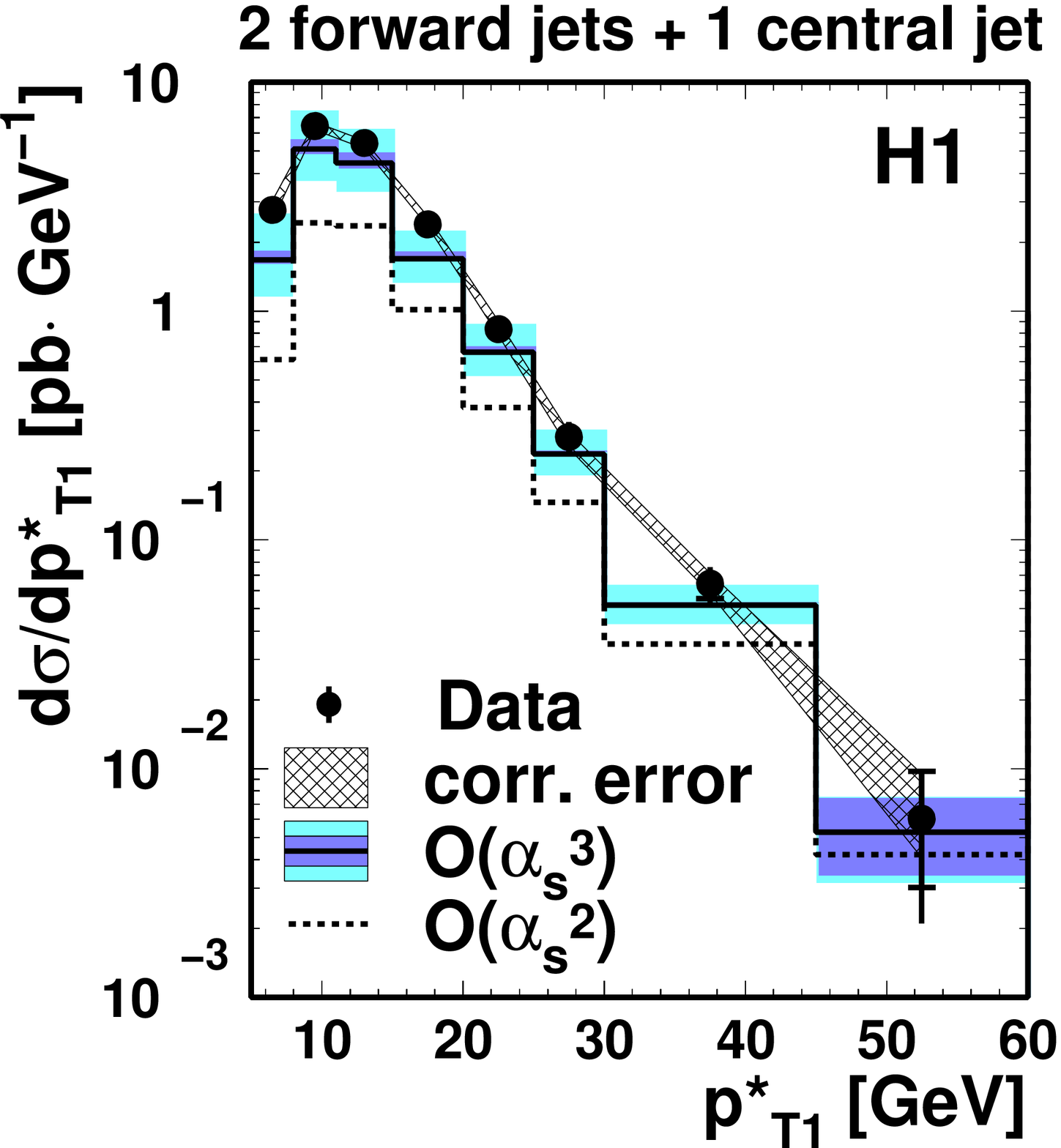}
\end{center}
\caption{
Differential cross sections as a function of the pseudorapidity 
$\eta_1$ and the transverse momentum ${p^*_\pperp}_1$
of the leading jet for
the two selected subsamples of events with at least three jets:
the sample with one forward jet and two central jets (\tc) 
on the left, the sample with two
forward jets and one central jet (\tf) on the right.
The additional global normalisation errors
of the data ($^{+19}_{-14}\%$ for the \tc\/ sample and $^{+18}_{-15}\%$ for 
the \tf\/ selection)
are not displayed.
See the caption of figure \ref{fig:njet} for further details.}
\label{fig:fwdeta}
\end{figure}

The observed excess of data versus QCD predictions in the region of forward jet rapidities
and at low $x$ is investigated here in further detail.
The sample of selected events with at least three jets fulfilling the 
criteria presented in table~\ref{tab:xsdef}
is reduced to a subsample by requiring that
at least one of the three 
leading jets be forward and carry a large proton
momentum fraction:
$$ \eta > 1{.}73 \ \hbox {\rm and}
\  x_{\rm jet} \equiv \frac{E_{\rm jet}}{E_{\rm p,\,beam}} > 0.035\;.$$ 
Here $E_{\rm jet}$ is the jet energy in the laboratory frame
and $E_{\rm p,\, beam}=920\;{\rm GeV}$ the proton beam energy.
Further requirements are applied to obtain two 
separate subsamples.
%
%\begin{itemize}
%\item
In the sample with one forward jet and two central jets (\tc) 
the other two leading jets are required to lie in the
central pseudorapidity range $-1<\eta_{\rm jet}<1$.
%
%\item
In the sample with two forward jets and one central jet (\tf)
it is demanded that one of the three leading jets is a central jet  
with $-1<\eta_{\rm jet}<1$ and the remaining leading jet must be in the
more forward region $\eta_{\rm jet}>1$ (for this second forward jet
no cut is applied on $x_{\rm jet}$). 
%
%\end{itemize}
%
The fraction of jets due to gluon radiation
is expected to be
larger for forward jets than for central jets.
This is confirmed by a study of the parton composition of three-jets in the
CDM simulation. 
Therefore the \tc\/ sample will have many events with 
only a single radiated gluon (as for the left diagram in figure \ref{feynman})
while the \tf\/
selection will have a larger fraction of events with 
two radiated gluons (as for the right diagram in figure \ref{feynman}).
Cross section measurements as a function of  $x$, $\eta_1$ and
${p^*_\pperp}_1$ for the \tc\/ and \tf\/ samples are presented in figures 
\ref{fig:fwdx} and \ref{fig:fwdeta}.
The additional global normalisation errors of $^{+19}_{-14}\%$
for the \tc\/ sample and $^{+18}_{-15}\%$ for the \tf\/ selection
are not shown in the figures.
The fixed order NLO $\mathcal{O}(\alpha_{\rm s}^3)$ prediction gives a rather
good description of the \tc\/ sample.
The step from LO $\mathcal{O}(\alpha_{\rm s}^2)$ to NLO $\mathcal{O}(\alpha_{\rm s}^3)$
improves the agreement mainly at very low $x<2\cdot 10^{-4}$, where a remaining deficiency of 
$\sim$$30\%$ is observed.
Towards larger $x$ values both the LO and the NLO calculation
fall off somewhat less steeply than the data and are too high
for the largest covered $x$ values $x>2 \cdot 10^{-3}$.
For the \tf\/ selection 
an even more dramatic change is observed at low $x$ from $\mathcal{O}(\alpha_{\rm s}^2)$ to $\mathcal{O}(\alpha_{\rm s}^3)$:
the discrepancy at $x<2\cdot 10^{-4}$
is reduced from a factor of $10$ to $2{.}6.$
The large remaining deficiency exceeds the combined error
of prediction and data and is thus highly significant.
It can also not be explained by a possible additional maximal cross section
error of 25\% due to detector jets which cannot be matched with
hard partons (as discussed in the last paragraph of section \ref{sec:xsec}).
This data excess provides
a strong hint for missing higher order QCD corrections, 
i.e. beyond  $\mathcal{O}(\alpha_{\rm s}^3)$, 
in this forward gluon radiation dominated phase space.
Note that for processes with
two radiated gluons, the  $\mathcal{O}(\alpha_{\rm s}^3)$ calculation 
can only provide a leading order perturbative estimate.

Excesses which are probably related to the one reported here were observed in the forward jet analyses
from H1 \cite{Aktas:2005up} and ZEUS \cite{Chekanov:2007pa}.
In these analyses the topologies of three jets were investigated
for events containing a dijet system in addition to a forward jet.
The ${\mathcal{O}}(\alpha_{\rm s}^3)$
predictions were found to undershoot the data 
in the region where all three jets tend to go forward.
However, in these analyses the data were either integrated over a larger $x$ 
range or restricted to somewhat larger $x$ values, which might explain why the excesses are less
prominent than observed in the present measurement.
\subsection{Monte Carlo Program Predictions for Events with at Least Three Jets}
\begin{figure}[htbp]
\begin{center}
\hspace*{14mm}
\includegraphics[bb=108 53 498 512,width=6cm,keepaspectratio]{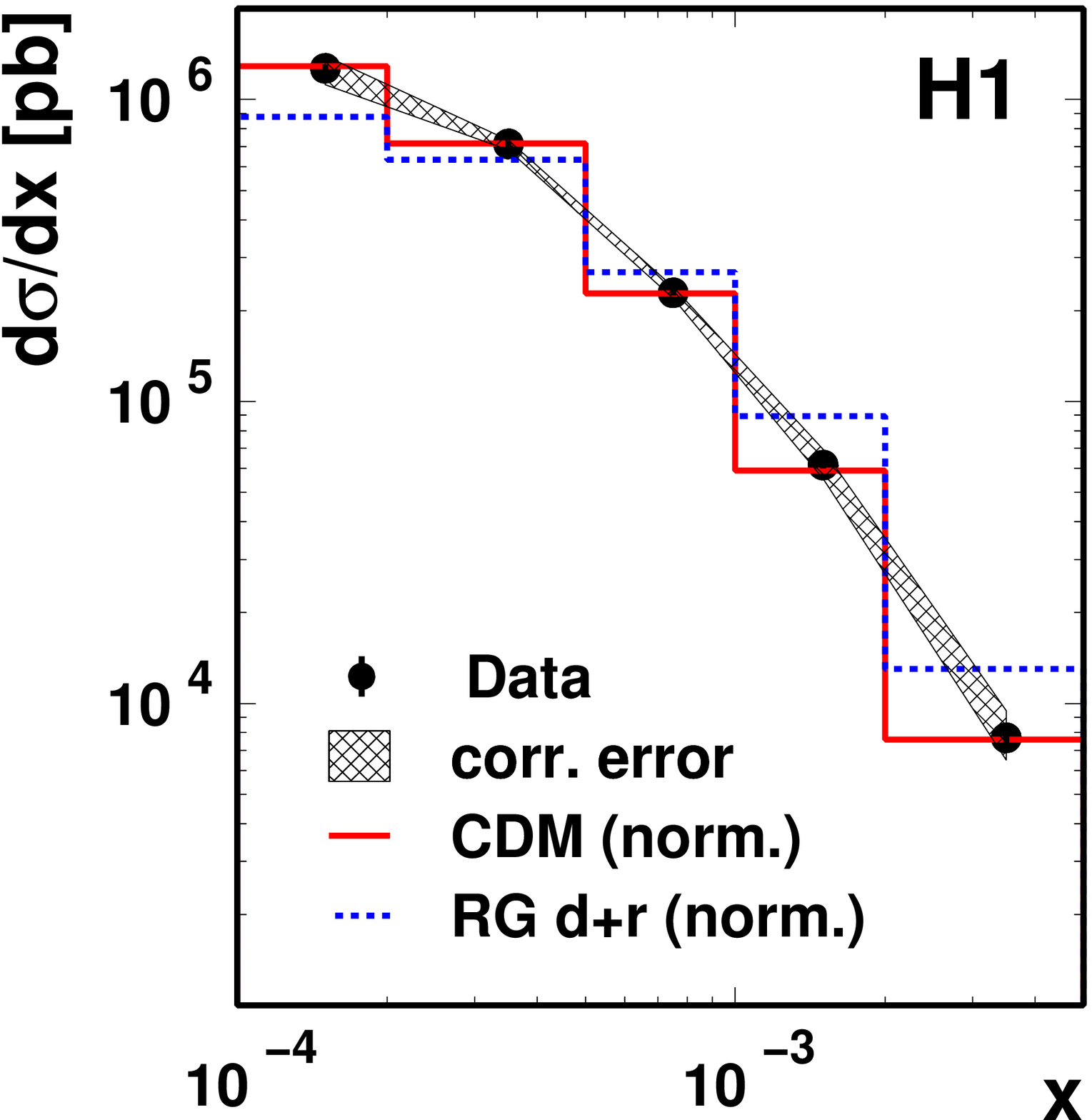}
\hspace*{20mm}
\includegraphics[bb=97 70 487 529,width=6cm,keepaspectratio]{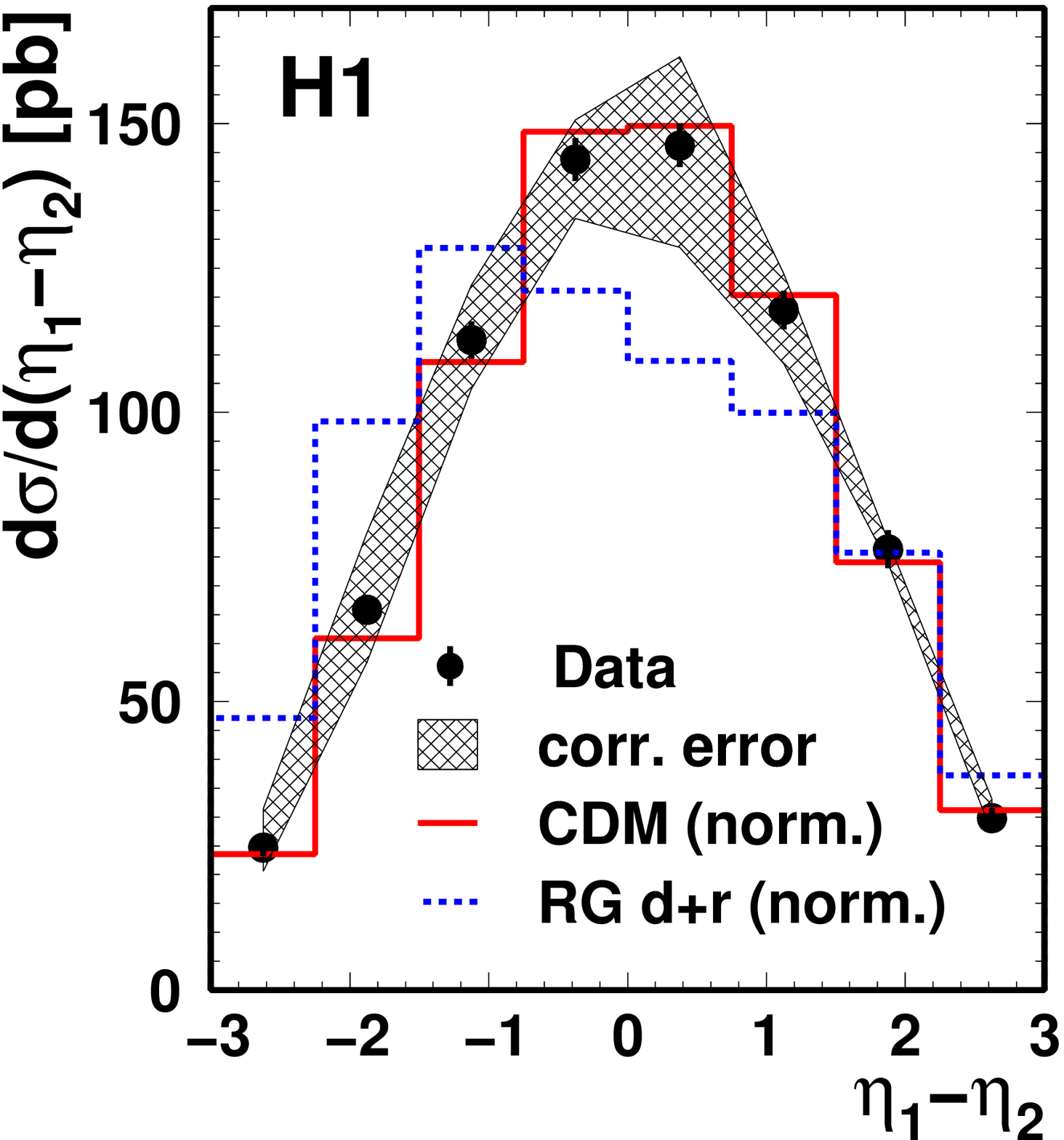}
\end{center}
\vspace*{11mm}
\caption{Differential cross sections 
as a function of the Bj{\o}rken scaling variable $x$ and
the difference of the pseudorapidities of the two leading jets
for the selected events with at least three jets 
in the kinematic range listed in table 
\ref{tab:xsdef}.
The cross sections are bin-averaged and plotted at the respective bin centers.
The inner error bars represent the statistical error of the data,
the total error bars correspond to the statistical and uncorrelated
systematic errors added in quadrature.
The correlated systematic errors are
shown by the hatched error band. 
The additional global normalisation error ($^{+16}_{-14}\%$)
of the data
is not displayed. The data are
compared to predictions from DJANGO (CDM) (solid line) and
RAPGAP (dashed line).
Both Monte Carlo predictions are scaled to match the total data cross section: CDM by a factor
 of
$1{.}05$ and RAPGAP by 1{.}55.}
%
%xxxOther details are as in the caption of figure \ref{fig:lo}.}
%
\label{fig:lo}
\end{figure}

\begin{figure}[htbp]
\begin{center}
\includegraphics[width=0.478\linewidth,keepaspectratio]{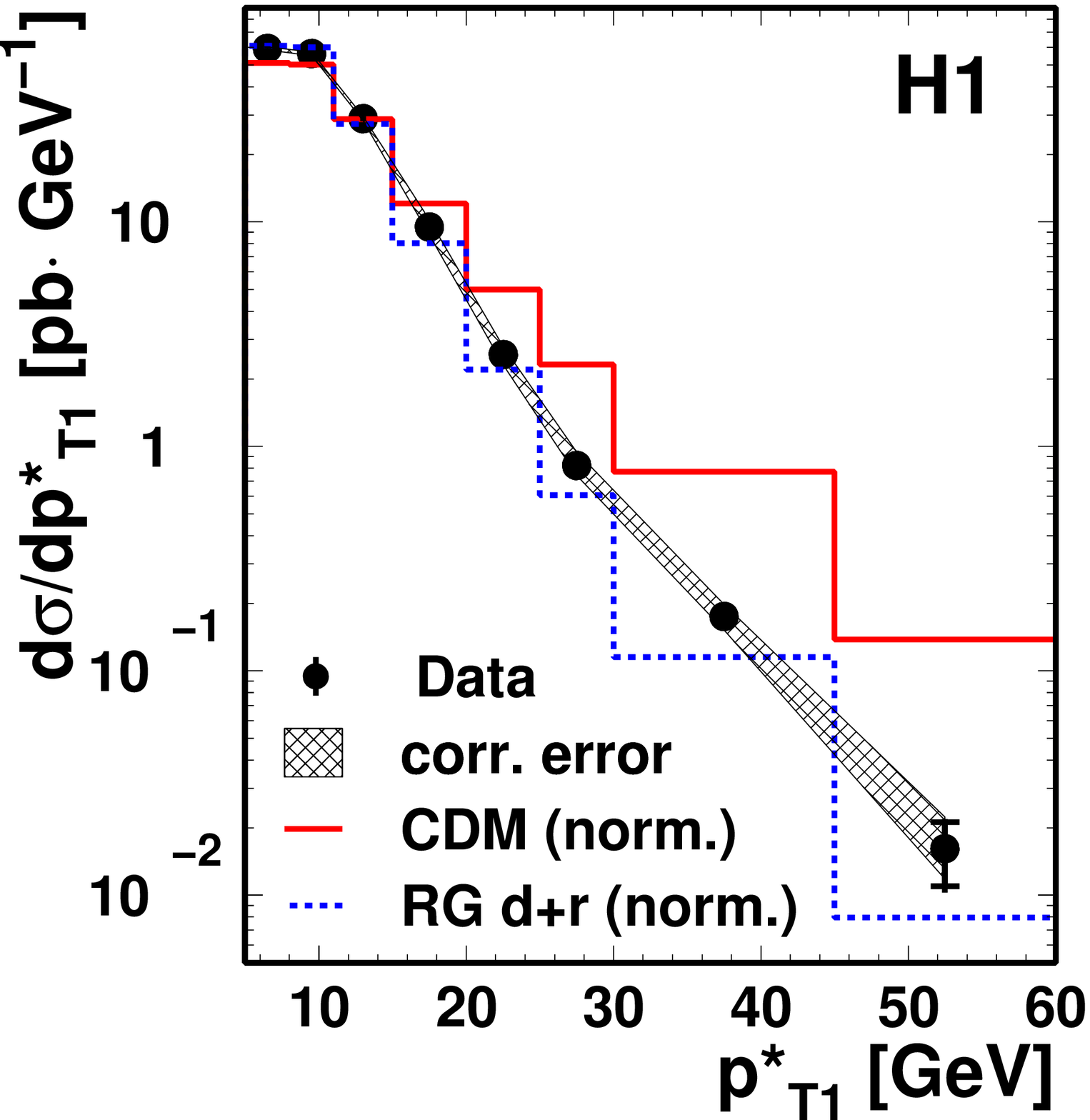}
\hspace*{4.5mm}
\includegraphics[width=0.473\linewidth,keepaspectratio]{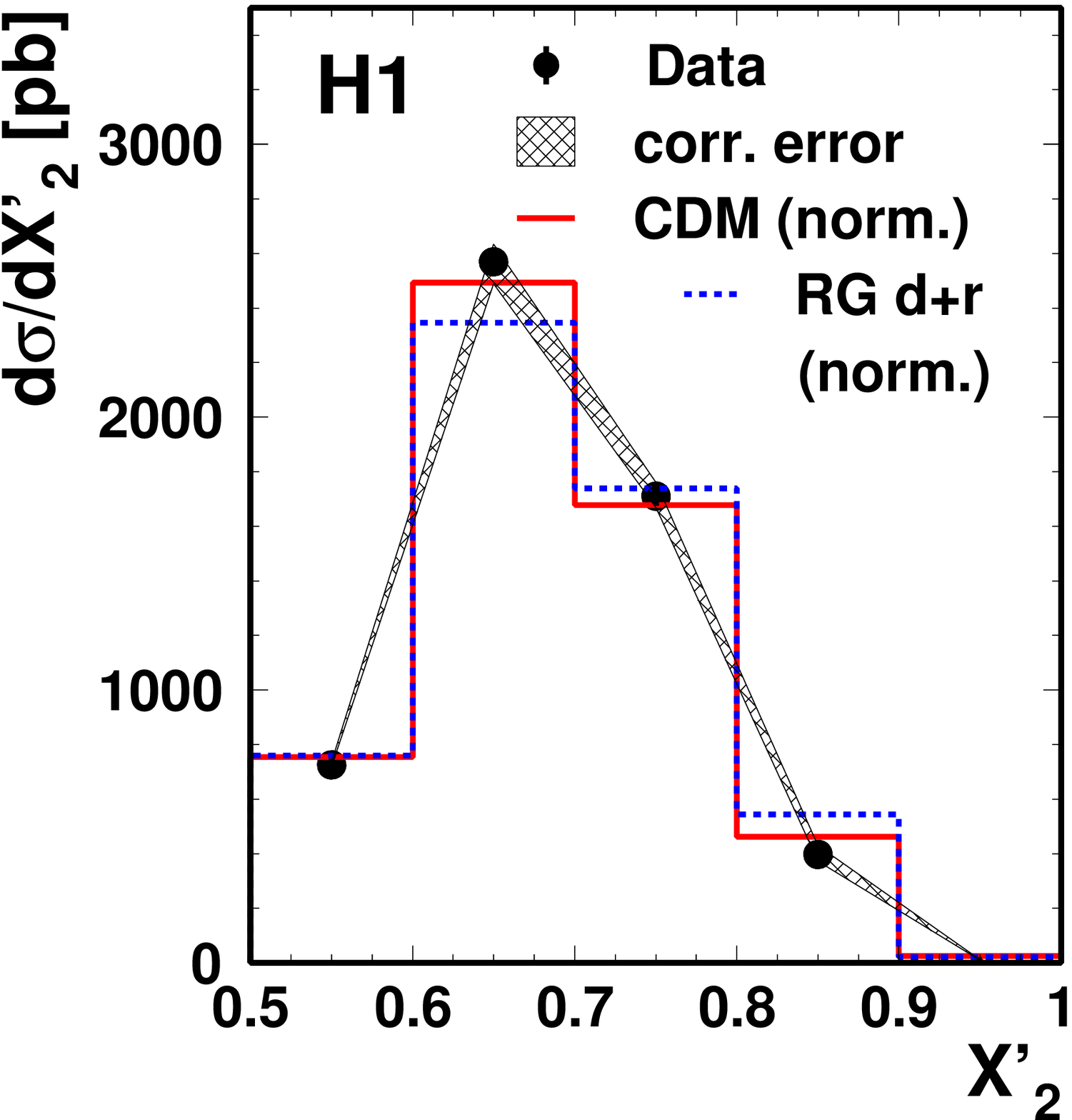}
\end{center}
\begin{center}
\hspace*{2mm}
\includegraphics[width=0.454\linewidth,keepaspectratio]{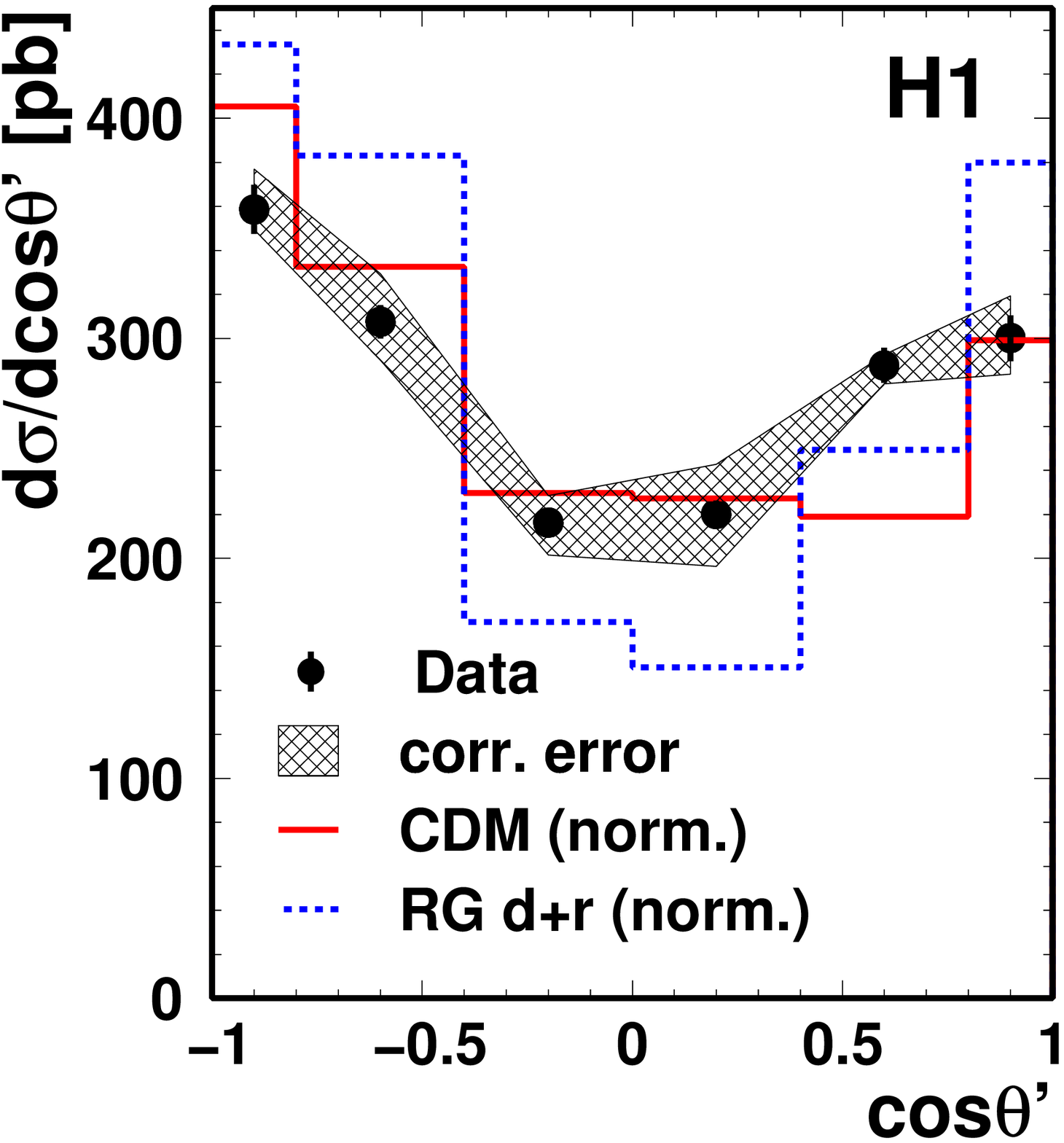}
\hspace*{9mm}
\includegraphics[width=0.454\linewidth,keepaspectratio]{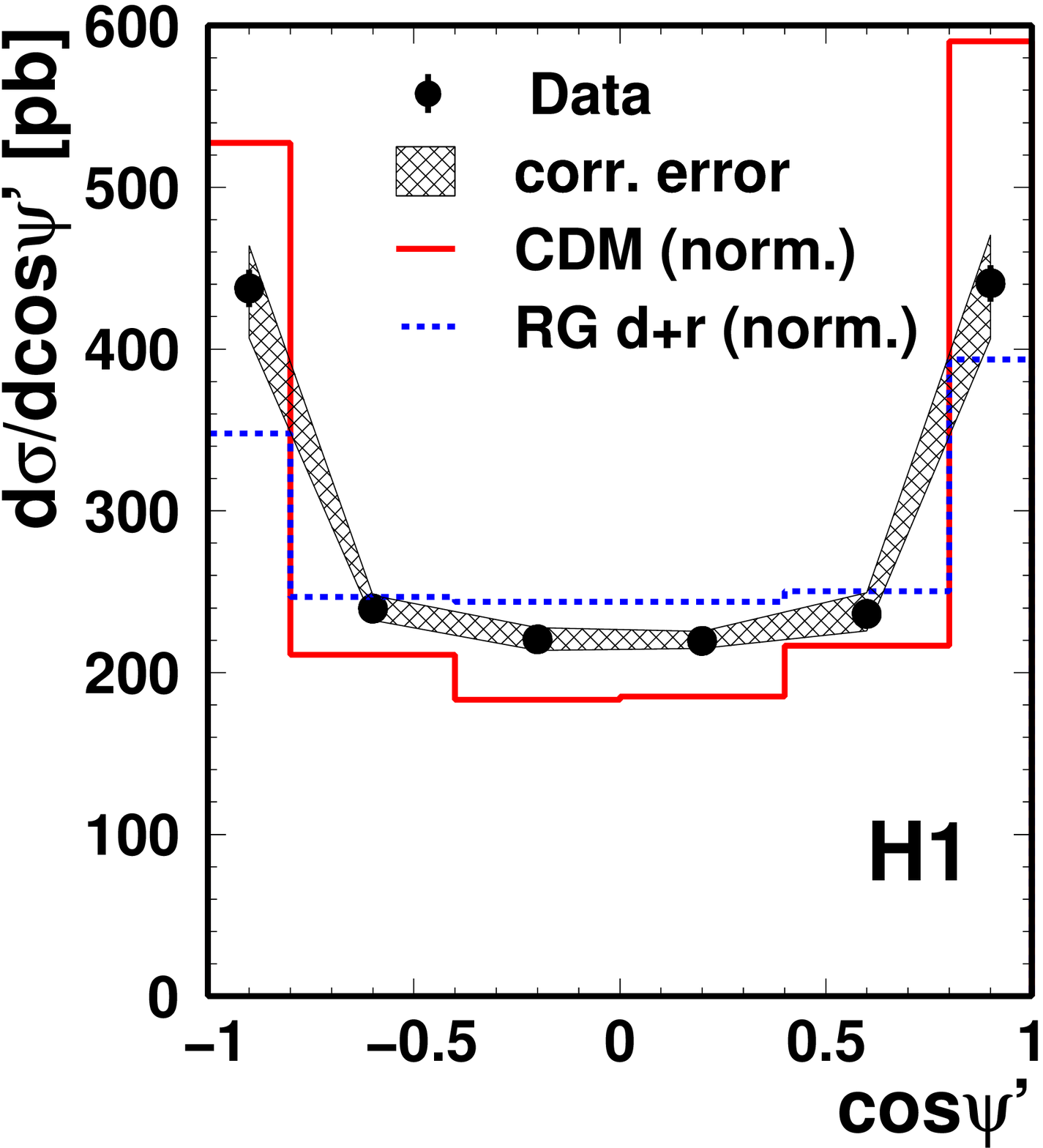}
\end{center}
\caption{Differential cross sections
for the selected events with at least three jets
as a function of the transverse momentum of the leading jet
$p^*_{\pperp 1}$, the scaled energy $X_2'$ of the
next to leading jet in the three-jet
centre of mass frame and the two angles $\theta'$ and $\psi'$ as defined in figure
\ref{Tevatronvar}. 
The data are
compared to the predictions from the Monte Carlo programs RAPGAP and CDM.
Both Monte Carlo predictions are scaled to match the total data cross section: CDM by a factor
 of
$1{.}05$ and RAPGAP by 1{.}55.
See the caption of figure \ref{fig:lo} for further details.}
\label{fig:loteva}
\end{figure}

\begin{figure}[htbp]
\begin{center}
\hspace*{12mm}
\includegraphics[bb=95 59 484 518,width=6cm,keepaspectratio]{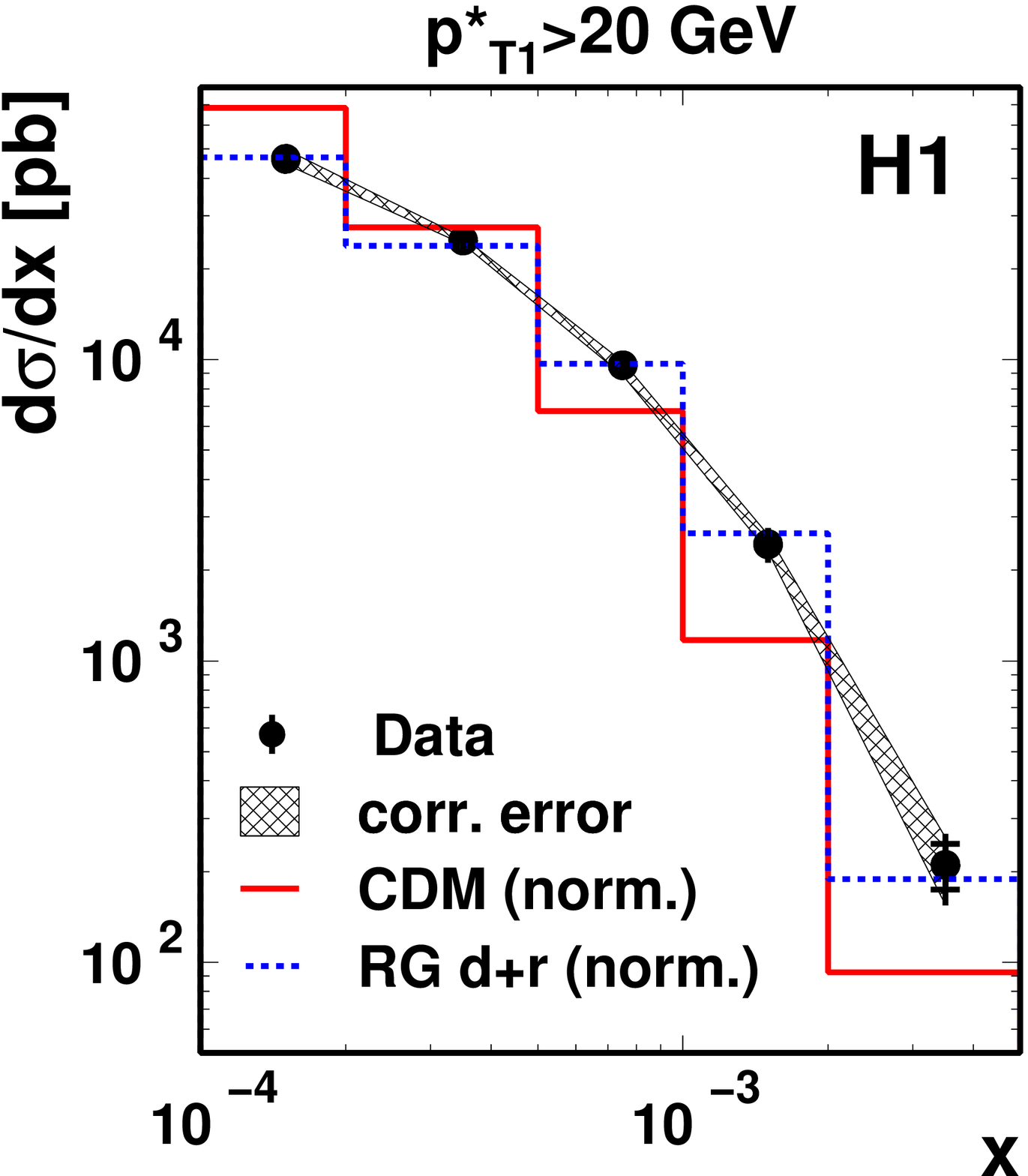}
\hspace*{20mm}
\includegraphics[bb=95 70 484 529,width=6cm,keepaspectratio]{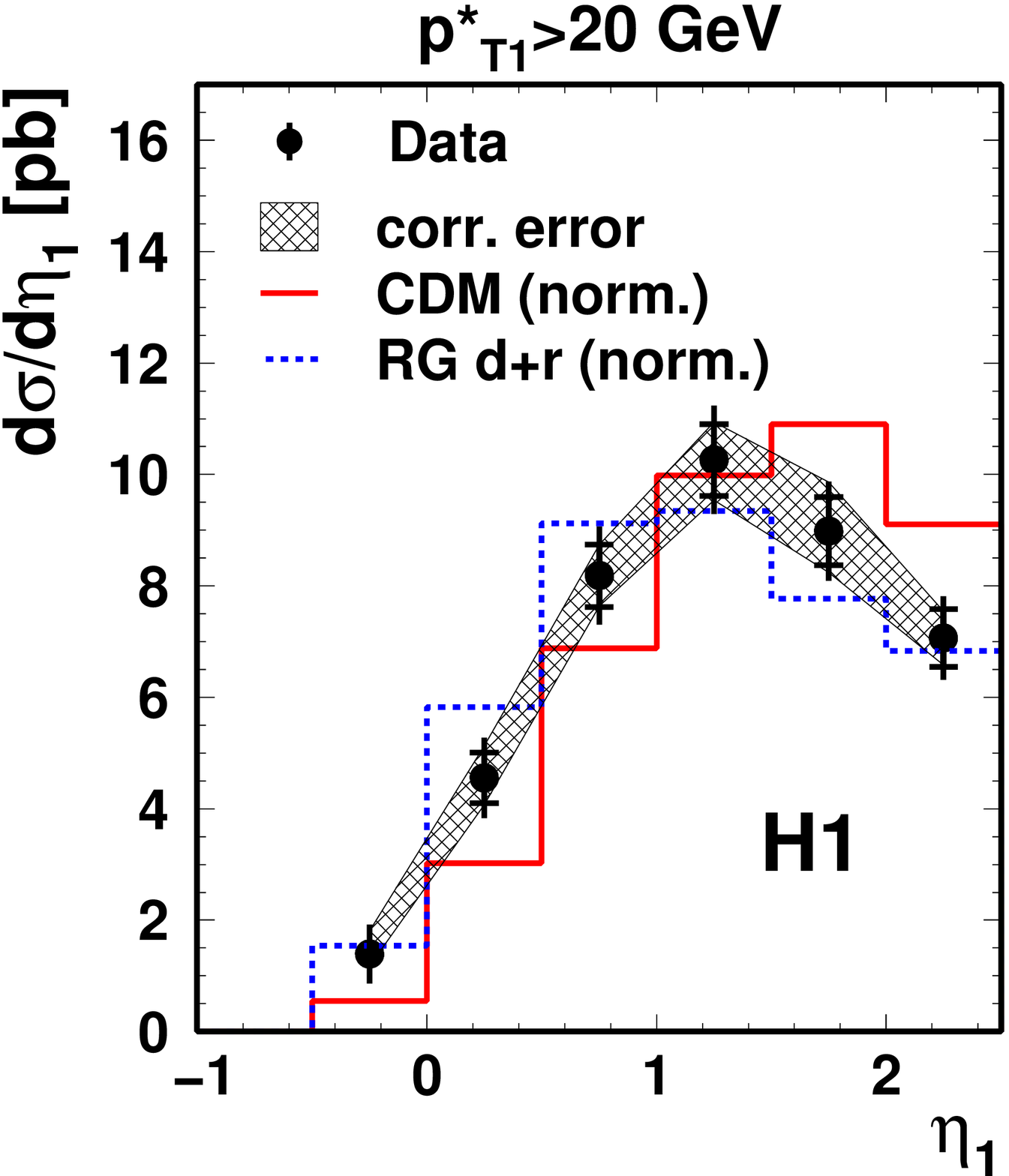}
\end{center}
\vspace*{11mm}
\caption{Differential cross sections %at hadron jet level
as a function of the Bj{\o}rken scaling variable $x$
and the leading jet pseudorapidity $\eta_1$ for the selected subsample of the
events with at least three jets: the leading jet is required to have 
a transverse momentum 
${p^*_\pperp}_1$ above $20\,{\rm GeV}$.
%{\bf CHECK: add remark on Global normalistion error}
The data are compared to the RAPGAP and CDM predictions.
Both Monte Carlo predictions are scaled to match the total data cross section: CDM by a factor
of $0{.}41$ and RAPGAP by 1.95.
See the caption of figure \ref{fig:lo} for further details.}
\label{threejet20}
\end{figure}

The cross sections for events with at least three jets are compared to 
predictions from RAPGAP
and CDM.
%
% which have both been able to describe rather well dijet~\cite{Breitweg:2001rq,Aktas:2003ja,Aktas:2004px} 
%and inclusive forward jet~\cite{Aktas:2005up,Chekanov:2005yb,Chekanov:2007pa}
%measurements.
%
The jet multiplicity shown in figure \ref{fig:njet} (left) is described well by
CDM while RAPGAP falls off too steeply.
The overall normalisation of the RAPGAP and CDM predictions is found to be too
low by 55\% and 5\% respectively. In the following the Monte Carlo predictions
are normalised to the total measured cross section
in order to compare only the shapes of the cross sections.
Figures \ref{fig:lo} and \ref{fig:loteva} show the comparison
as a function of
the Bj{\o}rken scaling
variable $x$, the difference of the pseudorapidity of the two leading $p^*_\pperp$ jets
($\Delta \eta = \eta_1-\eta_2$), the variables $p^*_{\pperp 1}, X'_2$  and the
two three-jet angles $\cos\theta'$ and $\cos\psi'$. RAPGAP
fails to describe the $x$
and $\Delta \eta$ distributions. CDM on the other hand gives a very good  description
of almost all observables besides $p^*_{\pperp 1}$, where
it predicts too many high momentum jets (${p^*_\pperp}_1>15\;{\rm GeV}$).
The three-jet angular distribution $\cos\psi'$ is also described
rather poorly by both RAPGAP and CDM.
A separate check of the cross sections for $p^*_{\pperp 1} > 20$
GeV reveals significant deviations to CDM for the shapes of various distributions, especially for $x$
and $\eta_1$, as presented in figure \ref{threejet20}. 
In this domain RAPGAP describes these distributions well.

\subsection{Cross Sections for Events with at Least Four Jets}
\begin{figure}[htbp]
\begin{center}
\includegraphics[width=0.45\linewidth,keepaspectratio]{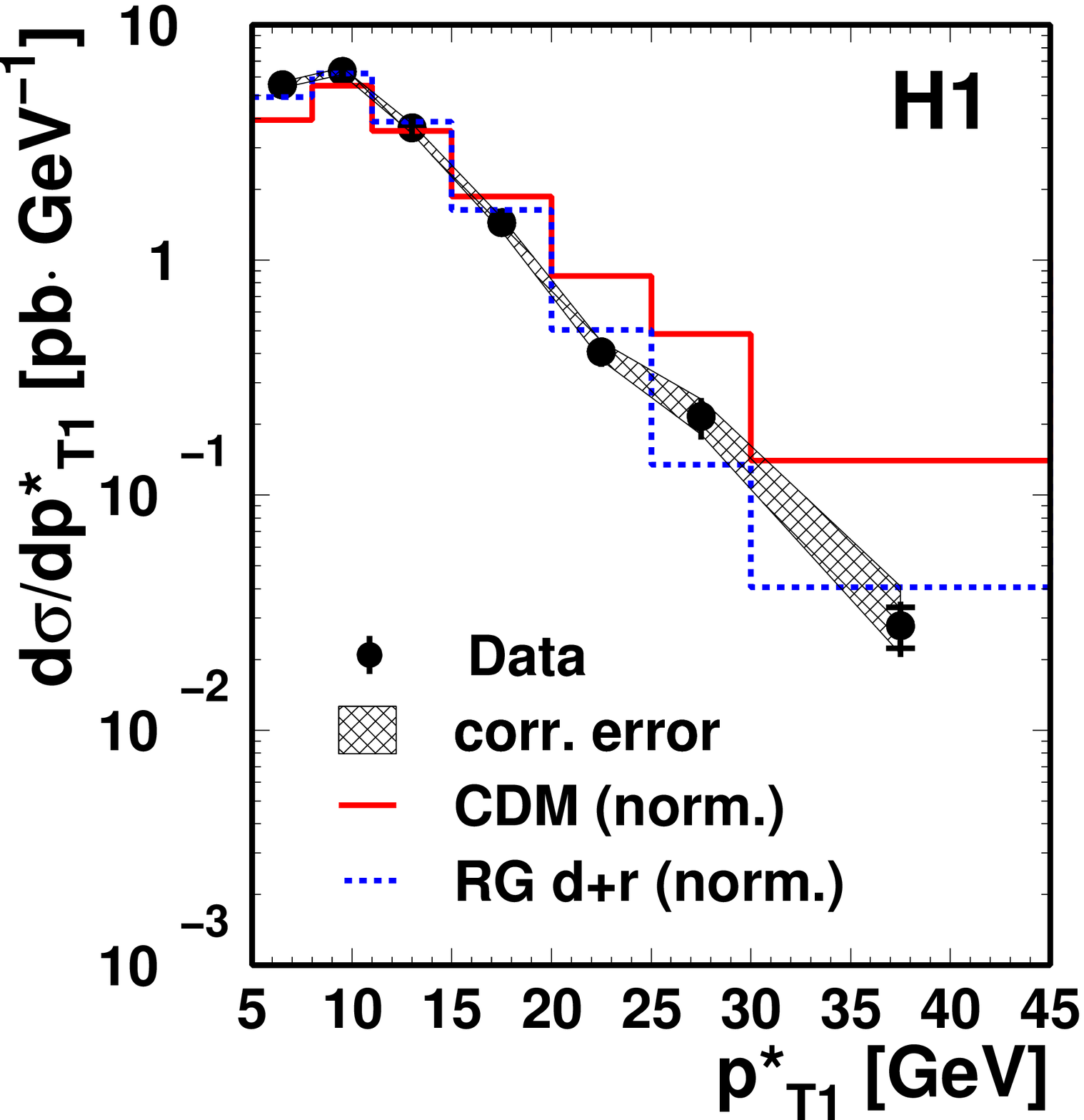}
\hspace*{5mm}
\includegraphics[width=0.413\linewidth,keepaspectratio]{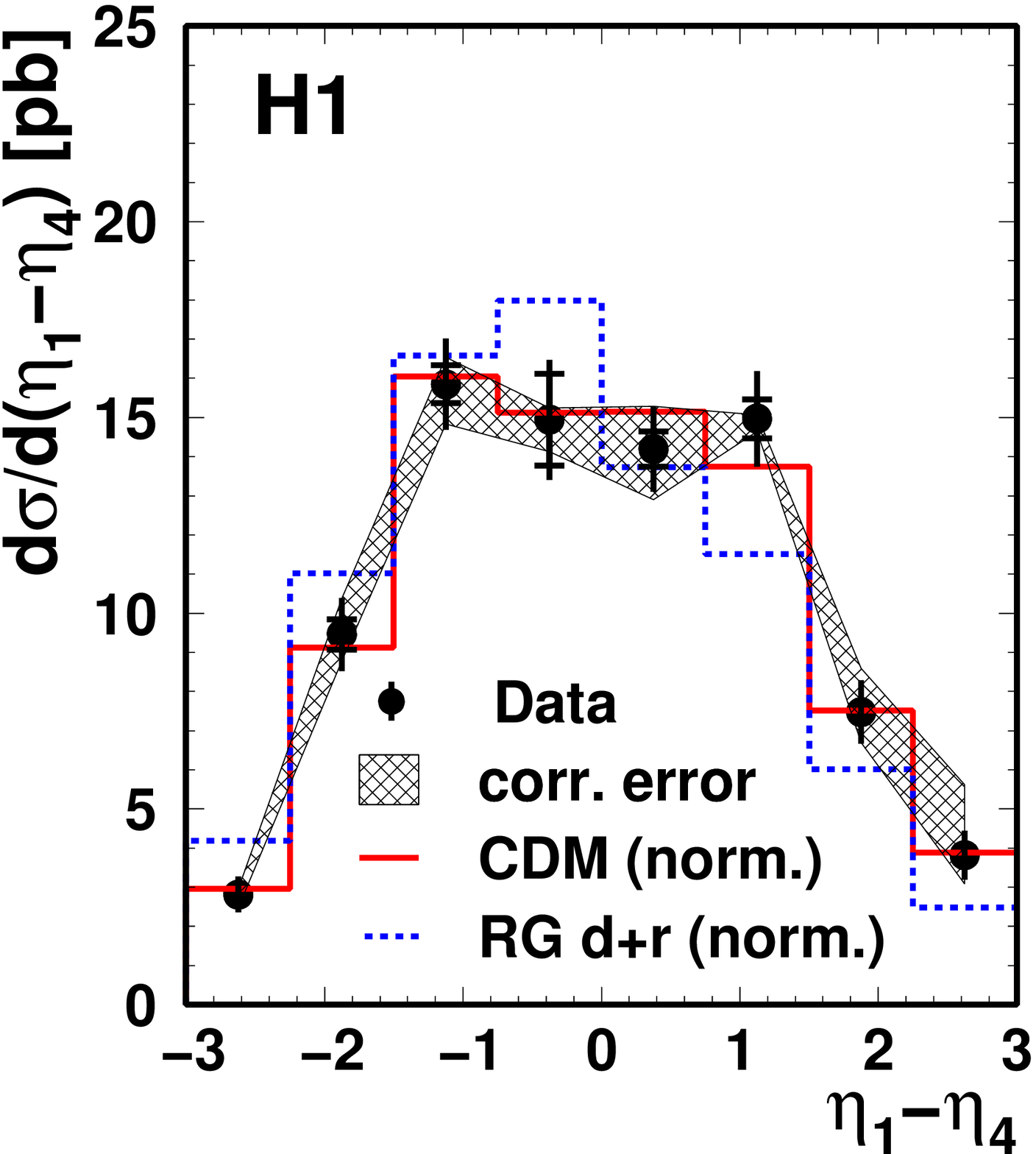}
\end{center}
\vspace*{3.5mm}
\begin{center}
\hspace*{14mm}
\includegraphics[bb=100 81 490 540,width=53mm,keepaspectratio]{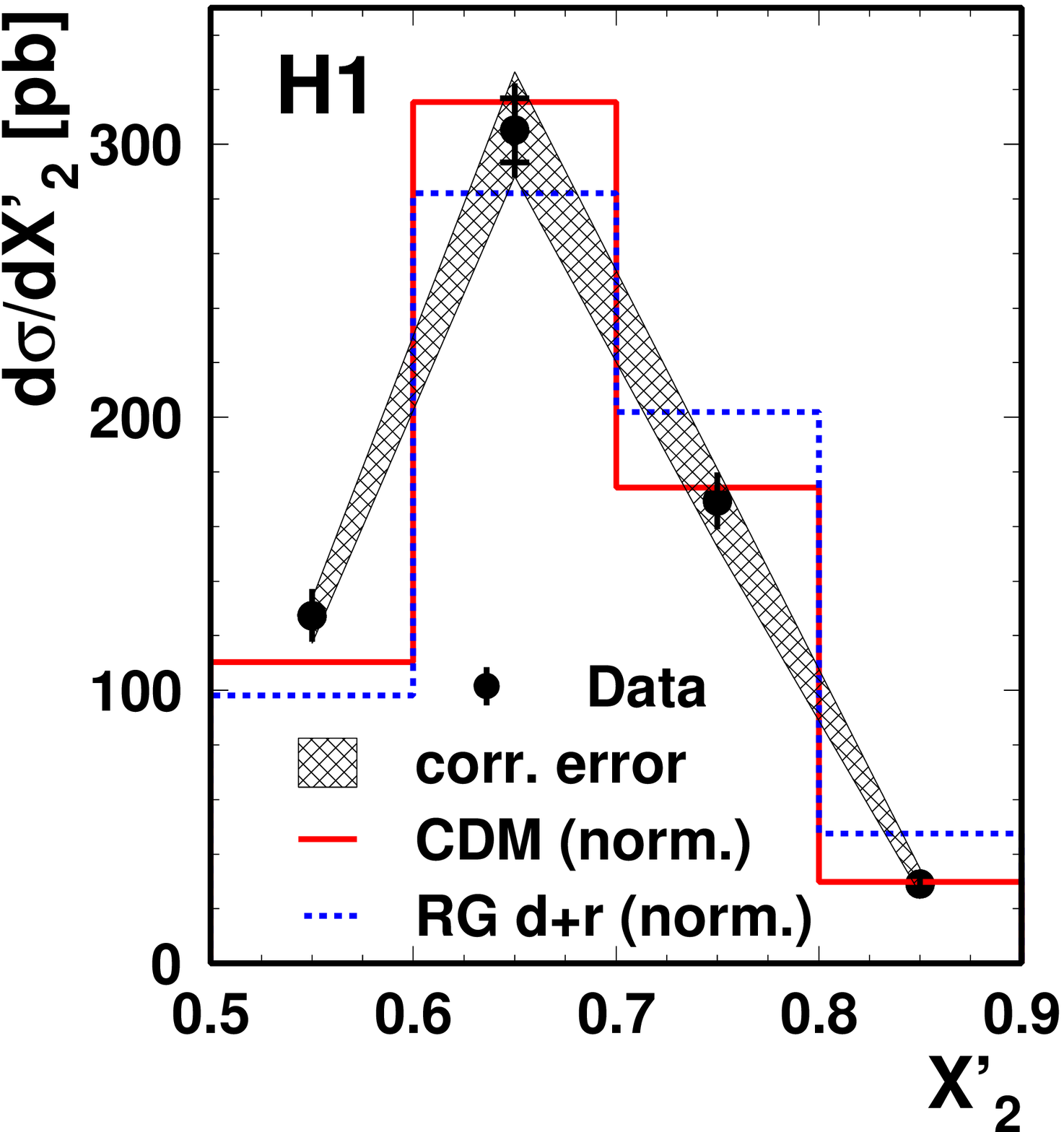}
\hspace*{20mm}
\includegraphics[bb=86 72 476 531,width=53mm,keepaspectratio]{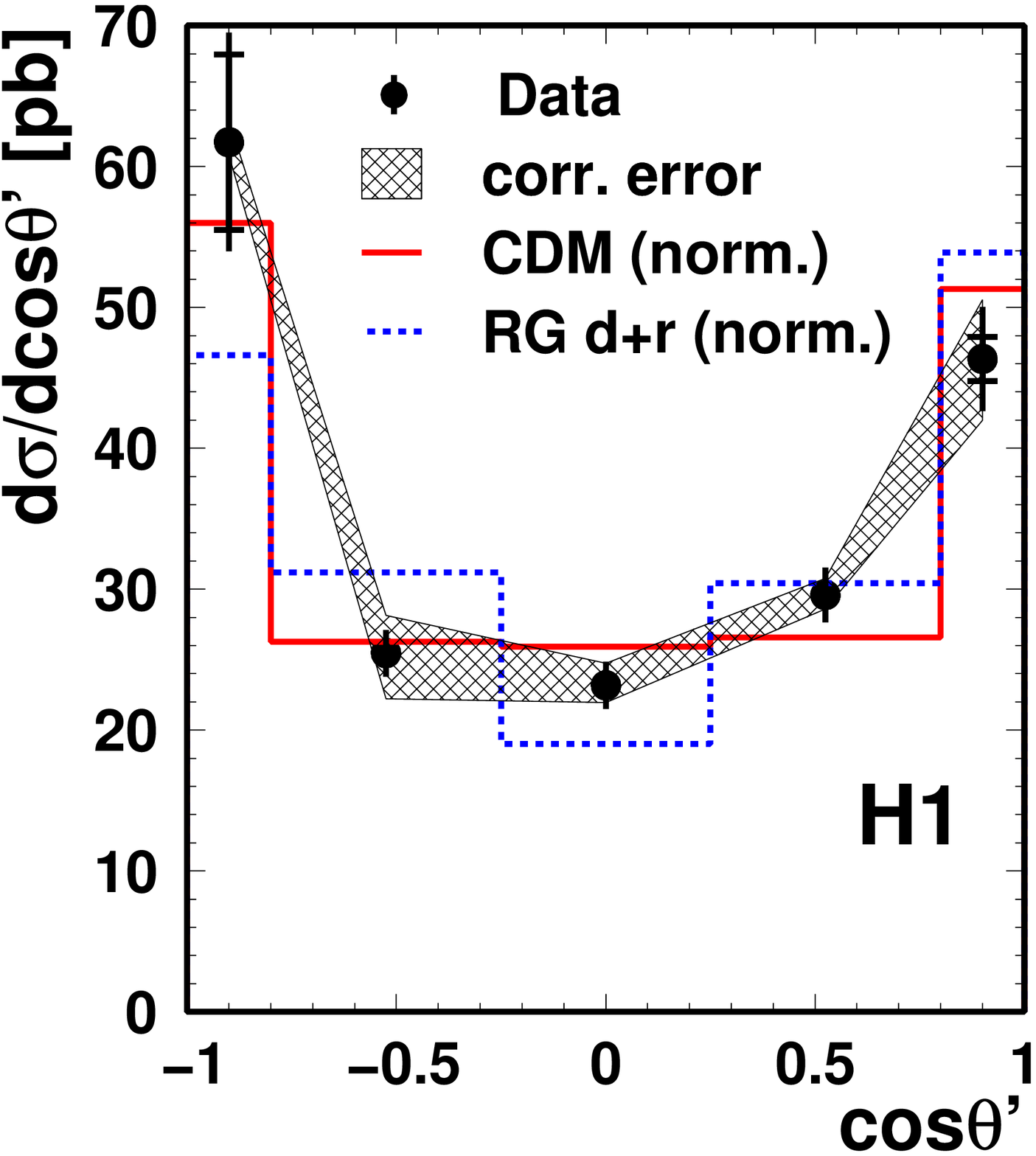}
\end{center}
\vspace*{11mm}
\caption{Differential cross sections for events with at least four jets as a function of the transverse momentum
$p^*_{\pperp 1}$ of the
leading jet, the pseudorapidity difference of the leading and the fourth
 jet $\eta_1-\eta_4$ 
(with ${p^*_\pperp}_1>{p^*_\pperp}_2>{p^*_\pperp}_3>{p^*_\pperp}_4$),
the scaled energy $X_2'$
and the angle $\theta'$ as defined in \mbox{figure \ref{Tevatronvar}.}
For the determination of $X'_2$ and
$\theta'$ the four jets are reduced to three by combining the two jets with the lowest
dijet mass. The resulting three jets (labelled with $''$) are then sorted 
with respect to their energy in
the three-jet centre of mass frame ($E''_1>E''_2>E''_3$) and both variables are calculated as 
for the three-jet case, for instance  $X_2'= \frac {E''_2}{E''_1+E''_2 + E''_3}$.
%xxx
For the details of the plotting of the data see 
the caption of figure \ref{fig:lo}.
%The cross sections are bin-averaged and plotted at the respective bin centers.
%The inner error bars represent the statistical error of the data,
%the total error bars correspond to the statistical and uncorrelated
%systematic errors added in quadrature.
%The correlated systematic errors are
%shown by the hatched error band. 
The additional global normalisation error of the data ($^{+22}_{-19}\%$) is not displayed.
The data are 
compared to the RAPGAP and CDM predictions. 
Both Monte Carlo predictions are normalised to
the total cross section of the data.}
%For further details see the caption of figure \ref{fig:lo}.}
\label{fourjet}
\end{figure}

A subsample of events with four or more jets
is also studied.
All selection criteria of the 
three-jet sample listed in table \ref{tab:xsdef}
have to be fulfilled.
In addition, at least one more jet has to be found which satisfies the
standard jet cuts 
$p_{\pperp}^* > 4\ {\rm GeV}$ and 
$-1 < \eta < 2.5$.
As already mentioned 
the NLOJET++ $\mathcal{O}(\alpha_{\rm s}^3)$
calculation can only provide 
a LO prediction for the final state with four jets 
which is by far too low as can be seen in
figure~\ref{fig:njet} (left).
Thus in the following the comparisons of the measured four jet cross sections
are restricted to the CDM and RAPGAP predictions,
where parton showers approximate higher orders and 
can lead to large jet multiplicities.
The total cross section 
predicted by CDM 
for events with four or more jets agrees well with 
the data while RAPGAP
is too low by a factor of $\sim 2.9$, 
as shown in figure~\ref{fig:njet} (left).
Differential cross sections as a function of
${p_\pperp^*}_1$, $\eta_1-\eta_4$, $X'_2$ and $\cos\theta'$
for events with at least four jets are shown in figure \ref{fourjet} 
and compared to the predictions by 
the two Monte Carlo generators normalised to the data.
RAPGAP fails to describe the shapes of the differential distributions,
again with the exception of the momentum
distributions of the jets. 
CDM on the other hand
disagrees with the data in  the $p^*_\pperp$ distributions but 
describes the scaled energies of the four jets correctly.
It also gives a very good description of all other distributions.

\section{Summary}
This paper presents a new measurement of three-jet production
in DIS at low $x$ and $Q^2$.
The measurement is carried out in an extended kinematic 
phase space covering lower jet transverse momenta compared to previous
three-jet analyses.
Very small $x$ values are reached down to $x=10^{-4}$.
The first measurement of four-jet production in DIS 
is also presented.
Three- and four-jet final states require the radiation
of at least one respectively two hard gluons from the initial state proton, 
in addition to the
$q\bar{q}$ pair from the dominating hard boson-gluon-fusion
scattering process $\gamma^{*}g \rightarrow q\bar{q}$, and are 
therefore well suited
to study parton dynamics at small $x$.

The measurements are compared with the 
NLOJET++ \cite{NLOjet} fixed order 
QCD calculations.
A remarkable result of the present analysis is the success of 
the next-to-leading order 
${\mathcal{O}}(\alpha_{\rm s}^3)$
calculation for the cross sections of events with at least three jets.
The inclusion of diagrams with
two radiated gluons improves dramatically the agreement with the data compared to the ${\mathcal{O}}(\alpha_{\rm s}^2)$
prediction which is far too low
especially at small $x$.
A similar improvement was already noticed in the previous H1 three-jet
analysis \cite{wobisch}.
In the present analysis, extending to lower invariant three-jet masses, 
an excess is observed of the data compared to the ${\mathcal{O}}(\alpha_{\rm s}^3)$
prediction at the lowest $x \sim 10^{-4}$.
This excess is 
found to be enhanced and to become highly significant for topologies with
two forward jets and one central jet.
Excesses which are probably related, albeit less significant, were observed in the forward jet analyses
from H1 \cite{Aktas:2005up} and ZEUS \cite{Chekanov:2007pa}.
The new analysis corroborates the hypothesis that the DGLAP leading log $Q^2$ approximation
starts to break down in the region of the lowest accessible $x \sim 10^{-4}$ for
$Q^2\ge4\,{\rm GeV}^2$, at least up to the order ${\mathcal{O}}(\alpha_{\rm s}^3)$
for which calculations are presently available.
In other words in a sizable fraction of events, 
which is much larger than predicted, two or more gluons are radiated from the
initial state proton which are unordered in their transverse momentum, i.e. they all have 
relatively large transverse momenta.
For events with at least four jets the 
${\mathcal{O}}(\alpha_{\rm s}^3)$ prediction is also too low, as expected, since the calculation can provide only a leading order estimate.

The new data presented here are also compared with two Monte Carlo simulation programs
implementing hard QCD 2$\rightarrow$2 processes 
complemented by parton showers modelling higher order effects.
The RAPGAP \cite{RAPGAP} program, using $k_\pperp$ ordered parton showers 
and including resolved photon processes,
fails to describe the data.
On the other hand the DJANGO \cite{DJANGOH} program with non $k_\pperp$-ordered gluon radiation as implemented
in the colour dipole model (CDM) 
gives a remarkably good description of the measured cross sections for events with at least
three and four jets and even for events with
higher jet multiplicities. 
The remaining discrepancies at high $p^*_\pperp$
require further studies.

The three- and four-jet production is investigated further 
by a detailed study of the jet topologies 
as represented by three-jet angles and scaled momenta.
The best description for the case of three-jet production 
is obtained by the ${\mathcal{O}}(\alpha_{\rm s}^3)$
NLOJET++ calculation, significantly better than that provided by CDM.
%

%%%%%%%%%%%%%%%%%%%%%%%%%%%%%%%%%%%%%%%%%%%%%%%%%%%%%%%%%%%%%%%%%%%%%%%%%%%%%%%%%%%%%
%\input{acknowledgements}
%%%%%%%%%%%%%%%%%%%%%%%%%%%%%%%%%%%%%%%%%%%%%%%%%%%%%%%%%%%%%%%%%%%%%%%%%%%%%%%%%%%%%
\section*{Acknowledgements}
We are grateful to the HERA machine group whose outstanding
efforts have made this experiment possible.
We thank
the engineers and technicians for their work in constructing and 
maintaining the H1 detector, our funding agencies for
financial support, the
DESY technical staff for continual assistance,
and the DESY directorate for the
hospitality which they extend to the non DESY
members of the collaboration. We thank J. Bartels for useful discussions. 

%%%%%%%%%%%%%%%%%%%%%%%%%%%%%%%%%%%%%%%%%%%%%%%%%%%%%%%%%%%%%%%%%%%%%%%%%%%%%%%%%%%%%
%\input{bibliography}
%%%%%%%%%%%%%%%%%%%%%%%%%%%%%%%%%%%%%%%%%%%%%%%%%%%%%%%%%%%%%%%%%%%%%%%%%%%%%%%%%%%%%

\newpage
%%%%%%%%%%%%%%%%%%%%%%%%%%%%%%%%%%%%%%%%%%%%%%%%%%%%%%%%%%%%%%%%%%%%%%%%%%%%%%%%%%%%%
%\input{figures}
%%%%%%%%%%%%%%%%%%%%%%%%%%%%%%%%%%%%%%%%%%%%%%%%%%%%%%%%%%%%%%%%%%%%%%%%%%%%%%%%%%%%%


\newpage
\clearpage
\begin{appendix}
\section{Cross Section Measurement Tables}
%
%%%%%%%%%%%%%%%%%%%%%%%%
%
% Cross Sections for three jet events
%
%%%%%%%%%%%%%%%%%%%%%%%%
%
%
%$\cos\psi'
%$N_{\rm Jet}$
%

\begin{table}[h]
\setlength{\tabcolsep}{0.2cm}
\renewcommand{\arraystretch}{1.25}
%\renewcommand{\arraystretch}{1.15}
%\scriptsize
\begin{center}
{\textbf{Cross sections for events with at least three jets}}
\vglue1em
\footnotesize
\begin{tabular}{|c|c|c|c|c|c|}
\hline
$N_{\rm Jet}$ & ${\rm d}\sigma/{\rm d} N_{\rm Jet}$  & $\delta_{\rm stat}$ & $\delta_{\rm uncorr}$ & 
$\delta_{\rm corr}$ & $c_{\rm had}$ \\
                                 &        (pb)    &  (\%)   &  (\%) &  (\%) &  \\
\hline
$3$  &   $4.8\cdot 10^2$  &      $1$    & $1$    & $^{+2}_{-3}$     & $0.71\pm 0.07$ \\
$4$  &   $5.6\cdot 10^1$  &      $2$    & $3$    & $^{+15}_{-14}$   & $0.65\pm 0.03$ \\
$5$  &   $6.0$   &               $4$    & $10$   & $^{+9}_{-6}$     & $0.70\pm 0.10$ \\
$6$  &   $0.25$  &               $11$   & $36$   & $^{+55}_{-52}$   & $0.64\pm 0.30$ \\
\hline
$x$  & ${\rm d}\sigma/{\rm d} x$  & $\delta_{\rm stat}$ & $\delta_{\rm uncorr}$ & 
$\delta_{\rm corr}$ & $c_{\rm had}$ \\
                                 &        (pb)    &  (\%)   &  (\%) &  (\%) &  \\
\hline
$[0.0001,0.0002)$  & $1.3\cdot 10^6$   &  $2$  & $3$  & $^{+10}_{-12}$ & $0.74\pm 0.08$ \\
$[0.0002,0.0005)$  & $7.1\cdot 10^5$   &  $1$  & $2$  & $^{+3}_{-4}$ & $0.70\pm 0.08$ \\
$[0.0005,0.001)$   & $2.3\cdot 10^5$   &  $1$  & $2$  & $^{+5}_{-4}$ & $0.69\pm 0.07$ \\
$[0.001,0.002)$    & $6.2\cdot 10^4$   &  $2$  & $2$  & $^{+12}_{-10}$ & $0.66\pm 0.07$ \\
$[0.002,0.005]$    & $7.7\cdot 10^3$   &  $2$  & $3$  & $^{+23}_{-16}$ & $0.65\pm 0.04$ \\

\hline
\end{tabular}
\end{center}
\caption{
Bin-averaged differential cross sections for the selected events with at least
three jets in the kinematic range listed in table~\ref{tab:xsdef}.
The cross sections are defined at the hadron level and are given
as a function of the jet multiplicity $N_{\rm Jet}$ and the 
Bj{\o}rken scaling variable $x$. 
The following cross section uncertainties are indicated: statistical ($\delta_{\rm stat}$), uncorrelated systematic
$\delta_{\rm uncorr}$ and correlated systematic ($\delta_{\rm corr}$).
The additional global normalisation uncertainty of 
$^{+16}_{-14}$\% is not included in the table.
The correction factors $c_{\rm had}$ 
for the effect of hadronisation and the associated uncertainty are also given.
They are applied to the NLOJET++ parton level calculations.
}
\label{tab:3jet_1}
\end{table}

\begin{table}[h]
\setlength{\tabcolsep}{0.2cm}
\renewcommand{\arraystretch}{1.3}
\begin{center}
{\textbf{Cross sections for events with at least three jets}}
\vglue1em
\footnotesize
\begin{tabular}{|c|c|c|c|c|c|}
\hline
$\eta_1$  & ${\rm d}\sigma/{\rm d}\eta_1$  & $\delta_{\rm stat}$ & $\delta_{\rm uncorr}$ & 
$\delta_{\rm corr} $ & $c_{\rm had}$ \\
                                 &        (pb)    &  (\%)   &  (\%) &  (\%) &  \\
\hline
$[1.0,-0.5)$  & $78$   &   $2$ & $3$ & $^{+11}_{-8}$    & $0.64\pm 0.09$ \\
$[-0.5,0.0)$  & $150$   &  $2$ & $3$ & $^{+9}_{-7}$     & $0.69\pm 0.08$ \\
$[0.0,0.5)$   & $191$   &  $1$ & $2$ & $^{+2}_{-2}$     & $0.70\pm 0.09$ \\
$[0.5,1.0)$   & $204$  &   $1$ & $2$ & $^{+3}_{-3}$     & $0.74\pm 0.06$ \\
$[1.0,1.5)$   & $187$  &   $2$ & $2$  & $^{+2}_{-3}$    & $0.70\pm 0.07$ \\
$[1.5,2.0)$   & $136$  &   $2$ & $3$  & $^{+4}_{-5}$    & $0.69\pm 0.05$ \\
$[2.0,2.5]$   & $139$  &   $2$ & $3$  & $^{+4}_{-5}$    & $0.72\pm 0.07$ \\
\hline
$\eta_2$  & ${\rm d}\sigma/{\rm d}\eta_2$  & $\delta_{\rm stat}$ & $\delta_{\rm uncorr}$ & 
$\delta_{\rm corr}$ & $c_{\rm had}$ \\
                                 &        (pb)    &  (\%)   &  (\%) &  (\%) &  \\
\hline
$[1.0,-0.5)$  & $105$  &   $2$ & $3$ & $^{+4}_{-4}$     & $0.62\pm 0.08$ \\
$[-0.5,0.0)$  & $157$  &   $2$ & $3$ & $^{+6}_{-7}$     & $0.67\pm 0.09$ \\
$[0.0,0.5)$   & $188$  &   $1$ & $2$ & $^{+4}_{-5}$     & $0.73\pm 0.09$ \\
$[0.5,1.0)$   & $193$  &   $1$ & $2$ & $^{+3}_{-4}$     & $0.72\pm 0.05$ \\
$[1.0,1.5)$   & $179$  &   $1$ & $3$  & $^{+4}_{-3}$    & $0.72\pm 0.06$ \\
$[1.5,2.0)$   & $128$  &   $2$ & $3$  & $^{+6}_{-4}$    & $0.70\pm 0.07$ \\
$[2.0,2.5]$   & $130$  &   $2$ & $3$  & $^{+8}_{-6}$    & $0.72\pm 0.07$ \\
\hline
$\eta_3$  & ${\rm d}\sigma/{\rm d}\eta_3$  & $\delta_{\rm stat}$ & $\delta_{\rm uncorr}$ & 
$\delta_{\rm corr}$ & $c_{\rm had}$ \\
                                 &        (pb)    &  (\%)   &  (\%) &  (\%) &  \\
\hline
$[1.0,-0.5)$  & $81$   &   $2$ & $4$ & $^{+5}_{-4}$    & $0.60\pm 0.09$ \\
$[-0.5,0.0)$  & $122$  &   $2$ & $3$ & $^{+4}_{-3}$    & $0.67\pm 0.06$ \\
$[0.0,0.5)$   & $177$  &   $2$ & $3$ & $^{+3}_{-4}$    & $0.71\pm 0.06$ \\
$[0.5,1.0)$   & $186$  &   $1$ & $3$ & $^{+1}_{-1}$    & $0.74\pm 0.08$ \\
$[1.0,1.5)$   & $188$  &   $1$ & $2$  & $^{+4}_{-3}$   & $0.74\pm 0.19$ \\
$[1.5,2.0)$   & $159$  &   $1$ & $3$  & $^{+2}_{-1}$   & $0.69\pm 0.08$ \\
$[2.0,2.5]$   & $174$  &   $1$ & $2$  & $^{+3}_{-4}$   & $0.72\pm 0.04$ \\
\hline
\end{tabular}
\end{center}
\caption{
Bin-averaged differential cross sections for the selected events with at least
three jets in the kinematic range listed in table~\ref{tab:xsdef}.
The cross sections are defined at the hadron level and are given
as a function of the pseudorapidities $\eta_{i,\;i=1,2,3}$ of the three
leading jets in the lab frame.
The following cross section uncertainties are indicated: statistical ($\delta_{\rm stat}$), uncorrelated systematic
$\delta_{\rm uncorr}$ and correlated systematic ($\delta_{\rm corr}$).
The additional global normalisation uncertainty of 
$^{+16}_{-14}$\% is not included in the table.
The correction factors $c_{\rm had}$ 
for the effect of hadronisation and the associated uncertainty are also given.
They are applied to the NLOJET++ parton level calculations.
}
\label{tab:3jet_2}
\end{table}

\begin{table}
\setlength{\tabcolsep}{0.2cm}
\renewcommand{\arraystretch}{1.3}
\begin{center}
{\textbf{Cross sections for events with at least three jets}}
\vglue1em
\footnotesize
\begin{tabular}{|c|c|c|c|c|c|}
\hline
$X'_1$ & ${\rm d}\sigma/{\rm d}(X'_1)$ & $\delta_{\rm stat}$ & $\delta_{\rm uncorr}$ & 
$\delta_{\rm corr}$ & $c_{\rm had}$ \\
                                 &        (pb)    &  (\%)   &  (\%) &  (\%) &  \\
\hline
$[0.6,0.7)$  & $667$   &  $2$ & $3$  & $^{+3}_{-1}$ & $0.59\pm 0.08$ \\
$[0.7,0.8)$  & $1940$  &  $1$ & $2$  & $^{+1}_{-3}$ & $0.65\pm 0.08$ \\
$[0.8,0.9)$  & $2650$  &  $1$ & $2$  & $^{+1}_{-1}$ & $0.72\pm 0.06$ \\
$[0.9,1.0]$  & $1260$  &  $1$ & $2$ & $^{+5}_{-4}$ & $0.83\pm 0.06$ \\
\hline
$X'_2$ & ${\rm d}\sigma/{\rm d}(X'_2)$ & $\delta_{\rm stat}$ & $\delta_{\rm uncorr}$ & 
$\delta_{\rm corr}$ & $c_{\rm had}$ \\
                                 &        (pb)    &  (\%)   &  (\%) &  (\%) &  \\
\hline
$[0.5,0.6)$  & $726$   &  $2$ & $3$  & $^{+2}_{-1}$ & $0.78\pm 0.08$ \\
$[0.6,0.7)$  & $2570$  &  $2$ & $2$  & $^{+3}_{-3}$ & $0.68\pm 0.07$ \\
$[0.7,0.8)$  & $1710$  &  $1$ & $2$  & $^{+3}_{-3}$ & $0.69\pm 0.07$ \\
$[0.8,0.9)$  & $397$   &  $2$ & $4$ & $^{+9}_{-6}$ & $0.74\pm 0.04$ \\
$[0.9,1.0]$  & $11$    &  $12$ & $15$ & $^{+13}_{-12}$ & $0.70\pm 0.06$ \\
\hline
$\rm{cos\,} \theta'$ & ${\rm d}\sigma/{\rm d}(\rm{cos\,} \theta')$ & $\delta_{\rm stat}$ & $\delta_{\rm uncorr}$ & 
$\delta_{\rm corr}$ & $c_{\rm had}$ \\
                                 &        (pb)    &  (\%)   &  (\%) &  (\%) &  \\\hline
$[-1.0,-0.8)$  & $359$  &  $2$ & $3$  & $^{+5}_{-3}$ & $0.70\pm 0.06$ \\
$[-0.8,-0.4)$  & $307$  &  $2$  & $2$  & $^{+7}_{-6}$ & $0.71\pm 0.09$ \\
$[-0.4, 0.0)$  & $216$  &  $1$  & $3$  & $^{+6}_{-7}$ & $0.71\pm 0.07$ \\
$[ 0.0, 0.4)$  & $220$  &  $1$  & $3$  & $^{+10}_{-11}$ & $0.74\pm 0.04$ \\
$[ 0.4, 0.8)$  & $288$  &  $1$  & $3$  & $^{+2}_{-3}$ & $0.72\pm 0.08$ \\
$[ 0.8, 1.0]$  & $300$  &  $2$  & $3$  & $^{+6}_{-5}$ & $0.60\pm 0.08$ \\
\hline
$\rm{cos\,} \psi'$ & ${\rm d}\sigma/{\rm d}(\rm{cos\,} \psi')$ & $\delta_{\rm stat}$ & $\delta_{\rm uncorr}$ & 
$\delta_{\rm corr}$ & $c_{\rm had}$ \\
                                 &        (pb)    &  (\%)   &  (\%) &  (\%) &  \\\hline
$[-1.0,-0.8)$  & $437$  &  $2$ & $2$  & $^{+6}_{-7}$  & $0.70\pm 0.06$ \\
$[-0.8,-0.4)$  & $240$  &  $2$  & $3$  & $^{+4}_{-3}$ & $0.71\pm 0.06$ \\
$[-0.4, 0.0)$  & $221$  &  $1$  & $3$  & $^{+3}_{-3}$ & $0.72\pm 0.08$ \\
$[ 0.0, 0.4)$  & $220$  &  $1$  & $3$  & $^{+3}_{-2}$ & $0.69\pm 0.07$ \\
$[ 0.4, 0.8)$  & $236$  &  $2$  & $3$  & $^{+6}_{-5}$ & $0.70\pm 0.08$ \\
$[ 0.8, 1.0]$  & $441$  &  $1$  & $2$  & $^{+7}_{-8}$ & $0.70\pm 0.07$ \\
\hline
\end{tabular}
\end{center}
\caption{
Bin-averaged differential cross sections for the selected events with at least
three jets in the kinematic range listed in table~\ref{tab:xsdef}.
The cross sections are defined at the hadron level and are given
as a function of the three-jet topological observables 
$X'_1$, $X'_2$, $\rm{cos\,} \theta'$ and $\rm{cos\,} \psi'$
as defined in section \ref{sec:kinobs}. 
The following cross section uncertainties are indicated: statistical ($\delta_{\rm stat}$), uncorrelated systematic
$\delta_{\rm uncorr}$ and correlated systematic ($\delta_{\rm corr}$).
The additional global normalisation uncertainty of 
$^{+16}_{-14}$\% is not included in the table.
The correction factors $c_{\rm had}$ 
for the effect of hadronisation and the associated uncertainty are also given.
They are applied to the NLOJET++ parton level calculations.
}
\label{tab:3jet_3}
\end{table}

\begin{table}
\setlength{\tabcolsep}{0.2cm}
\renewcommand{\arraystretch}{1.3}
\begin{center}
{\textbf{Cross sections for events with at least three jets}}
\vglue1em
\footnotesize
\begin{tabular}{|c|c|c|c|c|c|}
\hline
${p^*_\pperp}_1$ & ${\rm d}\sigma/{\rm d}({p^*_\pperp}_1)$ & $\delta_{\rm stat}$ & $\delta_{\rm uncorr}$ & 
$\delta_{\rm corr}$ & $c_{\rm had}$ \\
                   (GeV)              &   ($\rm{GeV}^{-1}\cdot{\rm pb}$)    &  (\%)   &  (\%) &  (\%) &  \\
\hline
$[5,8)$    & $60$  &  $1$    & $2$  & $^{+2}_{-2}$ & $0.67\pm 0.09$ \\
$[8,11)$   & $56$  &  $1$    & $2$   & $^{+2}_{-2}$ & $0.69\pm 0.07$ \\
$[11,15)$  & $29$  &  $1$    & $2$   & $^{+2}_{-2}$ & $0.73\pm 0.06$ \\
$[15,20)$  & $9.5$  &  $2$    & $3$   & $^{+8}_{-8}$ & $0.77\pm 0.05$ \\
$[20,25)$  & $2.6$  &  $4$  & $5$  & $^{+8}_{-10}$ & $0.79\pm 0.01$ \\
$[25,30)$  & $0.82$  &  $7$  & $7$  & $^{+15}_{-10}$ & $0.80\pm 0.02$ \\
$[30,45)$  & $0.18$  &  $9$ & $8$  & $^{+13}_{-13}$ & $0.81\pm 0.01$ \\
$[45,60]$  & $0.016$  &  $32$ & $17$  & $^{+39}_{-26}$ & $0.80\pm 0.09$ \\
\hline
$\eta_1 - \eta_2$  & ${\rm d}\sigma/{\rm d}(\eta_1-\eta_2)$  & $\delta_{\rm stat}$ & $\delta_{\rm uncorr}$ & 
$\delta_{\rm corr}$ & $c_{\rm had}$ \\
                                 &        (pb)    &  (\%)   &  (\%) &  (\%) &  \\
\hline
$[-3.0,-2.3)$  & $25$  &  $3$ & $6$ & $^{+27}_{-16}$ & $0.70\pm 0.05$ \\
$[-2.3,-1.5)$  & $66$  &  $2$ & $4$  & $^{+20}_{-14}$ & $0.72\pm 0.07$ \\
$[-1.5,-0.8)$  & $113$ &  $1$ & $3$  & $^{+9}_{-8}$ & $0.72\pm 0.09$ \\
$[-0.8, 0.0)$  & $144$ &  $1$ & $2$  & $^{+5}_{-7}$ & $0.71\pm 0.07$ \\
$[-0.0, 0.8)$  & $146$ &  $1$ & $2$  & $^{+11}_{-12}$ & $0.69\pm 0.07$ \\
$[ 0.8, 1.5)$  & $118$ &  $1$ & $3$  & $^{+5}_{-8}$ & $0.70\pm 0.05$ \\
$[ 1.5, 2.3)$  & $76$  &  $3$ & $3$ & $^{+2}_{-3}$ & $0.68\pm 0.08$ \\
$[ 2.3, 3.0]$  & $30$  &  $3$ & $5$ & $^{+11}_{-9}$ & $0.70\pm 0.05$ \\
\hline
\end{tabular}
\end{center}
\caption{
Bin-averaged differential cross sections for the selected events with at least
three jets in the kinematic range listed in table~\ref{tab:xsdef}.
The cross sections are defined at the hadron level and are given
as a function of the transverse momentum ${p^*_\pperp}_1$ of the leading
jet in the $\gamma^*p$ centre of mass frame
and the pseudorapidity difference $\eta_1 - \eta_2$ of the
two leading jets in the lab frame.
The following cross section uncertainties are indicated: statistical ($\delta_{\rm stat}$), uncorrelated systematic
$\delta_{\rm uncorr}$ and correlated systematic ($\delta_{\rm corr}$).
The additional global normalisation uncertainty of 
$^{+16}_{-14}$\% is not included in the table.
The correction factors $c_{\rm had}$ 
for the effect of hadronisation and the associated uncertainty are also given.
They are applied to the NLOJET++ parton level calculations.
}
\label{tab:3jet_4}
\end{table}
%
%
%
%%%%%%%%%%%%%%%%%%%%%%%%%%%%%%%%%%%%%%%%%%%%%%%%%%%%%%%%%
%
%  Three-jet Events with one Forward and two Central Jets
%
%%%%%%%%%%%%%%%%%%%%%%%%%%%%%%%%%%%%%%%%%%%%%%%%%%%%%%%%%
%

\begin{table}
\setlength{\tabcolsep}{0.2cm}
\renewcommand{\arraystretch}{1.3}
\begin{center}
{\textbf{Cross sections for events with one forward jet and two central jets}}
\vglue1em
\footnotesize
\begin{tabular}{|c|c|c|c|c|c|}
\hline
$x$  & ${\rm d}\sigma/{\rm d} x$  & $\delta_{\rm stat}$ & $\delta_{\rm uncorr}$ & 
$\delta_{\rm corr}$ & $c_{\rm had}$ \\
                                 &        (pb)    &  (\%)   &  (\%) &  (\%) &  \\
\hline
$[0.0001,0.0002)$  & $1.0\cdot 10^5$   &  $5$   & $9$   & $^{+7}_{-5}$  & $0.82\pm 0.10$ \\
$[0.0002,0.0005)$  & $6.0\cdot 10^4$   &  $3$   & $6$   & $^{+2}_{-5}$  & $0.72\pm 0.07$ \\
$[0.0005,0.001)$   & $2.1\cdot 10^4$   &  $5$   & $7$   & $^{+3}_{-2}$  & $0.69\pm 0.04$ \\
$[0.001,0.002)$    & $4.9\cdot 10^3$   &  $7$   & $9$   & $^{+17}_{-13}$  & $0.66\pm 0.08$ \\
$[0.002,0.005]$    & $4.0\cdot 10^2$   &  $12$  & $16$  & $^{+14}_{-17}$  & $0.74\pm 0.10$ \\
\hline

$\eta_1$  & ${\rm d}\sigma/{\rm d}\eta_1$  & $\delta_{\rm stat}$ & $\delta_{\rm uncorr}$ & 
$\delta_{\rm corr}$ & $c_{\rm had}$ \\
                                 &        (pb)    &  (\%)   &  (\%) &  (\%) &  \\
\hline
$[-0.5,0.0)$  & $8.4$   &   $7$  & $13$  & $^{+14}_{-10}$   & $0.76\pm 0.07$ \\
$[0.0,0.5)$   & $12.9$   &  $6$  & $10$  & $^{+24}_{-15}$   & $0.77\pm 0.20$ \\
$[0.5,1.0)$   & $16.9$  &   $5$  & $8$   & $^{+5}_{-8}$     & $0.76\pm 0.03$ \\
$[1.0,1.5)$   & $0.0$  &    $0$  & $0$   & $^{+0}_{-0}$     & $0.00\pm 0.00$ \\
$[1.5,2.0)$   & $10.4$  &   $6$  & $10$  & $^{+6}_{-7}$     & $0.78\pm 0.04$ \\
$[2.0,2.5]$   & $39.0$  &   $3$  & $6$   & $^{+4}_{-7}$     & $0.72\pm 0.08$ \\
\hline
${p^*_\pperp}_1$ & ${\rm d}\sigma/{\rm d}({p^*_\pperp}_1)$ & $\delta_{\rm stat}$ & $\delta_{\rm uncorr}$ & 
$\delta_{\rm corr}$ & $c_{\rm had}$ \\
                   (GeV)              &   ($\rm{GeV}^{-1}\cdot{\rm pb}$)    &  (\%)   &  (\%) &  (\%) &  \\
\hline
$[5,8)$    & $1.9$  &  $6$    & $12$  & $^{+10}_{-8}$ & $0.70\pm 0.20$ \\
$[8,11)$   & $4.4$  &  $4$    & $7$   & $^{+4}_{-6}$ & $0.71\pm 0.05$ \\
$[11,15)$  & $3.9$  &  $4$    & $7$   & $^{+14}_{-2}$ & $0.74\pm 0.07$ \\
$[15,20)$  & $1.4$  &  $6$    & $9$   & $^{+14}_{-17}$ & $0.76\pm 0.09$ \\
$[20,25)$  & $0.36$  &  $11$  & $13$  & $^{+29}_{-34}$ & $0.82\pm 0.01$ \\
$[25,30)$  & $0.11$  &  $21$  & $19$  & $^{+18}_{-15}$ & $0.87\pm 0.10$ \\
$[30,45]$  & $0.027$  &  $29$ & $22$  & $^{+28}_{-16}$ & $0.82\pm 0.07$ \\
\hline
\end{tabular}
\end{center}
\caption{
Bin-averaged differential cross sections 
for the selected events with at least three jets  
in the kinematic range listed in table~\ref{tab:xsdef}.
In addition one of the three leading jets is required to
be a forward jet with $\theta_{\rm Jet}<20^\circ$ and to
carry a large fraction of the proton beam energy $x_{\rm Jet}>0{.}035$.
The two other jets are required to lie in the central region of the H1 detector $-1<\eta_{\rm Jet}<1$ (sample with one
forward jet and two central jets).
The cross sections are defined at the hadron level and are given
as a function of the
Bj{\o}rken scaling variable $x$ and the leading jet observables: 
pseudorapidity $\eta_1$ in the lab frame and transverse momentum ${p^*_\pperp}_1$ in the 
$\gamma^*p$ centre of mass frame.
The following cross section uncertainties are indicated: statistical ($\delta_{\rm stat}$), uncorrelated systematic
$\delta_{\rm uncorr}$ and correlated systematic ($\delta_{\rm corr}$).
The additional global normalisation uncertainty of 
$^{+18}_{-14}$\% is not included in the table.
The correction factors $c_{\rm had}$ 
for the effect of hadronisation and the associated uncertainty are also given.
They are applied to the NLOJET++ parton level calculations.
}
\label{tab:3jet_1f2c}
\end{table}
%
%
%%%%%%%%%%%%%%%%%%%%%%%%%%%%%%%%%%%%%%%%%%%%%%%%%%%%%%%%
%
%  Three-jet Events with two Forward and one Central Jet
%
%%%%%%%%%%%%%%%%%%%%%%%%%%%%%%%%%%%%%%%%%%%%%%%%%%%%%%%%
%
%
\begin{table}
\setlength{\tabcolsep}{0.2cm}
\renewcommand{\arraystretch}{1.3}
\begin{center}
{\textbf{Cross sections for events with two forward jets and one central jet}}
\vglue1em
\footnotesize
\begin{tabular}{|c|c|c|c|c|c|}
\hline
$x$  & ${\rm d}\sigma/{\rm d} x$  & $\delta_{\rm stat}$ & $\delta_{\rm uncorr}$ & 
$\delta_{\rm corr}$ & $c_{\rm had}$ \\
                                 &        (pb)    &  (\%)   &  (\%) &  (\%) &  \\
\hline
$[0.0001,0.0002)$  & $1.3\cdot 10^5$   &  $4$  & $8$  & $^{+9}_{-7}$ & $0.76\pm 0.09$ \\
$[0.0002,0.0005)$  & $8.9\cdot 10^4$   &  $3$  & $5$  & $^{+7}_{-9}$ & $0.77\pm 0.03$ \\
$[0.0005,0.001)$   & $3.0\cdot 10^4$   &  $3$  & $5$  & $^{+8}_{-6}$ & $0.72\pm 0.05$ \\
$[0.001,0.002)$    & $9.3\cdot 10^3$   &  $4$  & $6$ & $^{+11}_{-10}$ & $0.70\pm 0.06$ \\
$[0.002,0.005]$    & $1.6\cdot 10^3$   &  $6$  & $8$  & $^{+13}_{-13}$ & $0.69\pm 0.04$ \\
\hline
$\eta_1$  & ${\rm d}\sigma/{\rm d}\eta_1$  & $\delta_{\rm stat}$ & $\delta_{\rm uncorr}$ & 
$\delta_{\rm corr}$ & $c_{\rm had}$ \\
                                 &        (pb)    &  (\%)   &  (\%) &  (\%) &  \\
\hline
$[1.0,-0.5)$  & $2.9$   &  $11$ & $20$ & $^{+29}_{-22}$  & $0.70\pm 0.01$ \\
$[-0.5,0.0)$  & $7.7$   &   $7$ & $14$ & $^{+6}_{-9}$    & $0.82\pm 0.01$ \\
$[0.0,0.5)$   & $9.3$   &   $6$ & $12$ & $^{+27}_{-17}$  & $0.92\pm 0.01$ \\
$[0.5,1.0)$   & $10.4$  &   $5$ & $11$ & $^{+10}_{-9}$   & $0.88\pm 0.04$ \\
$[1.0,1.5)$   & $15.3$  &   $5$ & $8$  & $^{+6}_{-5}$    & $0.72\pm 0.10$ \\
$[1.5,2.0)$   & $30.0$  &   $4$ & $6$  & $^{+5}_{-7}$    & $0.71\pm 0.04$ \\
$[2.0,2.5]$   & $62.6$  &   $2$ & $4$  & $^{+3}_{-4}$    & $0.73\pm 0.05$ \\
\hline
${p^*_\pperp}_1$ & ${\rm d}\sigma/{\rm d}({p^*_\pperp}_1)$ & $\delta_{\rm stat}$ & $\delta_{\rm uncorr}$ & 
$\delta_{\rm corr}$ & $c_{\rm had}$ \\
                   (GeV)              &   ($\rm{GeV}^{-1}\cdot{\rm pb}$)    &  (\%)   &  (\%) &  (\%) &  \\\hline
$[5,8)$    & $2.8$  &  $4$ & $10$     & $^{+10}_{-10}$ & $0.72\pm 0.05$ \\
$[8,11)$   & $6.4$  &  $3$ & $6$      & $^{+4}_{-4}$ & $0.68\pm 0.05$ \\
$[11,15)$  & $5.4$  &  $3$ & $5$      & $^{+3}_{-5}$ & $0.76\pm 0.06$ \\
$[15,20)$  & $2.4$  &  $4$ & $6$      & $^{+8}_{-2}$ & $0.78\pm 0.04$ \\
$[20,25)$  & $0.8$  &  $7$ & $9$      & $^{+8}_{-7}$ & $0.82\pm 0.09$ \\
$[25,30)$  & $0.28$  &  $11$ & $11$   & $^{+5}_{-11}$ & $0.79\pm 0.01$ \\
$[30,45]$  & $0.065$  &  $14$ & $11$  & $^{+11}_{-10}$ & $0.78\pm 0.01$ \\
$[45,60]$  & $0.006$  &  $60$ & $22$  & $^{+57}_{-32}$ & $0.86\pm 0.30$ \\
\hline
\end{tabular}
\end{center}
\caption{
Bin-averaged differential cross sections 
for the selected events with at least three jets  
in the kinematic range listed in table~\ref{tab:xsdef}.
In addition one of the three leading jets is required to
be a forward jet with $\theta_{\rm Jet}<20^\circ$ and to
carry a large fraction of the proton beam energy $x_{\rm Jet}>0{.}035$.
One of the two other jets has to fulfil $\eta_{\rm Jet}>1$ and the other 
 $-1<\eta_{\rm Jet}<1$ 
(sample with two forward jets and one central jet).
The cross sections are defined at the hadron level and are given
as a function of the
Bj{\o}rken scaling variable $x$ and the leading jet observables: 
pseudorapidity $\eta_1$ in the lab frame and transverse momentum ${p^*_\pperp}_1$ in the 
$\gamma^*p$ centre of mass frame.
The following cross section uncertainties are indicated: statistical ($\delta_{\rm stat}$), uncorrelated systematic
$\delta_{\rm uncorr}$ and correlated systematic ($\delta_{\rm corr}$).
The additional global normalisation uncertainty of 
$^{+19}_{-14}$\% is not included in the table.
The correction factors $c_{\rm had}$ 
for the effect of hadronisation and the associated uncertainty are also given.
They are applied to the NLOJET++ parton level calculations.
}
\label{tab:3jet_2f1c}
\end{table}
%
%%%%%%%%%%%%%%%%%%%%%%%%%%%%%%%%%%%%%%
%
% Three jets with one high pt jet
%
%%%%%%%%%%%%%%%%%%%%%%%%%%%%%%%%%%%%%%
%
%
\begin{table}
\setlength{\tabcolsep}{0.2cm}
\renewcommand{\arraystretch}{1.3}
\begin{center}
{\textbf{ Cross sections for events with at least three jets and $ {p^*_\pperp}_1>20\;{\rm \bf GeV}$ }}
\vglue1em
\footnotesize
\begin{tabular}{|c|c|c|c|c|c|}
\hline
$\eta_1$  & ${\rm d}\sigma/{\rm d}\eta_1$  & $\delta_{\rm stat}$ & $\delta_{\rm uncorr}$ & 
$\delta_{\rm corr}$ & $c_{\rm had}$ \\
                                 &        (pb)    &  (\%)   &  (\%) &  (\%) &  \\
\hline
$[-0.5,0.0)$  & $1.4$   &  $18$ & $33$ & $^{+30}_{-15}$ & $0.71\pm 0.40$ \\
$[ 0.0,0.5)$  & $4.6$   &  $10$ & $13$ & $^{+13}_{-11}$ & $0.88\pm 0.40$ \\
$[ 0.5,1.0)$  & $8.2$   &  $7$ & $8$ & $^{+7}_{-7}$     & $0.76\pm 0.40$ \\
$[ 1.0,1.5)$  & $10.3$  &  $6$ & $7$ & $^{+7}_{-7}$     & $0.80\pm 0.40$ \\
$[ 1.5,2.0)$  & $9.0$   &  $7$ & $7$ & $^{+10}_{-8}$    & $0.80\pm 0.40$ \\
$[ 2.0,2.5]$  & $7.1$   &  $7$ & $8$ & $^{+7}_{-7}$     & $0.79\pm 0.40$ \\
\hline
$x$  & ${\rm d}\sigma/{\rm d} x$  & $\delta_{\rm stat}$ & $\delta_{\rm uncorr}$ & 
$\delta_{\rm corr}$ & $c_{\rm had}$ \\
\hline
                                 &        (pb)    &  (\%)   &  (\%) &  (\%) &  \\
$[0.0001,0.0002)$  & $4.6\cdot 10^4$   &  $7$  & $7$  & $^{+8}_{-4}$ & $0.82\pm 0.40$ \\
$[0.0002,0.0005)$  & $2.5\cdot 10^4$   &  $5$  & $5$  & $^{+2}_{-3}$ & $0.77\pm 0.40$ \\
$[0.0005,0.001)$   & $9.6\cdot 10^3$   &  $7$  & $6$  & $^{+3}_{-7}$ & $0.81\pm 0.40$ \\
$[0.001,0.002)$    & $2.4\cdot 10^3$   &  $9$  & $10$ & $^{+8}_{-6}$ & $0.81\pm 0.40$ \\
$[0.002,0.005]$    & $2.0\cdot 10^2$   &  $17$ & $20$  & $^{+23}_{-26}$ & $0.83\pm 0.40$ \\
\hline
\end{tabular}
\end{center}
\caption{
Bin-averaged differential cross sections at the hadron level
as a function of the Bj{\o}rken scaling variable $x$ and the
pseudorapidity $\eta_1$ of the leading jet in the lab frame,
The cross sections are measured in the kinematic range listed in table~\ref{tab:xsdef}.
In addition the leading jet is required to have 
a large transverse momentum ${p^*_\pperp}_1>20\;\mbox{GeV}$ 
in the $\gamma^*p$ centre of mass frame.
The following cross section uncertainties are given: statistical ($\delta_{\rm stat}$), uncorrelated systematic
$\delta_{\rm uncorr}$ and correlated systematic ($\delta_{\rm corr}$).
The additional global normalisation uncertainty of 
$^{+19}_{-14}$\% is not included in the table.
The correction factors $c_{\rm had}$ 
for the effect of hadronisation and the associated uncertainty are also given.
They are applied to the NLOJET++ parton level calculations.
}
\label{tab:3jet_highpt}
\end{table}
%
%\hline
%\hline
%\label{ lab.tab.xbjhpt }
%\end{table}
%
%%%%%%%%%%%%%%%%%%%%
%
%  Four Jet xsecs:
%
%%%%%%%%%%%%%%%%%%%%
%
\setlength{\tabcolsep}{0.2cm}
\begin{table}
\renewcommand{\arraystretch}{1.3}
\begin{center}
{\textbf{Cross sections for events with at least four jets}}
\vglue1em
\footnotesize
\begin{tabular}{|c|c|c|c|c|}
\hline
$\eta_1 - \eta_2$  & ${\rm d}\sigma/{\rm d}(\eta_1-\eta_2)$  & $\delta_{\rm stat}$ & $\delta_{\rm uncorr}$ & $\delta_{\rm corr}$  \\
                                 &        (pb)    &  (\%)   &  (\%)  &  (\%) \\
\hline
$[-3.0,-2.3)$  & $2.8$  &  $7$ & $15$ & $^{+5}_{-12}$   \\
$[-2.3,-1.5)$  & $9.5$  &  $4$ & $9$  & $^{+10}_{-7}$  \\
$[-1.5,-0.8)$  & $15.8$ &  $3$ & $7$  & $^{+4}_{-6}$  \\
$[-0.8, 0.0)$  & $14.9$ &  $8$ & $7$  & $^{+2}_{-5}$  \\
$[-0.0, 0.8)$  & $14.2$ &  $3$ & $7$  & $^{+8}_{-9}$  \\
$[ 0.8, 1.5)$  & $15.0$ &  $3$ & $7$  & $^{+1}_{-2}$  \\
$[ 1.5, 2.3)$  & $7.5$  &  $5$ & $10$ & $^{+15}_{-10}$ \\
$[ 2.3, 3.0]$  & $3.8$  &  $7$ & $15$ & $^{+47}_{-19}$ \\
\hline
$X'_2$ & ${\rm d}\sigma/{\rm d}(X'_2)$ & $\delta_{\rm stat}$ & $\delta_{\rm uncorr}$ & 
$\delta_{\rm corr}$  \\
                                 &        (pb)    &  (\%)   &  (\%)  &  (\%) \\
\hline
$[0.5,0.6)$  & $127$  &  $3$ & $7$  & $^{+5}_{-8}$ \\
$[0.6,0.7)$  & $305$  &  $4$ & $4$  & $^{+7}_{-6}$ \\
$[0.7,0.8)$  & $169$  &  $3$ & $6$  & $^{+7}_{-10}$ \\
$[0.8,0.9]$  & $ 29$  &  $6$ & $13$ & $^{+20}_{-12}$ \\
\hline
$\rm{cos\,} \theta'$ & ${\rm d}\sigma/{\rm d}(\rm{cos\,} \theta')$ & $\delta_{\rm stat}$ & $\delta_{\rm uncorr}$ & 
$\delta_{\rm corr}$ \\
                                 &        (pb)    &  (\%)   &  (\%) &  (\%)   \\ \hline
$[-1.0,-0.8)$  & $61.7$  &  $10$ & $8$  & $^{+3}_{-2}$ \\
$[-0.8,-0.3)$  & $25.5$  &  $3$  & $6$  & $^{+10}_{-13}$ \\
$[-0.3, 0.3)$  & $23.2$  &  $3$  & $6$  & $^{+7}_{-5}$  \\
$[ 0.3, 0.8)$  & $29.6$  &  $3$  & $6$  & $^{+4}_{-3}$  \\
$[ 0.8, 1.0]$  & $46.3$  &  $3$  & $7$  & $^{+9}_{-9}$ \\
\hline
${p^*_\pperp}_1$ & ${\rm d}\sigma/{\rm d}({p^*_\pperp}_1)$ & $\delta_{\rm stat}$ & $\delta_{\rm uncorr}$ & 
$\delta_{\rm corr}$  \\
                   (GeV)              &   ($\rm{GeV}^{-1}\cdot{\rm pb}$)    &  (\%)   &  (\%) &  (\%)   \\\hline
$[5,8)$    & $5.6$  &  $5$ & $7$  & $^{+3}_{-3}$ \\
$[8,11)$   & $6.3$  &  $2$ & $6$  & $^{+4}_{-3}$ \\
$[11,15)$  & $3.7$  &  $3$ & $6$  & $^{+4}_{-6}$ \\
$[15,20)$  & $1.4$  &  $4$ & $8$  & $^{+6}_{-8}$ \\
$[20,25)$  & $0.41$  &  $8$ & $11$  & $^{+11}_{-8}$ \\
$[25,30)$  & $0.22$  &  $13$ & $16$  & $^{+19}_{-16}$ \\
$[30,45]$  & $0.028$  &  $20$ & $17$  & $^{+47}_{-24}$ \\
\hline
\end{tabular}
\end{center}
\caption{
Bin-averaged differential cross sections at the hadron level
as a function of the  pseudorapidity difference  $\eta_1-\eta_4$ of the leading and the fourth jet,
the topological observables $X'_2$, $\rm{cos\,} \theta'$ 
and the leading jet transverse momentum ${p^*_\pperp}_1$ 
in the $\gamma^*p$ centre of mass frame.
The cross sections are measured in the kinematic range listed in table~\ref{tab:xsdef}.
In addition it is required that a fourth jet 
is found in the events which fulfils the standard jet selection cuts
$p_{\pperp}^* > 4\ {\rm GeV}$ and $-1 < \eta < 2.5$.
The following cross section uncertainties are given: statistical ($\delta_{\rm stat}$), uncorrelated systematic
$\delta_{\rm uncorr}$ and correlated systematic ($\delta_{\rm corr}$).
The additional global normalisation uncertainty of 
$^{+22}_{-19}$\% is not included in the table.
}
\label{tab:4jet}
\end{table}
\end{appendix}

%\end{linenumbers}
\end{document}